\documentclass[onecolumn, numberedappendix]{aastex6}
\usepackage{graphicx}%{color}
\usepackage{epsfig}

\newcommand{\av}{$A_V$}

\newcommand{\eg}{{\it e.g.}}
\newcommand{\etal}{et~al.}

\newcommand{\ks}{$K_{\rm s}$}
\newcommand{\vmkz}{$(V-K_{\rm s})_0$}
\newcommand{\vmk}{$(V-K_{\rm s})$}

\newcommand{\msun}{M$_{\sun}$}

\newcommand{\teff}{$T_{\rm eff}$}

\newcommand{\mum}{$\mu$m}
\newcommand{\rop}{$\rho$ Oph}

\begin{document}

\title{Rotation of Low-Mass Stars in Upper Scorpius and Rho Ophiuchus with K2 }

\slugcomment{Version from \today}

\author{L.~M.~Rebull\altaffilmark{1,2},
J.~R.~Stauffer\altaffilmark{2},
A.~M.~Cody\altaffilmark{3,4},
L.~A.~Hillenbrand\altaffilmark{5}, 
T.~J.~David\altaffilmark{6,5},
M.~Pinsonneault\altaffilmark{7}
}

\altaffiltext{1}{Infrared Science Archive (IRSA), Infrared Processing
and Analysis Center (IPAC), 1200 E.\ California Blvd., California
Institute of Technology, Pasadena, CA 91125, USA; rebull@ipac.caltech.edu}
\altaffiltext{2}{Spitzer Science Center (SSC), Infrared Processing and
Analysis Center (IPAC), 1200 E.\ California Blvd., California
Institute of Technology, Pasadena, CA 9112, USA}
\altaffiltext{3}{NASA Ames Research Center, Kepler Science Office,
Mountain View, CA 94035, USA}
\altaffiltext{4}{Bay Area Environmental Research Institute, 625 2nd
St.\ Ste.\ 209, Petaluma, CA 94952, USA}
\altaffiltext{5}{Astronomy Department, California Institute of
Technology, Pasadena, CA 91125, USA}
\altaffiltext{6}{Jet Propulsion Laboratory, California Institute of
Technology, 4800 Oak Grove Drive, Pasadena, CA 91109}
\altaffiltext{7}{Astronomy Department, The Ohio State University,
Columbus, OH 43210, USA}

\begin{abstract}  

We present an analysis of K2 light curves (LCs) for candidate members
of the young Upper Sco (USco) association $\sim$8 Myr) and
the neighboring \rop\ embedded cluster ($\sim$1 Myr). We
establish $\sim$1300 stars as probable members, $\sim$80\% of which
are periodic. The phased LCs have a variety of shapes which can be
attributed to physical causes ranging from stellar pulsation and
stellar rotation to disk-related phenomena.  We identify and discuss a
number of observed behaviors. The periods are  $\sim$0.2-30 days with
a peak near 2 days and the rapid period end nearing break-up
velocity.  M stars in the young USco region rotate systematically
faster than GK stars, a pattern also present in K2 data for the older
Pleiades and Praesepe systems.  At higher masses (types FGK), the
well-defined period-color relationship for slowly rotating stars seen
in the Pleiades and Praesepe is not yet present in USco. Circumstellar
disks are present predominantly among the more slowly rotating Ms in
USco, with few disks in the sub-day rotators.  However, M dwarfs with
disks rotate faster on average than FGK systems with disks. For four
of these disked Ms, we provide direct evidence for disk-locking based
on the K2 LC morphologies.  Our preliminary analysis shows a
relatively mass-independent spin-up by a factor of $\sim$3.5 between
USco and the Pleiades, then mass-dependent spin-down between Pleiades
and Praesepe.

\end{abstract}

\section{Introduction}
\label{sec:intro}

Early empirical studies of the angular momentum evolution of
relatively low mass stars were based on spectroscopic projected
rotational velocities ($v \sin i$) of A and F stars in nearby open
clusters (e.g., Kraft 1965; Kraft 1967a,b; Abt \etal\ 1969) because
those were the stars amenable to study with photographic plates as
detectors. The striking dichotomy between high-mass rapid rotators and
low-mass slow rotators required strong angular momentum loss in Solar
analogs (Weber \& Davis 1967). Early theoretical models of angular
momentum evolution (e.g., Belcher \& MacGregor 1976), therefore,
concentrated on main sequence (MS) stars and ages $>$100 Myr. 
Subsequently, high resolution spectra using image-tube detectors of
pre-main-sequence stars (Kuhi 1978) suggested a range of rotational
velocities at few Myr ages, while rotation periods for K dwarfs in the
Pleiades (age $\sim$125 Myr) suggested a wide range in rotation on the
zero-age main sequence (ZAMS) for low mass stars (van Leeuwen \&
Alphenaar 1982).   This led both observers and theoretical modellers
to push their angular momentum studies to much younger ages.  On the
observational side, we now have rotation data for low mass, 1-2 Myr
old stars in Taurus (e.g., Bouvier \etal\ 1993, 1997a) and Orion
(e.g., Rebull 2001; Herbst \etal\ 2002).   On the theoretical side,
beginning with Endal \& Sofia (1981), theoretical models of the
angular momentum evolution of low mass stars generally began at ages
near 1 Myr and followed the evolution to the MS and beyond.  In order
to account for the wide range of rotation rates on the MS at low
masses, those models generally added extra free parameters related to
the range in lifetimes of primordial disks and  star-disk angular
momentum regulation mechanisms (e.g., Collier Cameron \etal\ 1995;
Bouvier \etal\ 1997b; Tinker \etal\ 2002); rapid rotators on the ZAMS
were ascribed to stars with very short-lived circumstellar disks
whereas the stars that arrive on the ZAMS as slow rotators were linked
to stars with the longest lived disks.

A problem with the existing angular momentum models is simply that
there are too many free parameters: observed pattern requires a model
of internal angular momentum transport, loss from magnetized
solar-like winds, and a treatment of star-disk interactions.  One way
to better confront the models would be to identify a pre-main-sequence
stellar population where the stars are still high up on their
evolutionary tracks but old enough that their disks are no longer
present -- that is, a nearby, populous, star-forming region with an
age of order 10 Myr. At such an age, whatever effects disks and
accretion have on pre-main-sequence rotation rates would have already
taken place; theoretical models that begin at such an age could have
many fewer free parameters and thus might hopefully be more amenable
to placing strong constraints on the remaining free parameters. Cool
main sequence stars (M dwarfs) are also either fully convective or
nearly so. This allows modelers of these stars to sidestep complex
issues around internal angular momentum transport, such as
core-envelope coupling timescales (Pinsonneault 1997). 

NASA's K2 mission (Howell \etal\ 2014) has recently provided high
quality, long duration, high cadence light curves (LCs) for more than
a thousand low mass members of the $\sim$8 Myr old Upper Scorpius
association.  Upper Sco (USco) provides a nearly ideal match to the
desired template post-disk pre-main sequence population needed to
better test the theoretical angular momentum evolution models.  It is
nearby ($\sim$ 140 pc), populous and at approximately the desired age,
though the precise age is likely between 3 and 10 Myr and 
still a subject of debate (see, e.g., Feiden 2016 [10 Myr],
Herczeg \& Hillebrand 2015 [5 Myr], Pecaut \etal\ 2012 [11 Myr],
Slesnick \etal\ 2008 [4 Myr], Preibisch \etal\ 2002 [5 Myr]).  Not
quite ideal is that a small fraction of the low mass members of Upper
Sco are still actively accreting and have not lost their primordial
disks (e.g., Carpenter \etal\ 2006, 2009; Cody \etal\ 2017, 2018
submitted).  Even for those stars, however, their disks will only last
a few Myr more and their rotation rate at 8 Myr will be a reasonable
reflection of what it will be when the disk goes away.

In this paper, we provide rotation periods derived from the K2
Campaign 2 data for both the Upper Sco association and for its nearby
1 Myr old neighbor \rop.  Much of our current analysis is very similar
to that we conducted in the Pleiades (Rebull \etal\ 2016a,b, Stauffer
\etal\ 2016b; papers I, II, and III, respectively) and Praesepe
(Rebull \etal\ 2017; paper IV). Somers \etal\ (2017) presented
an early version of these USco results, and provided a theoretical
discussion of the correlation between mass and rotation.

In Section~\ref{sec:obs}, we summarize the data we amassed, including
information about the K2 data, literature information collection for
the targets, member selection, dereddening, disk identification.
Section~\ref{sec:interp} begins with period identification and
interpretation, and comparison of our periods to those from the
literature. This section ends with the color-magnitude diagrams for
USco and \rop.  Section~\ref{sec:disks} discusses the influence of
disks on the period distribution of USco, including evidence for disk
locking in these LCs. Section~\ref{sec:rotationdistrib} presents the
distributions of periods and periods against color as a proxy for
mass. We also compare USco to the Pleiades (papers I-III) and Praesepe
(paper IV).  In Section~\ref{sec:linkage}, we include aspects of the
analysis of the USco and \rop\ LCs and periods in the same fashion as
we did for the Pleiades in papers I-III and for Praesepe in paper IV. 
Finally, we summarize our results in Section~\ref{sec:concl}.

\section{Data}
\label{sec:obs}

\subsection{K2 Data}
\label{sec:k2data}

\begin{figure}[ht]
\epsscale{0.7}
\plotone{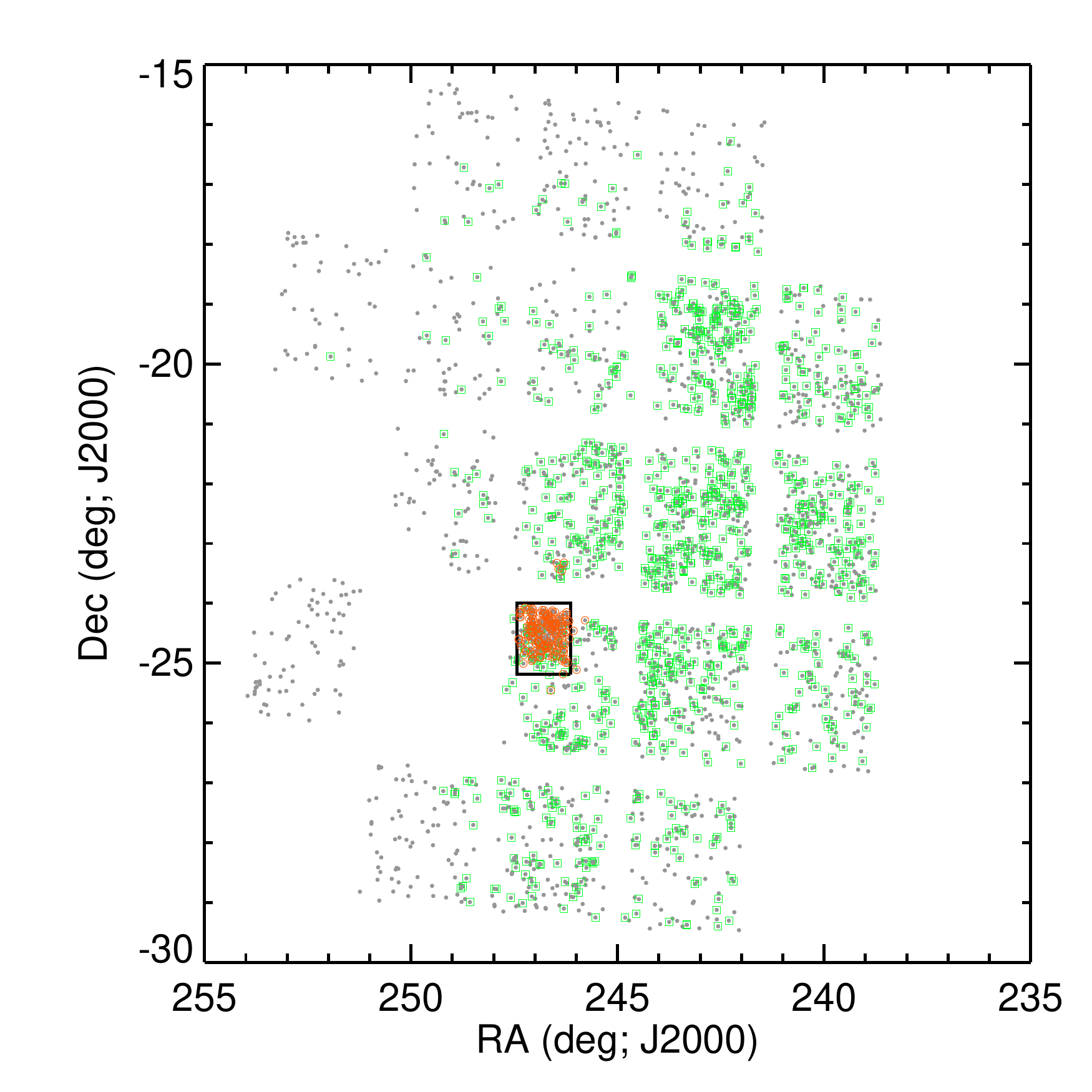}
\caption{All 2631 members or candidate members of USco or $\rho$ Oph 
with K2 LCs projected onto the sky. Note the gaps between K2
detectors. Small grey circles: objects in catalog; additional green
squares: USco members (see Sec~\ref{sec:membership}); orange $+$:
\rop\ members; black square: approximate spatial extent of \rop\
region. }
\label{fig:where}
\end{figure}

USco and $\rho$ Oph were observed in K2 campaign 2, from 2014 Aug 23
to 2014 Nov 13 (82 d). There are 2631 objects with K2 LCs that have
been claimed to be candidate members of USco or $\rho$ Oph. 
Figure~\ref{fig:where} shows the distribution of these objects with K2
LCs on the sky; note the gaps between detectors.  All of the LCs shown
were observed in the long-cadence ($\sim$30 min exposure) mode.  

As discussed in papers I and IV, K2 data can be challenging to reduce
because of the relatively large pixel sizes ($3.98\arcsec \times 3.98
\arcsec$) and because the whole spacecraft slowly drifts and then
repositions regularly every 0.245 d. Because Campaign 2 was relatively
early in the repurposed K2 mission, many artifacts were present in
these light curves that were not seen in later campaigns, presenting
additional challenges; for example, there is a 1.97 d period in about
1\% of the light curves that is most likely spacecraft-related. We
removed all of the periods that we belive are instrumental\footnote{We
did not find any 0.22 or 1.75 spurious periods, which were found by
Saylor \etal\ 2017.}.

For each target, we selected the best LC from up to three different
available LC versions. (1) A version with moving apertures with
magnitudes computed for several different aperture sizes, using custom
software developed by co-author Cody. (2) The `self-flat-fielding'
approach used by Vanderburg \& Johnson (2014) and the K2SFF pipeline
as obtained from MAST, the Mikulski Archive for Space Telescopes. (3)
The LCs from the EVEREST2 pipeline (Luger \etal\ 2016, 2017), which
uses pixel level decorrelation, as obtained from MAST. There are no
pre-search data conditioning (PDC) versions for this campaign.   We
removed any data points corresponding to thruster firings and any
others with bad data flags set in the corresponding data product. Many
of the LC versions, particularly the EVEREST2 version, had large-scale
structure remaining in the LCs that complicated the period searching;
for those LCs, we applied a least-squares (Savitzky-Golay) polynomial
smoothing filter (e.g., Press \etal\ 1992), as implemented in the
IDLastro library\footnote{https://idlastro.gsfc.nasa.gov}. 

We inspected LCs from each reduction approach, and we selected the
visually `best' LC from among the LC versions. Any periodic signals
are generally unambiguous, and are generally detected in all the LC
versions. 

Because this field is in the general direction of the Galactic Center
($l,b \sim 352\arcdeg,+19\arcdeg$), the surface density of targets is
very high and source confusion is a concern. By inspection of the
region using IRSA's Finder Chart
tool\footnote{http://irsa.ipac.caltech.edu/applications/finderchart},
as well as the diagnostic information provided by the various K2 data
reduction pipelines, $\sim$75\% (of the entire set of $\sim$2600 LCs)
are likely isolated enough that the LC is likely correctly tied to
that source. The rest are confused to varying degrees and some LC
versions wandered off the target; sometimes special handling (e.g.,
using a
very small aperture) was required to extract a viable period for the
target of choice.   In a very few cases, there is obvious source
confusion in the K2 aperture; when this resulted in two periods or two
targets close together whose LCs yielded functionally the same two
periods, we attempted to assign the period(s) to the appropriate
component. In a still smaller subset, we did not have enough
confidence in this process to assign periods to individual components.
The five EPIC numbers we omitted as duplicates or spatially unresolved
with K2 are 204514548, 204350687, 204986988, 204949182, and 203760606;
this leaves 2626 LCs to analyze.

\subsection{Literature Photometry}
\label{sec:litphotom}

We assembled information from the USco and \rop\ literature (Preibisch
\etal\ 1998, 2001, 2002,  Wilking \etal\ 2005, Erickson \etal\ 2011,
Slesnick \etal\ 2006, Kraus \& Hillenbrand 2007, Lodieu \etal\ 2011,
Luhman \& Mamajek 2012, Rizzuto \etal\ 2011, 2012, 2015).  To assemble
additional, uniform photometry, we queried many all-sky or large-scale
surveys, including the Gaia DR1 release (Gaia Collaboration 2016) for
their $G$ magnitudes, and the APASS database (Henden \etal\ 2016),
particularly for $V$ magnitudes. For more optical data, we queried the
recently released Pan-STARRS1 database (Chambers \etal\ 2016) and the
Sloan Digital Sky Survey (SDSS; e.g., Ahn \etal\ 2014).  We added to
this infrared data from the Two-Micron All Sky Survey (2MASS;
Skrutskie \etal\ 2006) and the Deep Near-Infrared Southern Sky Survey
(DENIS; Foqu\'e \& Bertin 1995). For mid-IR data from the Spitzer
Space Telescope (Werner \etal\ 2004), we included measurements from
the Spitzer Enhanced Imaging Products,
SEIP\footnote{http://irsa.ipac.caltech.edu/data/SPITZER/Enhanced/SEIP/overview.html},
as well as from FEPS (Meyer \etal\ 2006). We included data from the
Widefield Infrared Survey Explorer (WISE; Wright \etal\ 2010) at 3.5,
4.6, 12, and 22 \mum, and AKARI (Murakami \etal\ 2007) data at 9, 18,
65, 90, 140, and 160 \mum. Both WISE and AKARI are all-sky surveys,
but have different sensitivities; nearly all of the stars considered
here have WISE detections, but only $\sim$50 are detected with AKARI
at 9 and 18 \mum, and there are only $\sim$5 detections at AKARI's
longer wavelengths. Finally, we incorporated Herschel Space
Observatory Highly Processed Data Products (HPDP) for PACS 70 and 160
\mum\ (Marton \etal\ 2017). PACS did not conduct an all-sky survey;
there are detections for $\sim$80 targets.

Papers I-IV use $(V-K_s)$ as a proxy for mass, and we wish to
do the same here. We therefore need to either collect values of $V$
and $K_s$ or infer \vmk. We could obtain directly measured $K_s$ for
nearly everything from 2MASS; for a very small handful of stars, $K_s$
is only available from DENIS, not 2MASS. We could find $V$ for about
60\% of the stars, either from the literature (largely SIMBAD) or
APASS, so we use those where they exist. Generally, these measurements
are good to a few hundredths, though the provenance of the data and
intrinsic variablity of the stars may mean the uncertainty is larger.
If $V$ is not available and a Gaia $G$ magnitude is available, then
$(V-K_s)$ was interpolated from $(G-K_s)$ as in paper IV; we estimate
errors on these estimates to be $\sim$0.017-0.085 mag. For stars
redder than $(V-K_s)\sim5$, the relation from paper IV is linearly
extrapolated to $(V-K_s)\sim8$. Gaia-derived \vmk\ is used for about
30\% of the sample. Similarly, if no Gaia $G$ mag is available, but a
Pan-STARRS1 $g$ is available, then $(V-K_s)$ can be calibrated via an
empirical relation between $(g-K_s)$ and \vmk; this affects $\sim$8\%
of the sample, and errors on these estimates are probably comparable
to those from Gaia-derived colors. If there is still no estimate of
\vmk, then we use the $(V-K_s)$ is interpolated as part of the
membership analysis described in Appendix~\ref{app:member} (this is
done for $\sim$2\% of the sample). As a last resort, if the SED is
well-populated in the optical using literature photometry, a $V$
magnitude is interpolated from the SED, and compared to the observed
$K_s$  (this is done for $\sim$3\% of the sample).  Essentially all
(98.8\%) of the 2626 targets thus have a derived or interpolated
\vmk.

Table~\ref{tab:bigperiods} includes, for the probable members
(identified in the next section, \ref{sec:membership}), the relevant
supporting photometric data, including the observed or interpolated
\vmk, plus the periods we derive (in Section~\ref{sec:periods}) and
the IR excess assessments (Sec.~\ref{sec:finddisks}).  A similar table
with all the probable non-members (NM) appears in
Appendix~\ref{app:nm}.

\floattable
\begin{deluxetable}{cp{13cm}}
\tabletypesize{\scriptsize}
%\rotate
\tablecaption{Contents of Table: Periods and Supporting Data for
USco and \rop\ Members with K2 Light Curves\label{tab:bigperiods}}
\tablewidth{0pt}
\tablehead{\colhead{Label} & \colhead{Contents}}
\startdata
EPIC & Number in the Ecliptic Plane Input Catalog (EPIC) for K2\\
coord & Right ascension and declination (J2000) for target \\
othername & Alternate name for target \\
Vmag & V magnitude (in Vega mags), if observed\\
Kmag & \ks\ magnitude (in Vega mags), if observed\\
vmk-obs & \vmk, as directly observed, if $V$ and \ks\ exist\\
vmk-used & \vmk\ used (observed or inferred; see text)\\
ev-k & $E(V-K_s)$ adopted for this star (see \S~\ref{sec:dereddening}) \\
Kmag0 & dereddened $K_{s,0}$ magnitude (in Vega mags), as inferred (see \S\ref{sec:dereddening})\\
vmk-dered & $(V-K_s)_0$ (dereddened $V-K_s$), as inferred (see \S~\ref{sec:dereddening}; rounded to nearest 0.1 to emphasize the relatively low accuracy)\\
uncertaintycode & two digit code denoting origin of \vmk\ and \vmkz\
(see \S\ref{sec:litphotom} and \ref{sec:dereddening}):
First digit (origin of \vmk): 
1=$V$ measured directly from the literature (including SIMBAD) and $K_s$ from 2MASS; 
2=$V$ from APASS and $K_s$ from 2MASS;
3=\vmk\ inferred from Gaia $g$ and $K_s$ from 2MASS (see \S\ref{sec:litphotom});
4=\vmk\ inferred from Pan-STARRS1 $g$ and $K_s$ from 2MASS (see \S\ref{sec:litphotom});
5=\vmk\ inferred from membership work (see \S\ref{sec:membership}; rare); 
6=$V$ inferred from well-populated optical SED and $K_s$ from 2MASS (see \S\ref{sec:litphotom});
-9= no measure of \vmk.
Second digit (origin of $E(V-K_s)$ leading to \vmkz): 
1=dereddening from $JHK_s$ diagram (see \S\ref{sec:dereddening});
2=dereddening back to \vmkz\ expected for spectral type;
3=used median $E(V-K_s)$=0.7 (see \S\ref{sec:dereddening});
-9= no measure of  $E(V-K_s)$ \\
P1 & Primary period, in days (taken to be primary rotation period;
(see text)\\
P2 & Secondary period, in days\\
P3 & Tertiary period, in days\\
P4 & Quaternary period, in days\\
Membership & USco gold, silver, bronze, or ROph gold, silver, bronze
(see \S\ref{sec:membership}) \\
Disk & Whether an IR excess (a disk) is present or not (see
\S\ref{sec:disks})\\
DiskStart & Where the IR excess starts or the limit of our knowledge
of where there is no excess (see
\S\ref{sec:disks}) \\
dipper & indicator of whether LC matches dipper characteristics (see
\S\ref{sec:bursterdipper})\\
burster & indicator of whether LC matches burster characteristics (see
\S\ref{sec:bursterdipper})\\
single/multi-P & indicator of whether single or multi-period star \\
dd &  indicator of whether or not it is a double-dip LC (see
\S\ref{sec:LCPcats} and \ref{app:LCPcats})\\
ddmoving & indicator of whether or not it is a moving double-dip LC (see
\S\ref{sec:LCPcats} and \ref{app:LCPcats})\\
shch & indicator of whether or not it is a shape changer (see
\S\ref{sec:LCPcats} and \ref{app:LCPcats})\\
beat & indicator of whether or not the full LC has beating visible (see
\S\ref{sec:LCPcats} and \ref{app:LCPcats})\\
cpeak & indicator of whether or not the power spectrum has a complex,
structured peak and/or has a wide peak (see
\S\ref{sec:LCPcats} and \ref{app:LCPcats})\\
resclose & indicator of whether or not there are resolved close
periods in the power spectrum (see
\S\ref{sec:LCPcats} and \ref{app:LCPcats})\\
resdist & indicator of whether or not there are resolved distant
periods in the power spectrum (see
\S\ref{sec:LCPcats} and \ref{app:LCPcats})\\
pulsator & indicator of whether or not the power spectrum and period
suggest that this is a  pulsator (see
\S\ref{sec:LCPcats} and \ref{app:LCPcats})\\
\enddata
\end{deluxetable}

\subsection{Membership Summary}
\label{sec:membership}

Because of the source surface density in the direction of USco and
\rop, and because the net proper motion of the clusters ($-10,-25$ mas
yr$^{-1}$) is relatively small, it is not a trivial undertaking to
extract members from the background/foreground population. We
investigated several possibilities for sifting members (proper motions
from various sources, various color combinations); our final approach
is summarized in Appendix~\ref{app:member}. 

In the end, for USco, there are 1133 likely members; overall, 86\% are
periodic (see Sec.~\ref{sec:periods} below).  For \rop, there are 180
likely members, with 60\% periodic. We expect to find a higher
fraction of periodic LCs among the member stars, so the relatively
high fraction of periodic objects lends support to our membership
lists.  We also expect to find fewer periodic stars among the disked
sample (see Sec.~\ref{sec:finddisks} below); there are more disks
among \rop, so the lower periodic fraction there makes sense.

This process leaves 1313 LCs (50\% of the entire sample) that are tied
to objects that we believe are likely to be NM. However,
specifically because the membership lists were difficult to obtain,
and because upcoming data releases (e.g., Gaia) will shed more light
on membership, we analyzed all of these light curves in the same way,
and provide information on the remaining 1313 LCs in
Appendix~\ref{app:nm}. Future investigators coming to different
conclusions about membership can thus include the objects (and their
periods) as presented in the Appendix for a reanalysis.  

Unless explicitly indicated, the subsequent analysis in this paper
uses only the probable members of USco and \rop.
Table~\ref{tab:bigperiods} contains the probable members, and
Appendix~\ref{app:nm} lists the probable NM.

\subsection{Dereddening}
\label{sec:dereddening}

The reddening in the direction of USco and \rop\ is very uneven and
can be substantial. Spectral types are available for $\sim$60\% of the
members; generally these are biased towards the earlier (brighter)
types and USco members.  In order to deredden the ensemble,
specifically the $V-K_s$ colors, we took the approach summarized here.

$JHK_s$ magnitudes are available for 99\% of the entire sample, so we
can place nearly all of the stars on a $J-H$ vs.\ $H-K_s$ diagram.
Expected $JHK_s$ colors for young stars can be found in Pecaut \&
Mamajek (2013).  We can move about three-quarters of the sample back
along the reddening law derived by Indebetouw \etal\ (2008) to that
Pecaut \& Mamajek relation. For some red stars, the placement of the
star on the $JHK_s$ color-color diagram suggests that the star is
likely to be subject to an IR excess, and so we deredden back to the
T~Tauri locus from Meyer \etal\ (1997). Note that there is a
discontinuity between the end of the Pecaut \& Mamajek relation and
the beginning of the T~Tauri locus (noted in Meyer \etal\ 1997); this
results in a small gap in the dereddened $(J-K_s)_0$ distibution
between $\sim$0.9 and $\sim$1.0. Most stars have $(J-K_s)_0 \lesssim$
0.95; only the stars that have the largest disks result in  $(J-K_s)_0
\gtrsim$ 1, and few of those are periodic, so it does not affect our
results, and we have chosen to leave the colors as they are derived.

The reddening derived from the $JHK_s$ colors can be converted to
$E(V-K_s)$ via $A_K = 0.114 A_V$ (Cardelli \etal\ 1989).

For those stars with spectral types, the reddening derived in this
fashion is broadly consistent with the reddening based on a comparison
of spectral types and the expected colors from Pecaut \& Mamajek
(2013). The mode of the entirety of the spectral-type-derived
reddening values is $E(V-K_s)\sim$0.70.

If it was available, we took first the reddening derived from the
$JHK_s$ diagram ($\sim$64\% of the sample). If there is no value
available from the $JHK_s$ diagram, or the value derived from that is
consistent with zero or unphysical, but there is a spectral type, then
we take the reddening from an explicit comparison of the spectral type
and the expected colors from Pecaut \& Mamajek (2013). This is the
case for $\sim$15\% of the sample. For \rop\ members with spectral
types, because the reddening is so high, we explicitly enforced that
the reddening is drawn from a comparison to the Pecaut \& Mamajek
(2013) colors. Note that for all cases where the colors are forced to
match those from the spectral type (most obvious in the \rop\ sample),
the \vmkz\ values are `quantized' specifically because they are forced
to match the colors corresponding to that spectral type; this
manifests as `lines' of sources at a given \vmkz.  For the remaining
$\sim$20\% of the objects with no estimate of $E(V-K_s)$ to this
point, we assigned the modal reddening of $E(V-K_s)=0.70$.

The dereddened \vmkz\ we used for each object is included in
Table~\ref{tab:bigperiods} for the members and in
Appendix~\ref{app:nm} for the NM. However, to emphasize the net 
uncertainty, the ``vmk-dered'' column in Table~\ref{tab:bigperiods} 
has been rounded to the nearest 0.1 mag. The values used in plots 
here can be recovered by using the $E(V-Ks)$ (``ev-k'') and 
$(V-K_s)_{\rm observed}$ (``vk-used'') columns. 

Net errors are hard to quantify after all of these steps.
Table~\ref{tab:bigperiods} (and its analogous Table~\ref{tab:bignm}
for NM) include a 2-digit code indicating the origin of the \vmk\
value and the method by which the \vmk\ was dereddened to \vmkz\ (see
Table~\ref{tab:bigperiods} or \ref{tab:bignm} for specific
definitions). For most values of \vmk, the uncertainty in \vmk\ is
probably a few hundredths of a magnitude at most. For dereddening in
the $JHK_s$ diagram, via internal comparisons and uncertainties not
just on the assumed photospheric colors but also uncertainties in
spectral typing, we estimate that the typical uncertainty for USco
members could be conservatively $\sim$0.4 mag, and that for \rop\
members is likely worse at $\sim$0.9 mag. Dereddening based on
spectral type is likely comparable, as is assuming a median reddening;
uncertainties are larger in \rop, even when there is a spectral type,
because the extinction is large enough that there is likely to be
larger uncertainties on the (optical) spectral type estimates.

\subsection{Disk Indicators}
\label{sec:finddisks}

Both USco and \rop\ are young enough that a significant fraction of
the members have disks. One of our goals for this paper is to compare 
the rotation rates of the stars with disks to those without them.  For
that purpose, we prefer to have a list of stars with disks that is as
complete as possible.  To accomplish this, we tag a star as a disk
candidate if it has a plausibly real excess at any IR wavelength.  Our
process for doing this is detailed in this subsection. The wavelength
at which the IR excess begins is included in
Table~\ref{tab:bigperiods} (and in the Appendix in
Table~\ref{tab:bignm} for the NM).

For each of the targets, we constructed SEDs from the assembled
photometry. For WISE, we used IRSA's Finder Chart tool to inspect the
WISE images to see if the detections in the catalog reflect what can
be seen in the images. MIPS data are not widely available because MIPS
only observed about 300 of the USco/\rop\ members (and mostly just in
24 \mum), but are useful (and sensitive) when they exist. Just 14 were
observed with MIPS at 70 \mum; for 6 stars, the IR excess begins at
70 \mum. AKARI provides additional detections for about 70 member
sources that are consistent with the rest of the assembled SED, and
occasionally reveal IR excesses that start at 9 \mum. PACS provides
many more detections at 70 and 160 \mum, but did not reveal any new IR
excesses that were not already identified based on other IR data.

For the ensemble of true WISE detections (99\% of the members at 3.5
\mum; $\sim$50\% of the members at 22 \mum), we examined the
distributions of [3.4]$-$[22] and [3.4]$-$[12]. For those $\sim$400
objects for which there were MIPS 24 \mum\ detections ($\sim$300 of
the members), we examined the distribution of $K_s-[24]$. For all
three colors ([3.4]$-$[22], [3.4]$-$[12], $K_s-[24]$), we calculated
both the color and the significance, e.g., $\chi$ for 3.4 and 12 \mum\
is \begin{equation}  \chi = \frac{([3.4]-[12])_{\rm observed} -
([3.4]-[12])_{\rm expected}}{\sqrt{\sigma_{[3.4]}^2+\sigma_{[12]}^2}}
\end{equation}  For \vmkz$\lesssim$3.5 (early M), $([3.4]-[12])_{\rm
expected}$ can be taken to be 0; for 3.5$\lesssim$\vmkz$\lesssim$6.5,
$([3.4]-[12])_{\rm expected}$ is not zero. We took the set of all
USco/\rop\ stars and assessed the distribution of $[3.4]-[12]$ as a
function of  \vmkz, fitting a line to the distribution of non-disked
stars to predict $[3.4]-[12])_{\rm expected}$. We obtain a fit similar
to (and slightly larger than, e.g., more conservative than) Pecaut \&
Mamajek (2013). For the latest stars we have here, the intrinsic
photospheric color can be as much as $\sim$0.4-0.5 mag. We then
assessed the ensemble of information available for all sources (e.g.,
all points $>$2 \mum, shape of SED, etc.). For example, if the
significance of the excess ($\chi$) at 12 \mum\ is $>$5, and the
source looks ok (e.g., circular, unaffected by artifacts) in the
images at 12 \mum, then we took it as an excess. If the IR excess is
large enough (e.g., [3.4]$-$[12]$>$1.3 mag), and there is
corroborating information from another wavelength, then we took it as
an excess even if $\chi <$5.  There are some for which, at wavelengths
$>$10 \mum, we have only 12 and 22 \mum\ points, and  $\chi >$5 at 12
\mum, but $\chi <$5 from  [22]; those were not taken as disks because
it is unlikely that a dust disk could create an excess at 12 but not
22 \mum. Because the SEIP likely underestimates [24] errors, and
because the $K_s$ band was not observed at the same time as the [24]
band, we required a higher significance at 24 \mum.  If $\chi$  was
$>$10, then we took the excess as significant; if $K_s-[24] > 1$ then
there is a large enough excess to consider as real even if $\chi$
wasn't quite 10.

For the sample, then, we could identify clear disk candidates and
clear non-disk candidates (at least, non-disks given the available
data, which often extend to 12 or 22 \mum); for some, the data do not
extend very far into the IR. Finally, there are some for which it is
not clear whether or not there is a significant excesss.  Among the
USco members, 208 (18\%) are clear disk candidates, and 42 (4\%) may
have disks; this leaves 871 (77\%) for which there is no disk (and
12=1\% with no information). For \rop, 85 (47\%) are clear disks, and
15 (8\%) may have disks, leaving 78 (43\%) with no disk (and 2=1\%
with no data). These rates are all consistent with \rop\ being younger
than USco.  Note that the disk excess criteria are conservative and
that the non-disked sample will have contamination from weaker
($<5-10\sigma$ excess) disks. Note also that the lowest mass bin is
likely incomplete in the non-disks due to sensitivity issues (stars
with excesses are more likely to be detected than stars without
excesses). Note also that this disked sample is not statistically
rigorous, since the sample draws from many surveys and wavelengths,
and in order to be considered at all, they must be a detection in K2,
which is affected not only by extinction but pixel mask selection.

\section{Periods and Color-Magnitude Diagrams}
\label{sec:interp}

This section starts the analysis of data described in the prior
section. We first discuss period finding and interpretation. We end
with color-magnitude diagrams for various subsets of the two clusters.

\subsection{Finding Periods in the K2 LCs}
\label{sec:periods}

Our approach for finding periods was identical to that which we used
in the Pleiades and Praesepe (papers I, II, and IV). In summary, we
used the Lomb-Scargle (LS; Scargle 1982) approach as implemented by
the NASA Exoplanet Archive Periodogram
Service\footnote{http://exoplanetarchive.ipac.caltech.edu/cgi-bin/Periodogram/nph-simpleupload}
(Akeson \etal\ 2013). We also took advantage of the new Infrared
Science Archive (IRSA) Time Series
Tool\footnote{http://irsa.ipac.caltech.edu/irsaviewer/timeseries},
which uses the same underlying code as the Exoplanet Archive service,
but allows for interactive period selection. We looked for periods
between 0.05 and 35 d, with the upper limit being set by roughly half
the campaign length. Because the periods are typically unambigous,
false alarm probability (FAP) levels are calculated as exactly 0 for
97\% of the periods we present here (and the remaining FAP levels are
typically $<10^{-4}$).  

The periods we derive are in Table~\ref{tab:bigperiods} 
for the members and in Appendix~\ref{app:nm} for the NM.

\begin{figure}[ht]
\epsscale{1.0}
\plotone{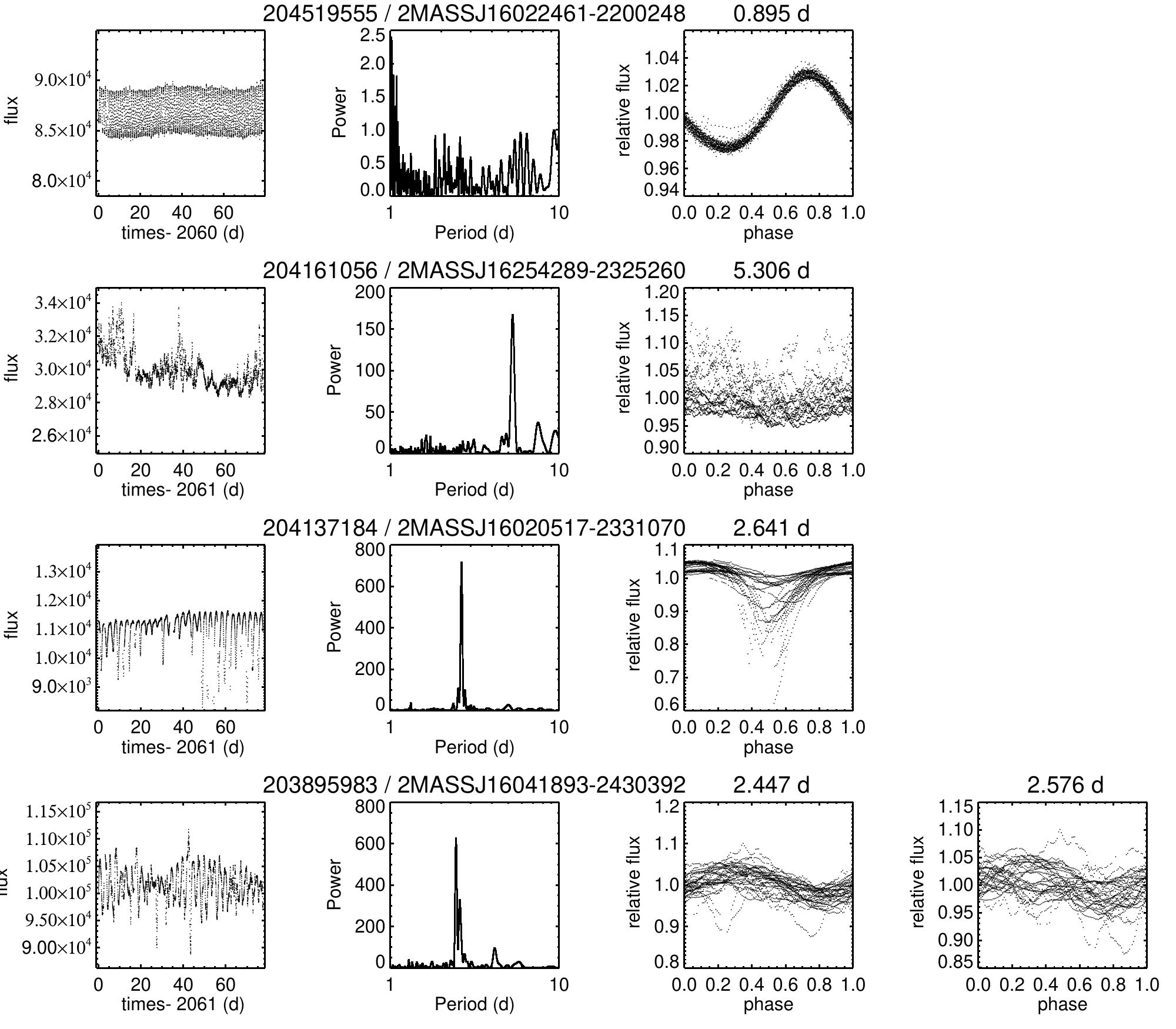}
\caption{Example LCs for a single-period sinusoid LC (first row), burster (second row), dipper (third row), and
disk-affected spot-modulated periods (fourth row). The panels are: best
LC, power spectrum, phased LC for first period, phased LC for second
period (only applies for last row). 
The objects are, top to bottom, EPIC 204519555/2MASSJ16022461-2200248, 204161056/2MASSJ16254289-2325260,
204137184/2MASSJ16020517-2331070, and
203895983/2MASSJ16041893-2430392. The first row is shown to demonstrate the 
contrast between it and the kinds of disk-affected LCs; 
are not found in older clusters studied by K2.}
\label{fig:lcexamples}
\end{figure}

\subsection{Interpretation of Periods}
\label{sec:interpofperiods}

In this section, we describe the different kinds of periodicities in
the LCs, and which we retain as likely rotation periods, $P_{\rm
rot}$. Each subsection below describes different subsets of the sample
of stars with detected periods and whether we interpret those periods
as the stellar rotation period or not.

For completeness, we note that there are some objects (18\% of the
probable members) that we do not detect to be periodic. Aside from
spot coverage and $\sin i$ effects, since we expect our membership
list to have some contamination by (older) field stars, it is likely
that some/many of the aperiodic stars are actually non-members.  As
for our work in Papers I and IV, there are some LCs with some repeated
patterns, but which seemed to be more `timescales' than rotation
periods (see Appendix~\ref{app:timescales}).

\subsubsection{Light Curves Consistent with Photospheric Spots}

Nearly three-quarters (76\%) of the periodic LCs from cluster members 
have LCs consistent with rotational modulation of
non-axi-symmetrically distributed star spots.  Most often, these light
curves are roughly sinusoidal in shape; see the first row of Fig.~\ref{fig:lcexamples}.  
However, other shapes are
also possible.  A star where a single spot dominates the light curve
and where that spot passes fully to the back-side of the star as seen
from Earth for part  of the period will have a phased light curve
showing a constant brightness then a broad flux dip.   A star with two
or more dominant spot groups at different longitudes (and latitudes)
can produce double peaked light curves, which happens in about 10\% of
the cluster members here (see, e.g., Davenport \etal\ 2015, or papers
I-IV for examples from the Pleiades and Praesepe).  For stars with one
period, the period we used was usually the strongest peak in the
periodogram. In certain cases (most notably the ones with two
peaks/dips per cycle), it was clear that a peak other than the
strongest was the most appropriate period to take as the primary
$P_{\rm rot}$. 

\subsubsection{Multiple Periods}

In about 20\% of the members, we found more than one period in the LC.
We retain up to four periods in our database, but for much of the
analysis here (e.g., plots of $P$ vs.\ \vmkz), only one period can be
used.  For stars with two (or more) periods that we believe are due to
rotation, we plot only one point at the period we believe corresponds
to the actual rotation period of the star dominating the \vmkz\
measurement.

Those periods that result in sinusoidal phased LCs are also most
likely to be spot-modulated stellar rotation rates.  In some cases, as
for papers II and IV, particularly for the M stars, two clear periods
in a LC reflect binaries, where each period corresponds to the
rotation rate of one of the stars in the binary (see
Sec.~\ref{sec:singleandmulti} below and in papers II and IV, as well
as Stauffer \etal\ in prep). In other cases, particularly for G and K
stars, latitudinal differential rotation may result in two distinct
periods if there are spot groups at significantly different latitudes
on the stellar surface (see discussion in paper II); the LC shapes
could also be due to starspot evolution. These LC shapes are less
common in these young stars than it was in the Pleiades and Praesepe
(papers II and IV).  

In a few cases, the LCs have very short periods and often there is a
forest of periodogram peaks; these stars are all earlier types.  As
discussed in paper II, these are most likely to be pulsators,
generally of the $\delta$ Scuti or $\gamma$ Dor type. The strongest
peak in some cases may be related to rotation, so those periods are
left in the distributions but flagged when necessary and flagged in
the tables (just as we did in papers II and IV).  If there is just one
period, we do not have enough information to classify it as a pulsator
beyond just its very short period; none of the objects like this are
members. Ripepi \etal\ (2015) identifies EPICs 203931628, 204175508,
and 204494885 as $\delta$ Scutis and EPIC 204054556 as a $\gamma$ Dor;
these are also identified as pulsators here. Ripepi \etal\ also
identify EPIC 204638251 as a $\delta$ Scuti, so we have tagged it as a
pulsator here. For EPIC 204760247 (=HD 142883, a USco member B star), the
period is likely to be a slow pulsation; see David \etal\ in
prep.

\subsubsection{Highly Structured LCs Not Due to Spots or Pulsation}

There are about 40 Upper Sco and \rop\ members which share very
unusual characteristics.  Those shared characteristics include: (a)
with possibly one or two exceptions, these stars show no evidence of
IR excess and hence no significant primordial circumstellar disk; (b)
all are inferred to be very low mass stars, with spectral types
generally M3.5 to M5; (c) all have phased LC shapes with much more
structure than can be explained by cold spots;  (d) all have short
periods, almost all less than 1.5 d  (with most of them having periods
$<$ 0.7 d). Furthermore, only young stars have been found to exhibit
these light curve morphologies.   We have discussed many of these
stars in two previous papers (Stauffer \etal\ 2017, 2018). While there
is no detailed physical model to explain these light curve
morphologies, their photometric variability almost certainly must be
due to gas and dust orbiting the star in a stable configuration at the
star's Keplerian co-rotation radius. We have therefore retained the
periods associated with these stars as indicative of the stellar
rotation periods.

Appendix~\ref{app:batwings} collects all of these objects together,
with a few example LCs.   Note that four of these stars are not
included in our list of probable cluster members (EPICs 204364515,
205046529, 205110559 and 204296148).  All of these stars failed our
proper motion membership criteria given the set of proper motions we
employed.   However, we note that two of these stars (EPICs 204364515
and 205046529) were considered as members by Preibisch \etal\ (2002),
Kraus \& Hillenbrand (2007) and Luhman \& Mamajek (2012), and both
have strong Li $\lambda$6708 absorption features (which at spectral
type M4 assures they are younger than $\sim$15 Myr).  We suspect the
other two are also likely members and simply have relatively poor
proper motion data.

\subsubsection{Bursters, Dippers, and Other Disk-Related Behavior}
\label{sec:bursterdipper}

Among the 350 USco/\rop\ stars we identify as having disks, we detect
periods for 276 stars.  Of those, 185 have light curves that appear
consistent with rotational modulation by photospheric spots.  Of the
remaining periodic disked stars, 74 have LC morphologies which
resemble the dippers from NGC 2264, and just 17 have LC morphologies
dominated by accretion bursts (Cody \etal\ 2014; Stauffer \etal\ 2014,
2015, 2016a).  In the remainder of this paper, we associate these
periods with the rotation period of the stars; our reasoning for this
is as follows.

Dippers are fading events.  The photometric variability in dippers
(such as the prototype AA Tau, Bouvier \etal\ 1999) can be
interpreted  as originating from variable extinction to our line of
sight linked to warps in their inner circumstellar disk (Terquem \&
Papaloizou 2000), to dust entrained in funnel flows (Blinova \etal\
2016), or to waves of various sorts excited by the interaction between
the stellar magnetosphere and the inner disk (Romanova \etal\ 2013). 
Stable disk warps, however, are only expected for the case where the
stellar rotation period and the inner disk rim orbital period are
locked to each other (Terquem \& Papaloizou 2000; Long \etal\ 2005),
leading to the expectation that most of the periodic dippers (AA Tau
analogs) we detect have $P_{\rm rot}$ = $P_{\rm dip}$ (Romanova \etal\
2013). Dippers in USco are discussed in Ansdell \etal\ (2016), Hedges
\etal\ (2018), and Cody \etal\ (2018). 

In general, bursters are sudden brightening events and are interpreted
as a result of accretion instabilities. When the bursting behavior is
periodic, it is thought to be due to hot spots on the stellar surface
tied to relatively long-lived accretion columns rotating into and out
of view (Blinova \etal\ 2016).   These hot spots may shift in position
with time on the surface of the star and thus may have somewhat
wandering periods (Romanova \etal\ 2013).  We assign $P_{\rm burst}$ =
$P_{\rm rot}$ for these stars, even though this may not always be the
case.  Because very few of our disked-star periods arise from stars
where bursts are the dominant variability type, this choice should
have no affect on any of our conclusions. Bursters in USco are
discussed in Cody \etal\ (2017, 2018). 

We have identified dippers and bursters in our sample
independently from the papers noted above. The objects selected to be
included in these categories differ from those in the other papers 
because we used different selection criteria, different light curves,
and a different set of candidate members.   Based on our criteria, we
identify about 2\% of the USco members as bursters and about 6\% of
the USco members as dippers; for \rop, the fractions are 7 and 9\%,
respectively.   An example burster and dipper appear in
Figure~\ref{fig:lcexamples}.

In \S\ref{sec:disklock} below, we provide direct evidence for
``disk-locking" in a small number of the USco disked stars that show
both sinusoidal waveforms from photospheric spots and narrow
extinction dips having the same period (also see Stauffer \etal\
2015).

There are USco/\rop\ stars for which there is a periodic
signal with shorter-timescale variations superimposed. These
additional variations are likely disk-related in that they are
probably superpositions of both accretion-related brightening and disk
occultation-related fading.  We interpret the period as being a result
of spot modulation, and so we retain these periods as rotation
periods. These kinds of LCs make up only about 2\% of the member
sample, and they all have disks. An example of this kind of LC is
given in Figure~\ref{fig:lcexamples}.

\subsubsection{Periods That Are Not Rotation Periods}

We removed from this distribution any periods that are unlikely to be
rotation or rotation-related, such as eclipsing binaries (see, e.g.,
David \etal\ 2016, 2018 submitted). However, the $P_{\rm rot}$ was
retained for those objects for which we can determine a $P_{\rm rot}$
(as distinct from the binary period).

There is one object, EPIC 203497438 (CD-25 11199), which is not likely
to be a USco or \rop\ member, which appears to have a burst every 23.5
d; this might be a `heartbeat binary' (e.g., Thompson \etal\ 2012).
Because this period is not likely to be rotation (and the star
unlikely to be a member), we have removed its period from the
dataset.

\subsection{Comparison to Literature Periods}
\label{sec:lit}

In this section, we identify stars in common between our study and two
literature studies (Mellon \etal\ 2017 and Ansdell \etal\ 2018), and
compare the resultant periods. Our periods match or can be explained
when they do not match.  Scholz \etal\ (2015) also present periods
from K2 data, but only for 16 brown dwarfs. Our periods match very
well the periods obtained there. We conclude that our approach to
obtaining periods is working at least as well as those published
elsewhere in the literature. 

\begin{figure}[h]
\epsscale{1}
\plottwo{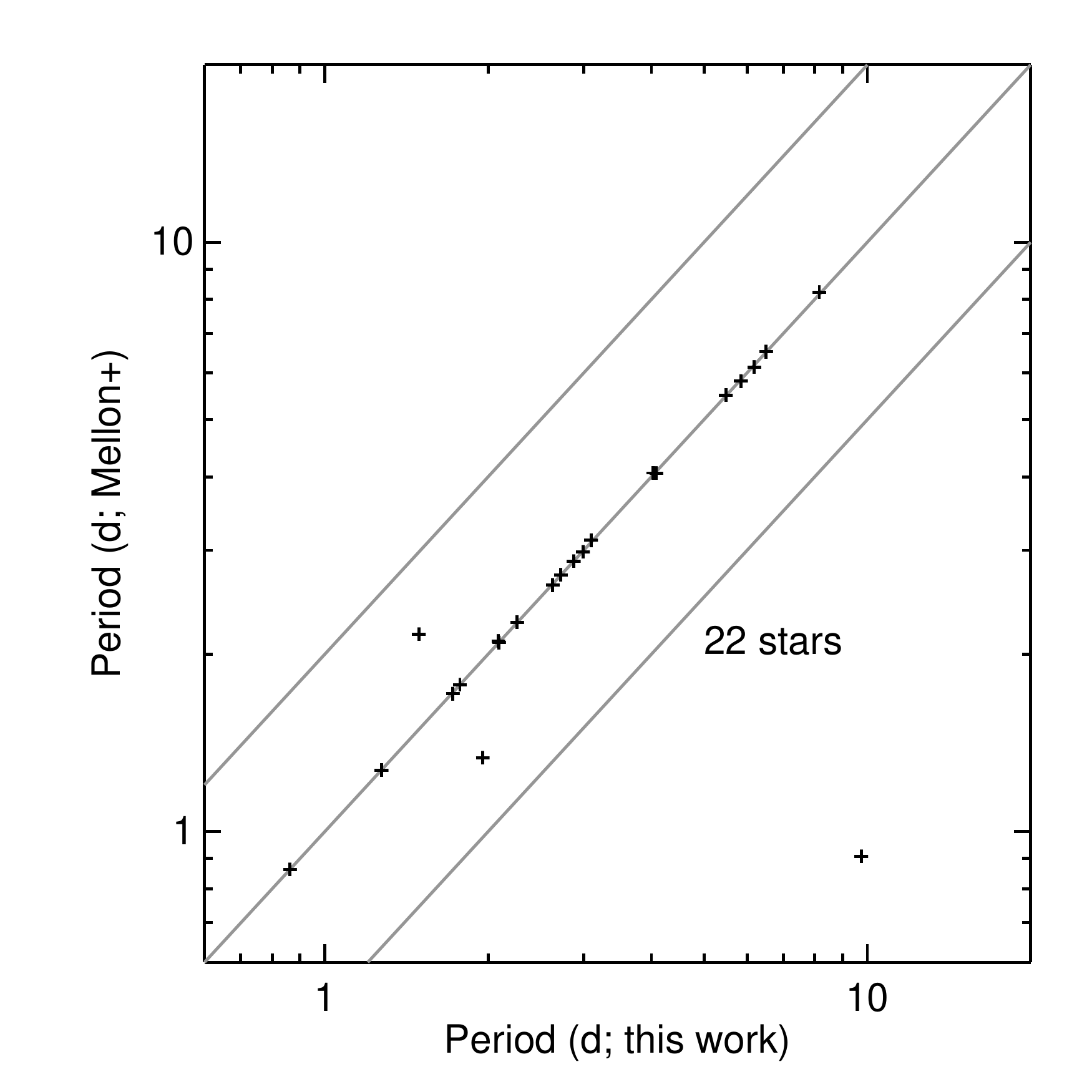}{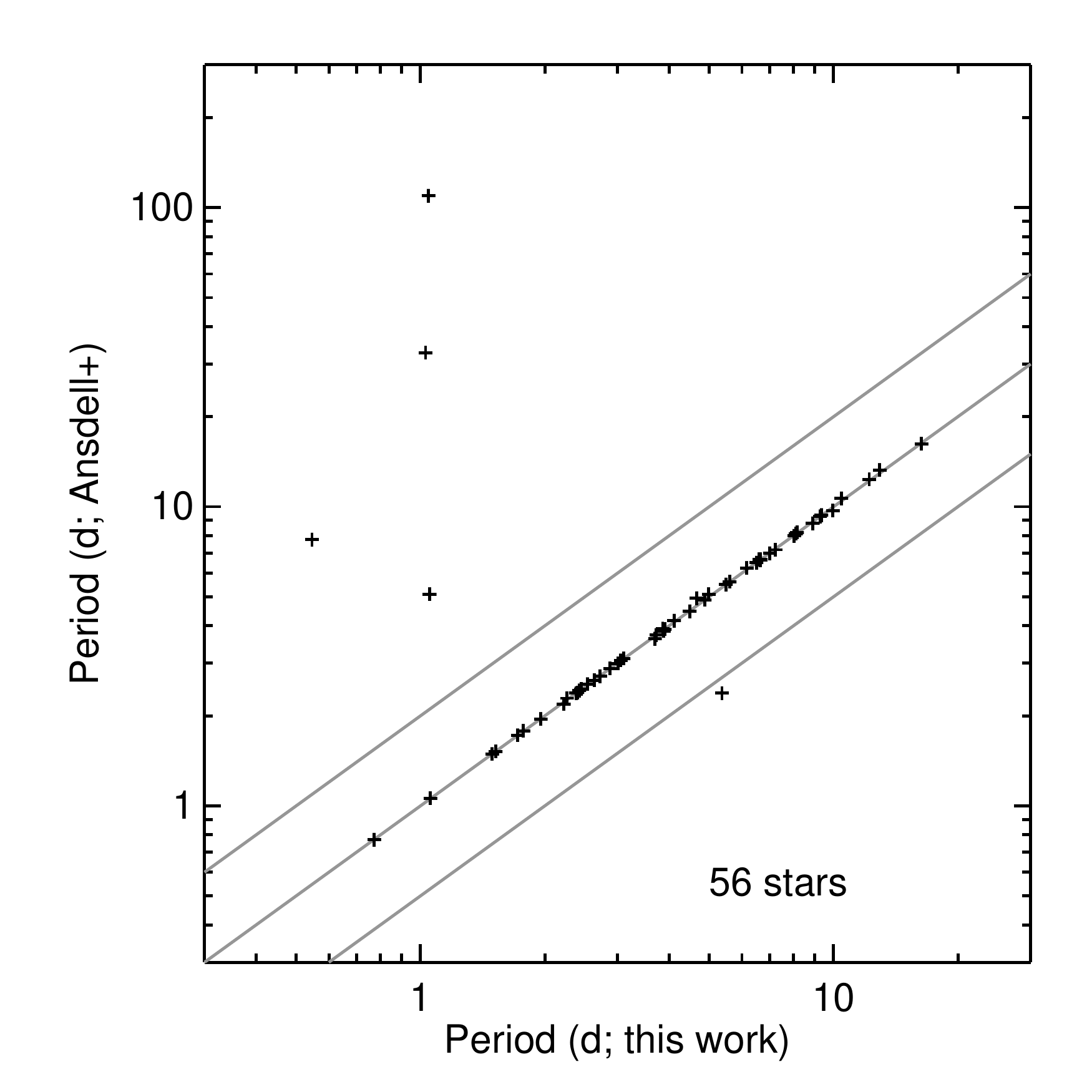}
\caption{Comparison of periods obtained here to periods obtained in
the literature. Left: Mellon \etal\ 2017, SuperWASP, 22 stars in common;
right: Ansdell \etal\ 2018, KELT, 56 stars in common. The
grey lines are at 1-to-1, $P$/2, and 2$P$. Most of the periods match
well (see text). }
\label{fig:compareliterature}
\end{figure}

Mellon \etal\ (2017) used SuperWASP to monitor stars in the Sco-Cen OB
association. While covering a much larger area on the sky than K2, the
SuperWASP data only provide periods for relatively bright stars,
leading to there only being 22 stars in common with our study.
Figure~\ref{fig:compareliterature} shows that there is good agreement
in derived periods between the two studies. There are only 3 stars
with discrepant periods. For EPIC 204794876 (2MASSJ16014743-2049457),
we report 2 periods (1.490 and 2.153 d), Mellon \etal\ report only one
(2.161 d), and Ansdell \etal\ (2018) obtain 1.49 d.  This target is
likely to be a binary, with the two periods we recover corresponding
one to each star. For EPIC 204894575 (2MASSJ16025396-2022480 2), we
report 1.954 and Mellon \etal\ report 1.333; Ansdell \etal\ report
1.95d. For our LC, our period is correct, and there is no evidence of
a 1.333 d period. This is a K6 star, so surface differential rotation
of this magnitude is unlikely; this could also be a binary where one
star did not have organized enough spots/spot groups to create a
periodic signature in our LC.  Lastly, EPIC 204447221
(2MASSJ16094098-2217594) is an interesting case; we report a 9.742 d
period and Mellon \etal\ report a period $\sim$10 times faster, 0.907
d. Our light curve does not seem to be subject to source confusion; it
has a very obvious, long period, though the waveform changes shape
over the campaign, and there is no evidence for oscillations less than
a day. The factor of 10 difference in periods is too large to be
explained by differential rotation.  It is not clear why these results
are so discrepant between the two data sets.

Ansdell \etal\ (2018) used KELT (as well as the K2 C2 data analyzed
here) to explore rotation in USco. There are 56 stars in common
between the studies. Figure~\ref{fig:compareliterature} shows that
there is again good agreement between the two studies. There are five
stars where there is disagreement, only three of which fall in the
boundaries of Fig.~\ref{fig:compareliterature}. The two outside the
boundaries are  EPIC 204819202 (2MASSJ15554141-2043150) and EPIC
204054556 (HD144729). For the former (204819202), we report a period
of 1.028d; they report a period of 32.720 d. Our LC has no evidence of
even a long-term trend, much less a period of $\sim$30 d; our period
is correct for our LC. For the latter (204054556), we find many peaks
in the periodogram and report the top four periods; this star is also
noted as a $\gamma$ Dor-type pulsator (Ripepi \etal\ 2015). We could
not have recovered the $\sim$100d period from Ansdell \etal\ because
our campaign is not long enough; in any case, a 100d period for an F3
dwarf would make it a very anomalously slowly rotating star for that
mass. EPIC 204637622 (2MASSJ16042097-2130415) is one of the remaining
three stars whose periods do not agree but are close enough to appear
in Fig.~\ref{fig:compareliterature}. In this case, there are several
stars in close proximity. Most of the K2 LCs are drawn off from the
target to a nearby brighter star (which is 204638512), and that
brighter star has a period of $\sim$5 d. When a LC extraction is done
using a much smaller aperture centered on the target star, different
periods are obtained for 204637622, 1.052 and 1.385 d. Ansdell \etal\
report the 5d period, which we believe belongs instead to 204638512. 
We find several periods for EPIC 205080616 (2MASSJ16082324-1930009),
but the LC is contaminated by nearby EPIC 205080360.  EPIC 205080360
has an unambiguous period of 2.381 d, which matches the period
reported by Ansdell \etal. We have removed the 2.38d period from EPIC
205080616, leaving just two periods.  Finally, for EPIC 205141287
(GSC06209-01215), we find multiple periods and the Ansdell \etal\
period is the second period we report.

\subsection{Color-Magnitude Diagrams}
\label{sec:cmd}

\begin{figure}[ht]
\epsscale{1.0}
\plotone{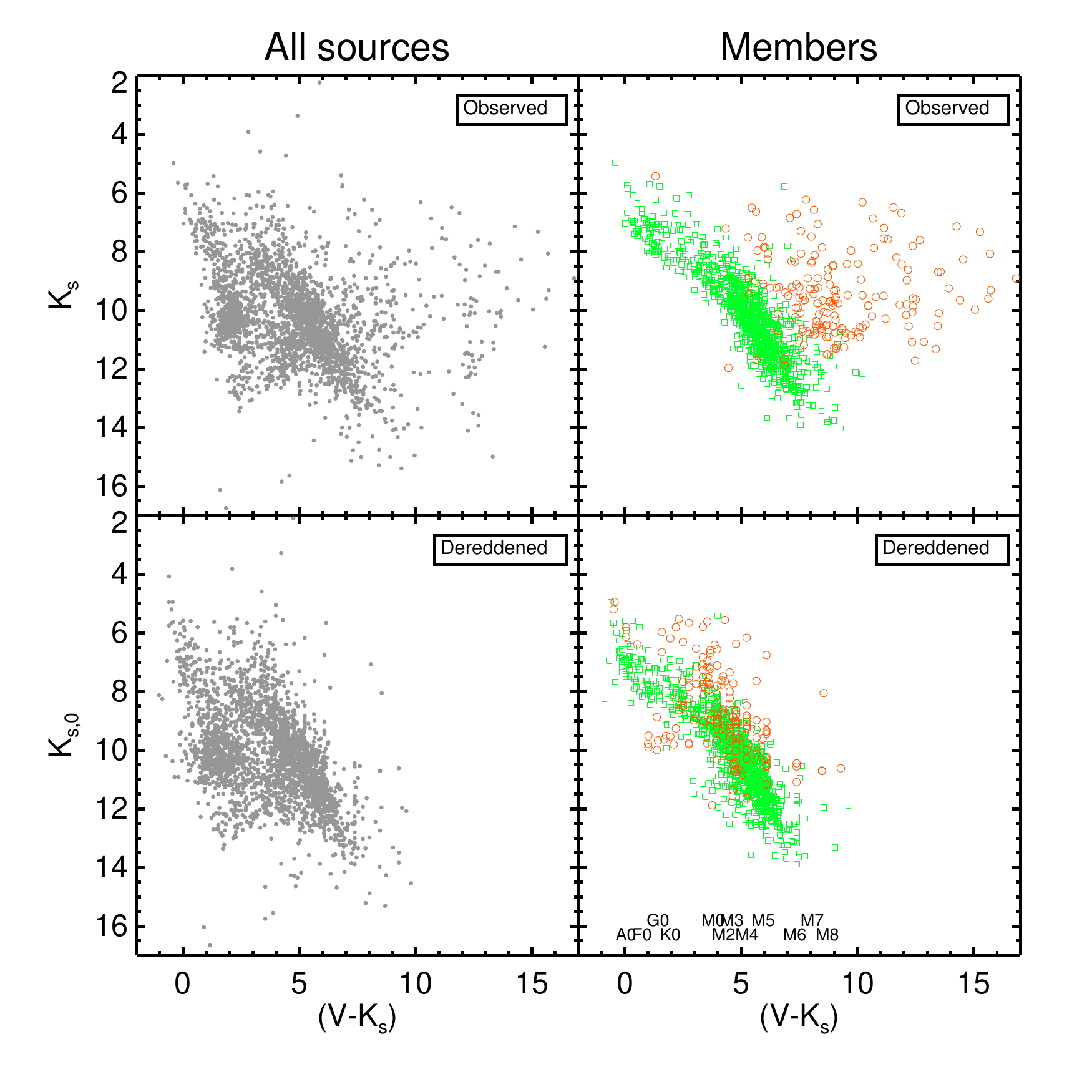}
\caption{Color-magnitude diagram (\ks\ vs.\vmk) for targets with K2
LCs and for which we had or could infer \vmk. Top row: observed;
bottom row: dereddened. Left column: all sources; right column:
members, where green squares denote USco members (see text) and orange
$+$ denote \rop\ members.   Note that some quantization can be seen
as a result of the dereddening approach for some stars with spectral 
types (see Sec.~\ref{sec:dereddening}). The USco cluster sequence is
better defined among the member sample than among the entire sample,
but the cluster sequence is not as crisply defined as the Pleiades
(paper I) or Praesepe (paper IV). The younger \rop\ stars are observed
to have a wider range of colors than USco and tend towards brighter
magnitudes at any given color.  Dereddening brings most of the \rop\
stars (highly reddened on average) much closer to the USco stars.}
\label{fig:optcmd1}
\end{figure}

\begin{figure}[ht]
\epsscale{1.0}
\plotone{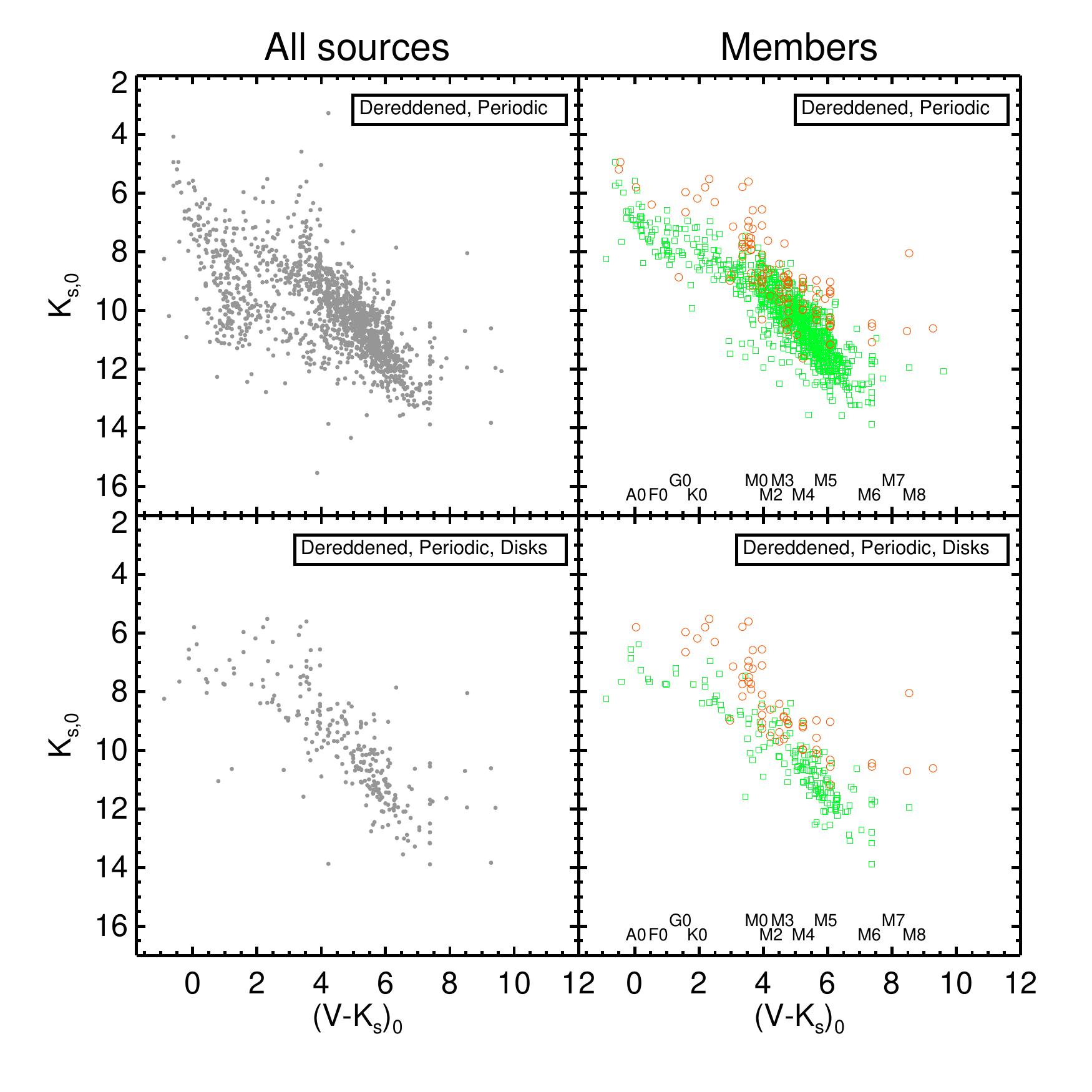}
\caption{Color-magnitude diagram (\ks$_0$\ vs.\vmk$_0$) for targets
with K2 LCs, for which we had or could infer \vmk, and for which we
could determine periods. Top row: periodic; bottom row: periodic with
obvious IR excess, e.g., unambiguous disk candidates.  Left column:
all sources; right column: all members where USco (green squares) and
\rop\ (orange $+$) members are highlighted. Note that the $x$-axis
range is smaller here than in Fig.~\ref{fig:optcmd1}. A high fraction
of the USco member stars are periodic (86\%); a lower fraction of the
\rop\  members are periodic (60\%).  Among the periodic sample, 18\%
of USco periodic members have unambiguous disks, and 56\% of \rop\
periodic members have unambiguous disks.}
\label{fig:optcmd2}
\end{figure}

Figure~\ref{fig:optcmd1} shows color-magnitude diagrams (CMDs) for the
entire sample and for just the subset of members of USco and \rop. The
members span a broader swath of the CMD than the Pleiades or Praesepe
did, because the USco and \rop\ stars are young and for the most part
still above the main sequence (MS).  Members of \rop\ are further
above the MS than most of the USco stars, even in the dereddened
version of the diagram. The ensemble of all stars has much more
scatter in the CMDs, consistent with there being many more NM stars
included. Figure~\ref{fig:optcmd2} has just the subset of objects that
are periodic. A high fraction of the USco members are periodic 
(85\%); a lower fraction of the \rop\ members are periodic (60\%)
because disks are more common there and can obscure periodicities. 
Among the periodic sample, 18\% of USco periodic members have
unambiguous disks, and 56\% of \rop\ periodic members have unambiguous
disks.

In papers I and IV, we omitted stars that were too bright or faint to
result in viable K2 LCs.  As seen in Figs~\ref{fig:optcmd1} and
\ref{fig:optcmd2}, the limits are not as clear-cut in USco and \rop.
Objects with \ks$\lesssim$5 and \ks$\gtrsim$14 effectively are dropped
by the member selection in USco; for \rop, there is an additional
restriction that $J<$14, with the result that there are few members
with $K_{s,0}>$11.

\clearpage

\section{The Influence of Disks}
\label{sec:disks}

\subsection{Context}

When it first became possible to measure the rotation rates of the G,
K, and early M stars in young open clusters like the Pleiades and
Alpha Persei (Stauffer \& Hartmann 1987; Stauffer \etal\ 1989), it was
discovered that those stars showed a bimodal rotation distribution.  
Many of the stars in that mass range were relatively slow rotators,
but there also existed a population of very fast rotators.   That
bimodal rotation distribution was initially a mystery.   Eventually,
the consensus explanation for that distribution was that it arose
because of interaction between the star and its primordial
circumstellar disk during contraction onto the MS.  While the star
continues to  accrete from its primordial disk, the stellar surface is
forced to rotate at the same rate as the inner edge of its disk (Ghosh
\& Lamb 1977; K\"onigl 1991); this process, which prevents the star
from spinning up as it contracts, is commonly referred to as
disk-locking.   When the disk dissipates, the rotation lock is removed
and the star is then free to spin up.   Thus, the slowly rotating
population in young clusters like the Pleiades is linked to stars with
relatively long-lived primordial disks, whereas the rapid rotators are
more likely to have lost their disk much earlier. The rapidly rotating
population required a modification of existing angular momentum loss
prescriptions. Simple Skumanich-style laws predicted no very rapid
rotators on the main sequence (Pinsonneault \etal\ 1990), a problem
solved by introducing a saturation threshhold (MacGregor \& Brenner
1991).

Because primordial disks are believed to have lifetimes generally less
than 10 Myr, one has to go to young, star-forming regions to search
for direct evidence for the disk-locking hypothesis.  There have been
many papers devoted to that goal, primarily using data from the Orion
Nebula Cluster or NGC~2264 (\eg, Rebull \etal\ 2006; Herbst \& Mundt
2005; Cieza \&  Baliber 2007;  Biazzo \etal\ 2009; Rodriguez-Ledesma
\etal\ 2010;  Dahm \etal\ 2012; Davies \etal\ 2014; Venuti \etal\
2017).   Those studies generally find evidence claimed to be
supportive of disk-locking.  This has most often been illustrated in
plots of period versus some measure related to presence of absence of
a primordial disk, with stars lacking disks showing a wide range in
rotation (including rapid rotators) while stars with disks show a
narrower period distribution weighted towards slow rotators.  Another
way of expressing this correlation has been to plot the fraction of
stars with disks as a function of measured period, with diskless stars
predominating at short periods and disked stars predominating at long
periods.   These previous studies have essentially assumed that the
periods that have been measured for young disked stars are rotation
periods; for the ground-based data used in those papers, that
assumption was necessary because the quality of the light curves was
sufficient to detect periodicity but not to separately identify
spotted stars and (for example) AA Tau-type analogs.  Two of the above
papers (Biazzo \etal\ 2009 and Dahm \etal\ 2012) instead used
high-resolution spectra to determine spectroscopic rotation rates ($v
\sin i$ values). Their finding of highly significant correlations
betweeen rotation rate and IR excess provide support to the assumption
that the measured periods for young disked stars are indeed stellar
rotation periods.   The CoRoT light curves for NGC~2264 (Venuti \etal\
2017) and  The K2 light curves for USco/\rop\ provide the first
photometric time series dataset where, in most cases, one can separate
periodic variability due to spots on the star's photosphere from
periodic variability due to variable (disk-related) extinction.  As
discussed in \S\ref{sec:bursterdipper}, when searching for
correlations between IR excess and period in our data, we adopt the
assumption that the periods we identify as due to variable extinction
are equivalent to the stellar rotation period.   However, in the
following section, we also use our K2 Campaign 2 data to provide new,
direct evidence that at least for some YSOs with disks the stellar
rotation period is indeed the same as the inner disk orbital period.

\subsection{Direct Evidence for Disk Locking from our K2 Light Curves}
\label{sec:disklock}

Direct evidence in favor of disk locking can be established from high
quality light curves if one can identify stars with disks whose light
curves show signatures arising separately from the disk and from the
stellar photosphere and where both sets of features share the same
period.  Using CoRoT light curves for stars in the $\sim$2 Myr old
NGC~2264 star forming region, Stauffer \etal\ (2014)  found two disked
stars (Mon-21 and Mon-56) that showed well-defined spotted-star light
curves superposed on which were periodic, narrow flux dips best
interpreted as arising from dust structures near their inner disk
rims.  The periods associated with both signatures were the same,
thereby identifying these systems as stars whose photospheric rotation
rate are locked  to the Keplerian rotation rate of their inner disks.

\begin{figure}[ht]
\epsscale{1.}
\plotone{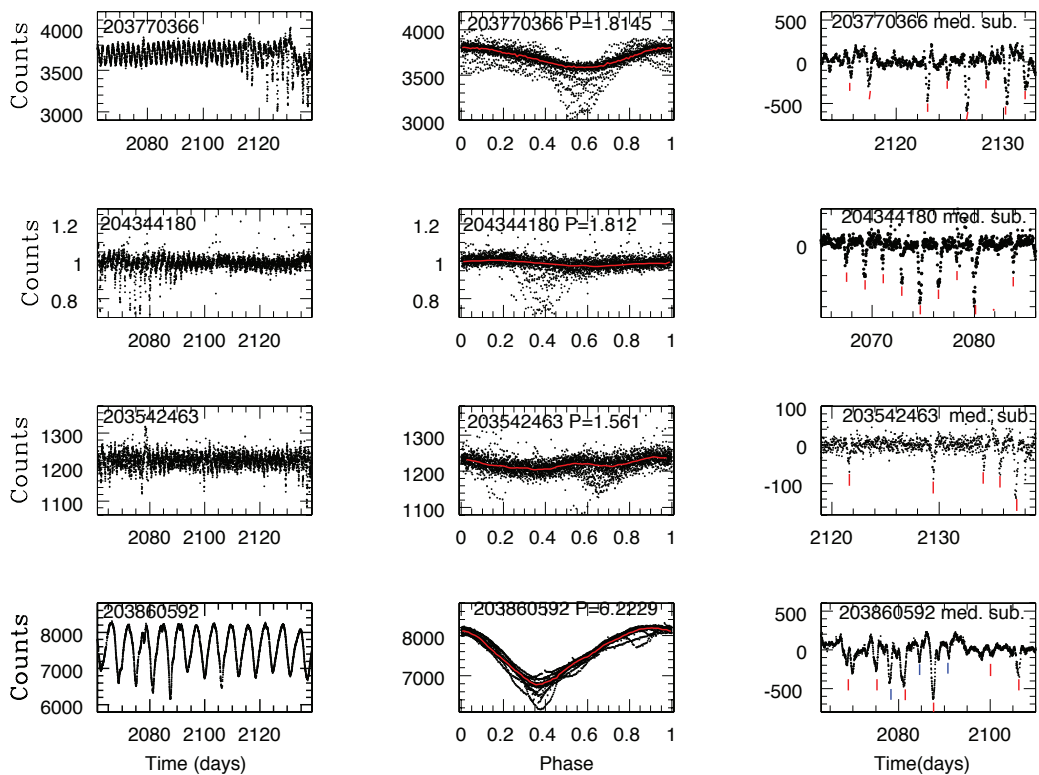}
\caption{K2 LCs for four disked stars from K2 Campaign 2.  See
Sec.~\ref{sec:disklock} for a more detailed description of these plots. 
Left-hand column:  detrended light curve for each star showing data
for the entire campaign. The light curve for EPIC 204344180 is the
Vanderburg \& Johnson (2014) reduction, which are delivered to users
in counts normalized to the median; all other light curves are
delivered with counts as the ordinate value. Middle column:
Phase-folded light curves for each star, based on the period (in days)
noted in the plot.  The red curve is a median fit to the light curve,
outside the time period of the strongest flux dips. Right-hand column:
Median subtracted light curves for each star.  The red vertical bars
are separated from each other by an exact, integral number of
periods, and the blue vertical bars are displaced in phased by 0.5;
they are not simply positioned to mark the centroid of the flux dips.
In all of these cases, the spot-modulated signal is the same period 
as the dipper portion of the signal.
}
\label{fig:dipperrot}
\end{figure}

\begin{figure}[ht]
\epsscale{1.}
\plotone{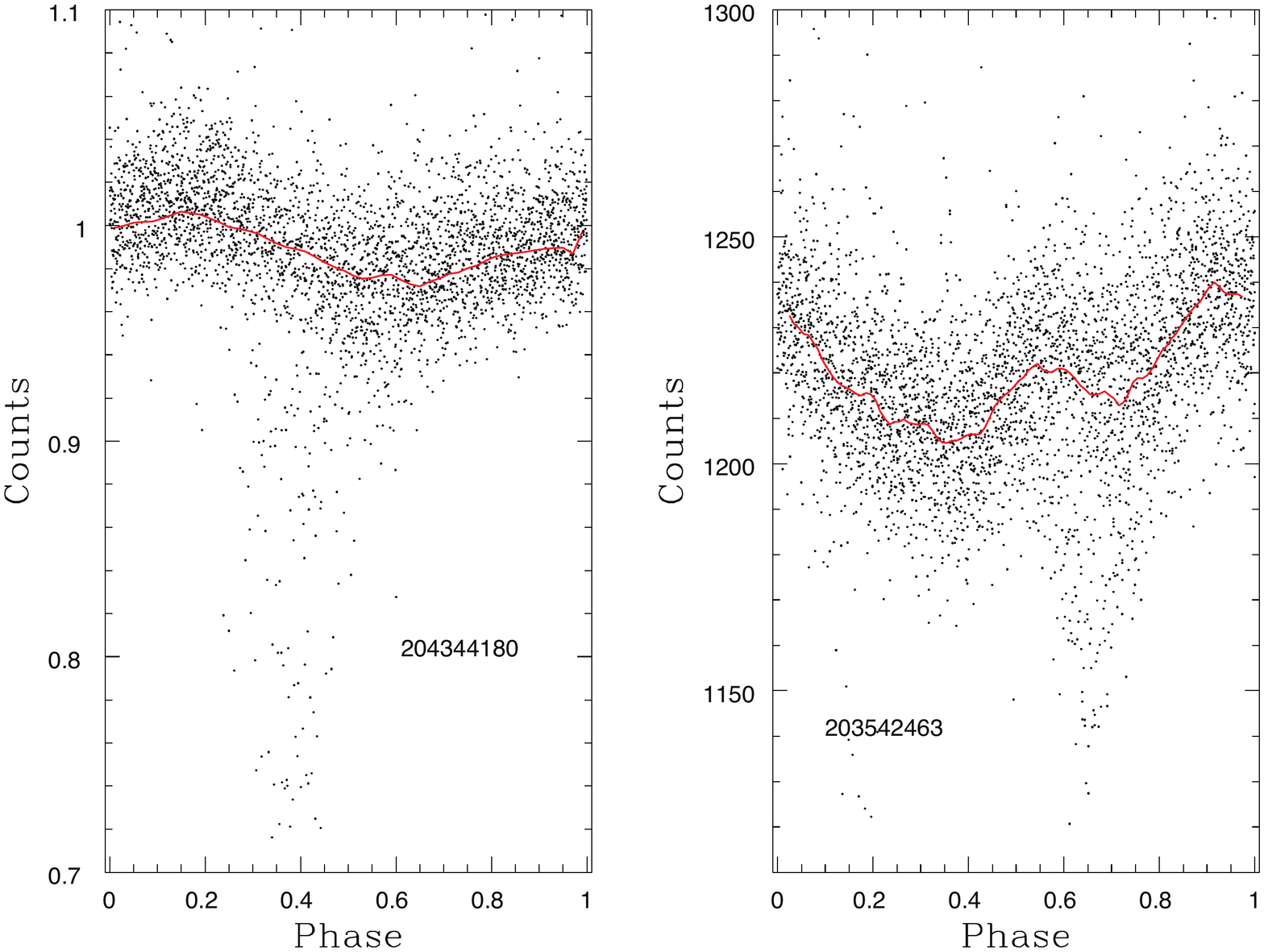}
\caption{Phase-folded light curves for the two stars from
Fig.~\ref{fig:dipperrot} with the smallest variability amplitude for
their spotted-star waveform. These plots show the texture in these LCs
more clearly than the prior figure. }
\label{fig:dipperrot2}
\end{figure}

We have $\sim$290 young stars with IR excesses in our member catalogs
for  USco and \rop, 77\% of which show at least one significant period
in  their Lomb-Scargle  periodogram.  We find that $\sim$10 of these
systems show possible evidence for disk locking; we discuss four of
the best examples and illustrate their light curve morphologies here;
see Figure~\ref{fig:dipperrot}.   These four stars are EPICs 203770366,
204344180\footnote{The K2 light curve for EPIC 204344180 had been
previously noted to contain both spot and flux-dip signatures by
Scholz \etal\ 2015; they also concluded this provided evidence in
favor of disk locking for that star.}, 203542463, and 203860592. The
first three of these stars are late-type (M5 to M6.5) members of USco;
the last star (EPIC 203860592) is a member of \rop\ with spectral type
K5.  Fig.~\ref{fig:dipperrot} shows three versions of the K2 light
curves for each of these stars.   The left panel of each row shows the
complete, detrended light curve for the star.  Each star shows a
stable, periodic pattern extending over the full duration of the K2
campaign, consistent with that expected for a spotted star --
sinusoidal for three of the stars; ``double-dip" for EPIC 203542463.
Amplitudes for the spotted-star variability range from 2.5\% to 15\%,
and periods from 1.6 to 6 days -- typical values for stars of their
spectral type and age.  In all four cases, superposed on the stable
spotted-star variability are intermittently occurring, narrow flux
dips whose amplitudes are comparable to the spotted star signature, or
in some cases much larger than that.

The middle panels of Fig.~\ref{fig:dipperrot} show phase-folded light
curves for these four stars, where the black points are the data and
the red curve is a median fit to the phased light curve.  For all four
stars, the narrow flux dips align in phase, indicating that they
share the same period as the spot waveform.  Because the spotted-star
light curve amplitude is small for EPIC 204344180 and 203542463, we
provide an expanded version of their phased light curves in
Figure~\ref{fig:dipperrot2}.  

The rightmost panels of Fig.~\ref{fig:dipperrot} show the result of
subtracting the median fit to the spotted-star waveform from the
original light curve, now emphasizing the narrow dips that are present
for all four stars.  The red vertical bars are displaced from each
other by intervals of $N$ periods from the first vertical bar.  Only a
portion of the light curves are shown in order to better illustrate
that the narrow flux dips are well-aligned with the bars marking the
period cadence.  For EPIC 203860592, an additional set of blue
vertical bars mark points displaced 0.5 in phase from the red bars
(corresponding to the dimples at the tops of the light curve near day
2080 in the left-panel of the figure for 203860592).  For this star,
we speculate that we are seeing extinction dips due to accretion
columns intersecting our line of sight to the star directed towards
both of its magnetic poles.

If disk locking is ubiquitous for stars with disks, why do we not see
more examples of these types of light curves?   We believe the answer
is that for most of the other stars with disks, the signatures due to
accretion bursts or variable extinction from disk warps are much
higher amplitude than that due to photospheric spots, and so the spot
signatures are masked.

\subsection{Evidence for Period Locking from Period Distributions}

\begin{figure}[ht]
\epsscale{1.0}
\plottwo{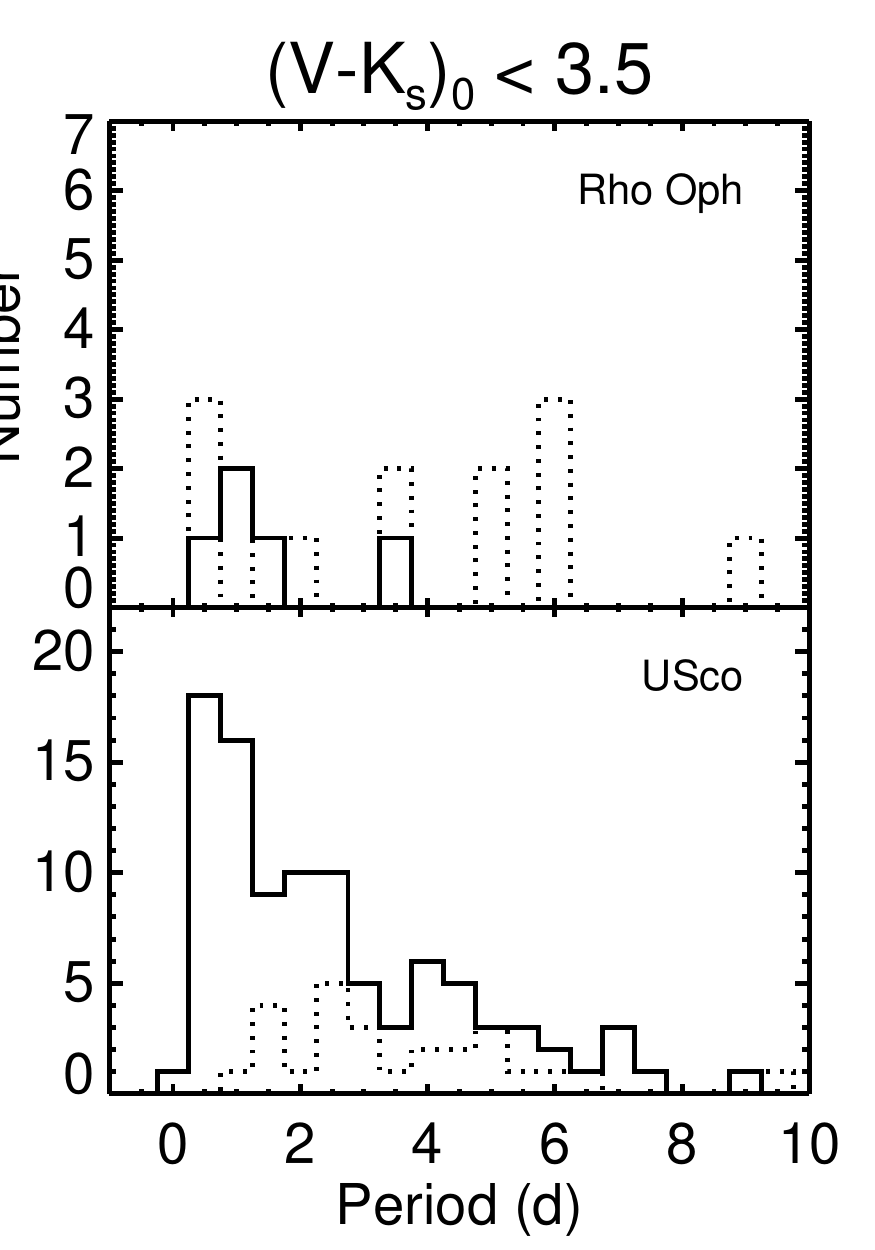}{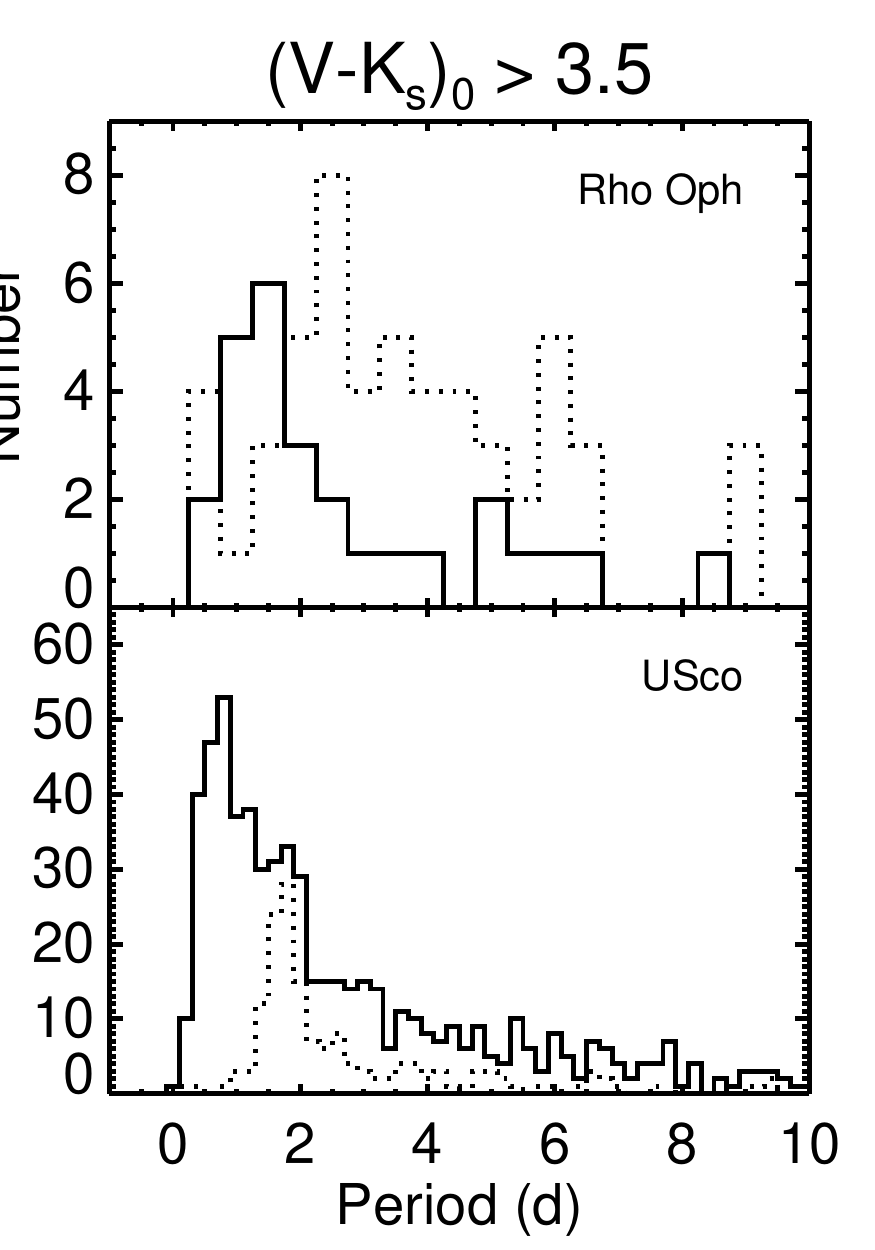}
\caption{Histograms of the (member) periods up to 10 days for USco
(top) and \rop\ (bottom), plotting the distribution of stars clearly
without disks (solid line) separately from the stars clearly with
disks (dotted line), for \vmkz$<$ 3.5 (left) and \vmkz$>$3.5 (right). 
(Note that ambiguous disks or stars without IR data are omitted from
this plot.) Stars with disks on average rotate slightly slower than
those without disks. }
\label{fig:phisto2sepdisks}
\end{figure}

The simplest method to use the period data to search for evidence of
disk locking is to make histograms comparing the period distributions
of the disked and disk-free coeval populations. 
Figure~\ref{fig:phisto2sepdisks} provides such histograms for our K2
data for both \rop\ and USco. Note that only secure disks and secure
non-disks are included in these plots; stars for which there were
insufficent data or ambiguous evidence for an IR excess are omitted. 
Our sample for \rop\ is too sparse to reach a definitive conclusion,
but at least for the late-type stars, the disk-free stars  are
preferentially faster rotating than the disked stars, though not by a
large amount. This is perhaps not unexpected since at this very young
age, disk locking (even if present) would not have had long to
operate.  In contrast, the USco histograms show a much more
significant difference. For both the cooler and warmer stars, the
disk-free stars show a distribution peaking near $P$ = 1 day, with a
broad tail to longer periods; a KS test finds that the high mass and
low mass USco disk-free stars are consistent with being drawn from the
same population. The disked stars in either color range show few or no
stars with $P \sim$1 day, and have a mean period larger than for the
disk-free stars.  Both KS and AD tests of the later \rop\ stars
suggests that there is a 0.2-0.4\% chance that the periods from the
disked and non-disked populations are from the same distribution;
similarly, for USco, there is a $10^{-3}-10^{-8}$\% chance that the
disked and non-disked populations are from the same distribution.
Thus, the K2 USco period data confirms the results found previously in
Orion and other clusters: stars with disks rotate on average more
slowly.

\begin{figure}[ht]
\epsscale{1.0}
\plotone{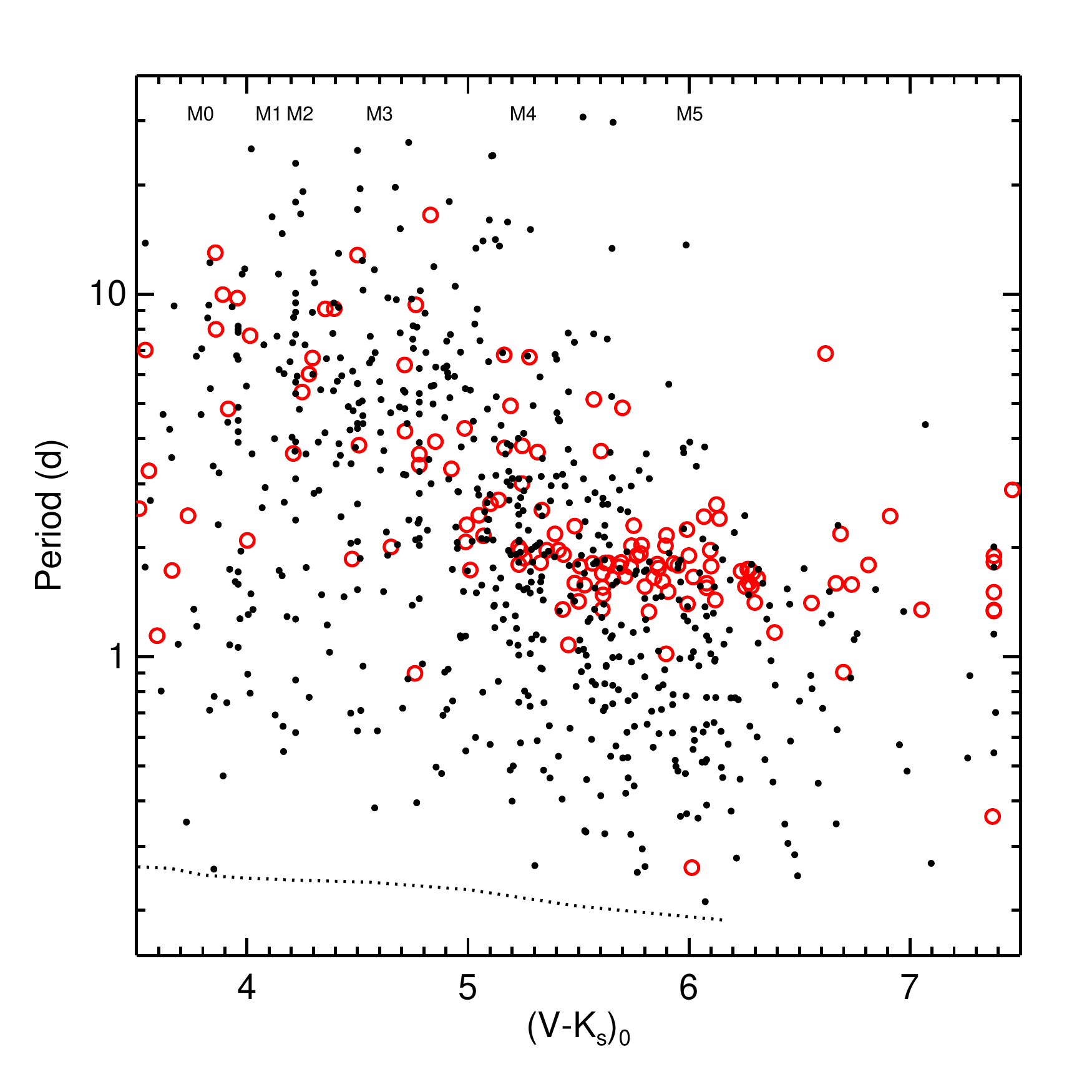}
\caption{$P$ vs.\ \vmkz\ for the M stars in USco, differentiating
between the unambiguous disked stars (red, open circles) and the
unambiguous non-disked stars  (small black dots). Dotted line=breakup,
as discussed in more detail in later Figures; spectral types also
approximately indicated near the top.  There are relatively few disked
stars with periods faster than $\sim$1 d. }
\label{fig:pvmk1mstars}
\end{figure}

\begin{figure}[ht]
\epsscale{1.0}
\plotone{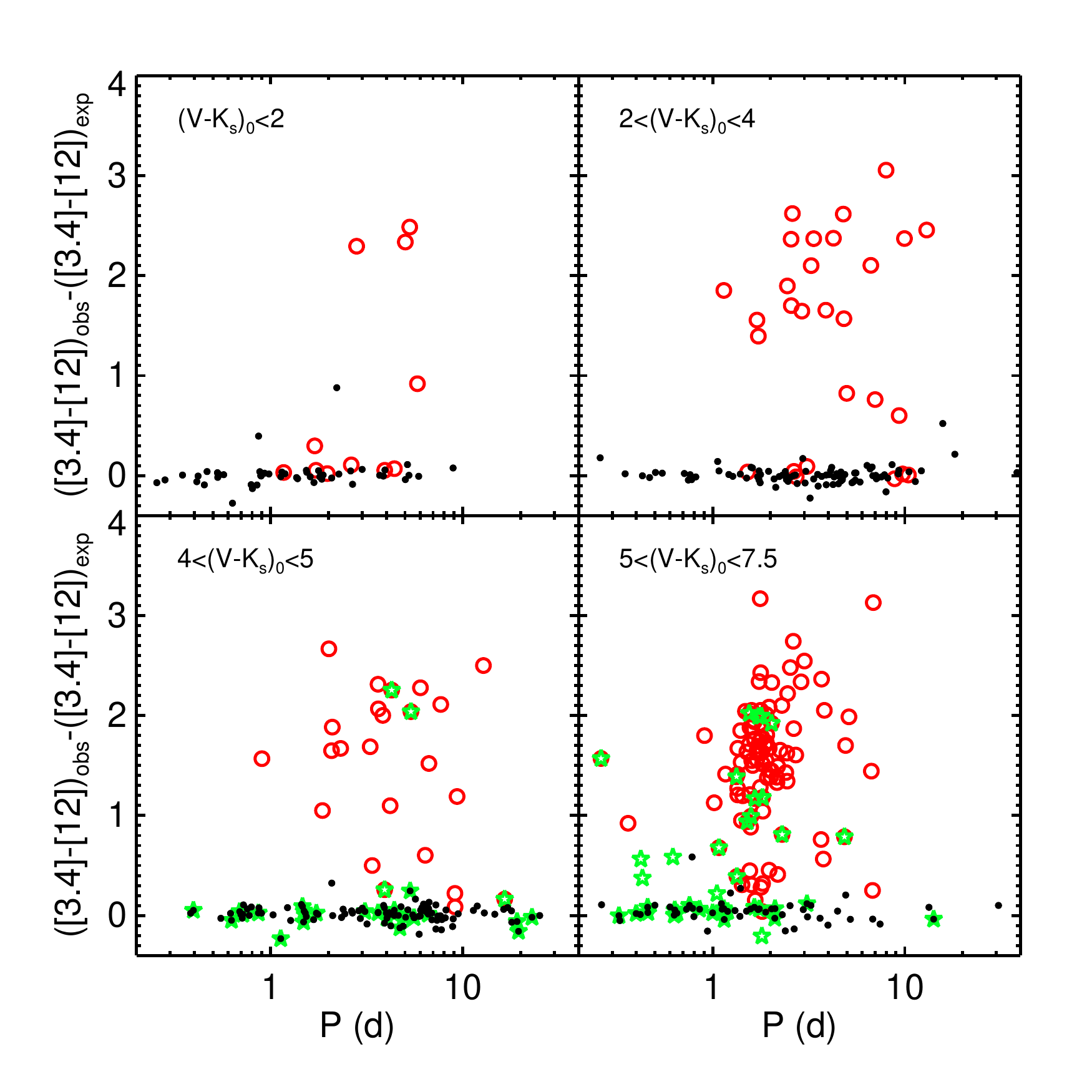}
\caption{Observed [3.4]$-$[12] minus the expected photospheric
[3.4]$-$[12] (see \S\ref{sec:finddisks}) vs.\ $P$ for stars in USco,
differentiating between the high-confidence disked stars (red, open
circles) and all of the non-disked stars (small
black dots).  M stars are the bottom row; left, \vmkz\ 4 to 5, is
roughly M0-M3 and right, \vmkz\ 5-7.5,  is roughly M4-M5. An
additional green star in these panels denotes that it has more than
one period (and is not tagged a pulsator), e.g., a likely binary; see
text. The well-defined disk-free (no IR excess) population  has a wide
range of $P$, whereas stars with disks (with IR excess) have a
narrower $P$ distribution.  }
\label{fig:irx2}
\end{figure}

For the late-type USco stars, there is a prominent peak in the period
histogram of the disked stars at a period near 2d\footnote{We
checked all of the $\sim$2 d LCs extensively to be sure they are real.
We have removed all of the instrumental 1.97d periods as noted
above.}. 
Figure~\ref{fig:pvmk1mstars} illustrates this feature in another way,
comparing periods for the disked and non-disked stars of USco as a
function of  \vmkz\ color for just the M stars (the full range of
colors will be discussed in Sec.~\ref{sec:rotationdistrib}).  
Particularly among the latest type USco members (\vmkz$>$5, spectral
type $>$M4), the two groups of stars show a striking dichotomy in this
diagram, with the stars with $P<$ 1 day being almost entirely
disk-free and the stars with $P >$ 1.5 days being predominantly
disked.  The sharp peak in the disked stars $P$ distribution evident
in Fig.~\ref{fig:phisto2sepdisks} is also prominent here.   The three
late dM stars with K2 light curve evidence for disk-locking in
Fig.~\ref{fig:dipperrot} are part of this group, linking the direct
evidence for disk-locking from the light curves to the somewhat
indirect evidence provided by the period distributions.

Rebull \etal\ (2006, Figure 3) combined data from Spitzer's IRAC
camera and ground-based rotation periods for Orion stars to
demonstrate a correlation between IR excess (hence disk
presence/absence) and rotation at $\sim$1-2 Myr.  We provide similar
diagrams for four mass ranges in USco using IR data from WISE and K2
rotation periods in Figure~\ref{fig:irx2}.  The USco plots mirror the
Orion results very well, showing the well-defined disk-free (no IR
excess) population with a wide range of rotation periods, extending
well below $P$=1 day, and the stars with disks (with IR excesses)
having a narrower distribution of period and avoiding the period range
$P <$ 1.5 day.  Our plots highlight something not obvious in previous
studies of young star rotation.  There is a narrow peak in periods
between 1.5 and 2 d for disked stars with \vmkz$>$ 5 (mass $\lesssim$
0.2 \msun), compared to the broader distribution (weighted to larger
periods) at higher mass.  There is more that can be gleaned from this
Figure; see Sec.~\ref{sec:singleandmulti}. 

Despite the fact that USco should be old enough that (primordial) disk
locking should no longer dominate the stellar rotation, our results
show that at low masses, a fairly large fraction of USco members still
have disks.  For the four color ranges given in Fig.~\ref{fig:irx2},
and considering only those objects with either strong evidence for no
IR excess or strong evidence for an IR excess (at any band), the disk 
fractions are 18$^{+4}_{-6}$, 24$^{+3}_{-4}$, 19$^{+3}_{-4}$, and
64$\pm4$\%, respectively (where errors are calculated as per the 
appendix in Burgasser \etal\ 2003). Disks are both significantly more
common among, and evidently more important for the rotation rates of,
the latest stars here. However, it is important to note that there is
likely a bias against disk-free stars being detected at these lowest
masses, just because of the sensitivity of the longer-wavelength data;
moreover this sample is biased in other ways as well, in that the
stars have to have been selected to have a K2 LC, be bright enough to
have a viable K2 LC (bias against high extinction), and have a
periodic signal (likely bias against disks).

\subsection{Comparison to Previous Studies}

Several previous studies have claimed to find evidence that the
correlation between disk presence and rotation was strongest for
``high mass" stars and was weaker or not present for very low mass
stars or brown dwarfs (Lamm \etal\ 2005; Rodriguez-Ledesma \etal\ 2010;  Davies
\etal\ 2014; Scholz \etal\ 2015).   Our data do not support that
conclusion. If anything, Figure~\ref{fig:irx2} shows the strongest
correlation between IR excess and rotation for the lowest mass group
of stars.   The very low mass, disked stars in USco do have
comparatively short periods (relative to the higher-mass disked stars
in the above papers), but that is probably indicative that their inner
disk edges are comparatively nearer to the star rather than that disk
locking for them is ineffective (see also discussion of this point in
Davies \etal\ 2014).   The conversion from observed color or spectral
type to mass also remains a significant problem for pre-MS
populations, and care must be taken when comparing different studies
that the same or consistent mass scales are being used.

Based on VLT FLAMES high resolution spectra for a sample of $M <$ 0.25
\msun\ Orion stars, Biazzo \etal\ (2009) concluded that disked stars
in that mass range had been disk-locked in the past but no longer were
locked by Orion age (and presumably would not be locked at later but
still young ages).  For the same mass range (our
Figures~\ref{fig:dipperrot} and \ref{fig:irx2}), we find strong
evidence that disked stars at 8 Myr do exhibit disk-locking.   We
infer that our stars in this mass range are still accreting based on
their strong IR excesses; it would be useful to confirm that fact
using high-resolution spectroscopy, and thereby more directly confront
the Biazzo \etal\ conclusion.

\clearpage

\section{Period-Color Distributions}
\label{sec:rotationdistrib}

Now, we explore the distribution of $P$ as a function of \vmkz\ as a
proxy for mass over the full range of stars we have, FGKM.  We compare
these K2 results for \rop\ ($\sim$2 Myr) and USco ($\sim$8 Myr) with
those from papers I-IV on the Pleiades ($\sim$125 Myr; papers I-III)
and Praesepe ($\sim$790 Myr; paper IV). Note that for stars with more
than one period, we have taken the first period and the measured
\vmkz\ as representative of the same star (likely the primary if it is
a multiple); both the assumed \vmkz\ and first period are listed in
Table~\ref{tab:bigperiods}. Even if the star is a multiple identified
only from additional periods and position in the CMD, we don't include
subsidiary companions separately in this analysis.

\subsection{Distribution of $P$ vs.\ $(V-K_s)_0$: \rop\ and USco}
\label{sec:pvmkdistrib}

\begin{figure}[ht]
\epsscale{1.0}
\plotone{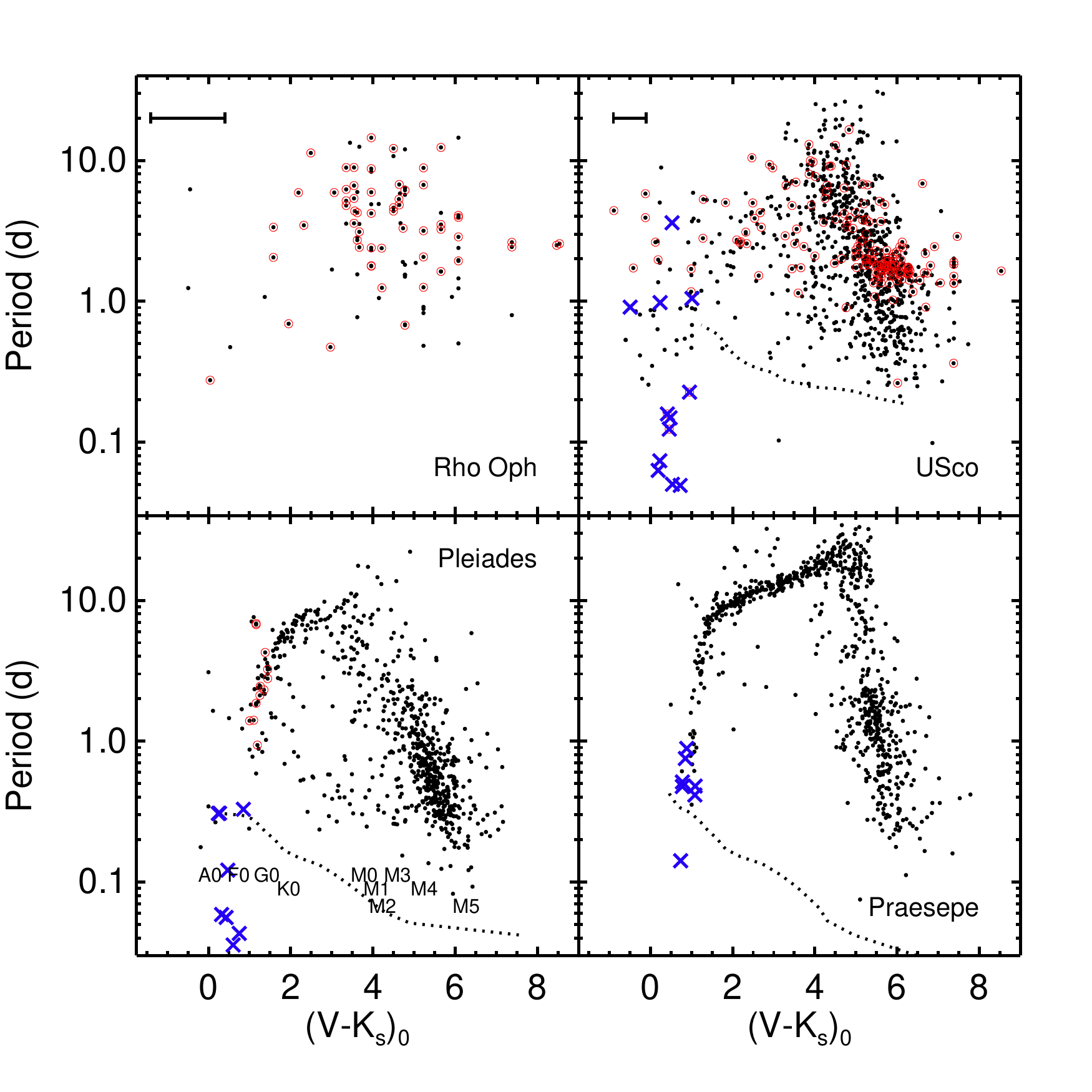}
\caption{$P$ vs.\  \vmkz\ for stars in : \rop\ (upper right, $\sim$1
Myr), USco (upper left, $\sim$8 Myr), Pleiades (lower left, from
papers I-III, 125 Myr), and Praesepe (lower right, from paper IV, 790
Myr). Approximate spectral types are indicated for reference in the
Pleiades panel. Errors on colors are conservatively estimated to be
$\sim$0.9 mag for \rop, $\sim$0.4 mag for USco, and smaller than the
points for the Pleiades and Praesepe; example error bars are given in
the upper left of the top 2 plots. An additional red circle around a
point denotes that the star is an unambiguous disk candidate; note
that Pleiades disks are all debris disks, while most of the disks in
USco and all of those in \rop\ are primordial disks. An additional
blue $\times$ symbol means that the star is likely a pulsator. Dotted
lines in USco, Pleiades, and Praesepe panels are calculated breakup
periods; see text. There are significant changes in this plot across
clusters, and more stars in USco are near the breakup line than in the
other clusters; see text. }
\label{fig:pvmk1}
\end{figure}

Figure~\ref{fig:pvmk1} shows $P$ vs.\  \vmkz\ for stars in all four of
the clusters (\rop, $\sim$1 Myr; USco, $\sim$8 Myr; Pleiades, 125 Myr;
\& Praesepe, 790 Myr). These include all stars in the K2 FOV for which
we derive a period that we interpret as due to rotation.  We discuss
just \rop\ and USco in this subsection. Recall that the apparent
quantization of some stars' \vmkz\ (most apparent in \rop\ and in some
of the latest USco stars) is a result of our dereddening to the
expected \vmk\ color for that spectral type when our more frequently
used dereddening procedure was not possible; see
\S\ref{sec:dereddening}.

There are far fewer stars available in \rop\ than in the other
clusters, but it is clear that most of the stars are rotating
relatively slowly, and a large fraction of the periodic stars also
have disks.  There is significantly less obvious structure in the
\rop\ panel than the other panels, most likely as a result of the
smaller sample size, high extinction and resultant larger uncertainty
on \vmkz\ (see Sec.~\ref{sec:dereddening} above). There are some
apparent echoes of the structure from USco seen in \rop: the slowest
rotators are early M, and there is a steep decrease in $P$ for later
Ms and for earlier FGKs.   We do not discuss the \rop\ rotational data
further.  

USco, in contrast, will be discussed extensively in the rest of this
section. USco stars are rotating quickly, on average. Examination of
the period-color plot for USco shows that there are strong mass
dependent trends.  Later M stars rotate much faster than early M
stars.  At earlier spectral types, the USco stars show a larger
dispersion in period at a given color than the older clusters, but
with the same general trend for the mean period to increase as one
goes from F to G to K. The earliest types with the fastest periods may
be pulsation periods rather than rotation periods.

\subsection{Evolution of $P$ vs.\ $(V-K_s)_0$ Across Clusters}

We now have rotation periods for $\sim$1000 stars at each of three
important ages. USco, at $\sim$8 Myr, is about the time when most
primordial circumstellar disks go away. Pleiades, $\sim$125 Myr, is
roughly the age when low mass  stars arrive on the MS. Praesepe,
$\sim$790 Myr, is after MS angular momentum loss has significantly
altered the ZAMS rotational velocities.   These datasets improve upon
what was available up until now not only by providing larger samples,
but more importantly by providing rotation periods from F stars down
to nearly the hydrogen-burning mass limit, and doing so for nearly
complete sets of stars (within the K2 FOV) for each of the clusters.
In Somers \etal\ (2017), we have recently used the USco and Pleiades
rotational data for the M dwarfs in those two clusters to highlight
the strong dependence of rotation on mass for M$<$0.5 \msun\ at early
ages and how those data constrain angular momentum loss from winds
prior to the ZAMS.  Here we provide a qualitative summary of the
rotational velocity evolution of stars over the full mass range of our
data for the age sequence from USco to Pleiades to Praesepe.

Figure~\ref{fig:pvmk1} includes Pleiades and Praesepe. Throughout the
subsequent discussion, we assume that the stars in these three
clusters represent snapshots in time of the same population, though
that may not be the case (Coker \etal\ 2016). 

The primary conclusion we draw from Figure~\ref{fig:pvmk1} is that the
basic dependence of period on color (and hence, loosely speaking, on
mass) is already set in place by 8 Myr, but with the scatter in period
at a given mass decreasing as age increases.  For FGK stars (mass from
1.5 to 0.5 \msun; 1$<$\vmkz$<$ 3.5), this dependence corresponds to a
monotonically increasing period to lower mass. M stars ($<$0.5 \msun;
\vmkz$>$ 3.5) show the opposite trend, with a strongly decreasing
period as mass decreases. Between USco and Pleiades age, spin-up from
pre-MS contraction and angular momentum loss from winds compete
throughout the entire mass range. In the FGK range, angular momentum
loss from winds dominates and most stars spin down to longer periods;
for the M dwarfs, pre-MS contraction wins and the predominant effect
is for periods to decrease with time. Between Pleiades and Praesepe
ages, pre-MS contraction is only still in process at the lowest masses
and angular momentum loss from winds shifts the distribution to longer
periods at all masses except possibly for the lowest mass M dwarfs
here (\vmkz$\sim$6), where the mean rotation period appears to be very
similar at Pleiades and Praesepe ages.

The much larger scatter in period at a given color in USco is probably
due to a combination of astrophysical and observational influences.
The \vmkz\ colors in USco have larger uncertainties compared to the
other clusters due to the larger extinction corrections and possibly
due to variable extinction or accretion for the stars with long-lived
disks.  The non-simultaneity of the Gaia $G$ and 2MASS $K_s$ mags,
combined with the larger variability amplitudes at $\sim$8 Myr,
compared to the older clusters will also lead to larger uncertainties
in the inferred \vmkz\ for USco stars. A plausible age spread across
the K2 FOV in USco could also broaden the period distribution at a
given mass, whereas that same age spread at Pleiades or Praesepe age
would have a negligible effect on their period distributions.  The
significantly larger contamination of our USco catalog by NM will also
add scatter to its period distribution, particularly by adding slowly
rotating field stars. However, the decrease in scatter in the period
distribution with age likely has another astrophysical
component.  As originally noted by Skumanich (1972), a standard
(non-saturated) wind will cause faster rotating stars to spindown more
rapidly than their slowly rotating counterparts -- thereby causing an
initial spread in rotation rate at a given mass to decrease with time.

\begin{figure}[ht]
\epsscale{1.0}
\plotone{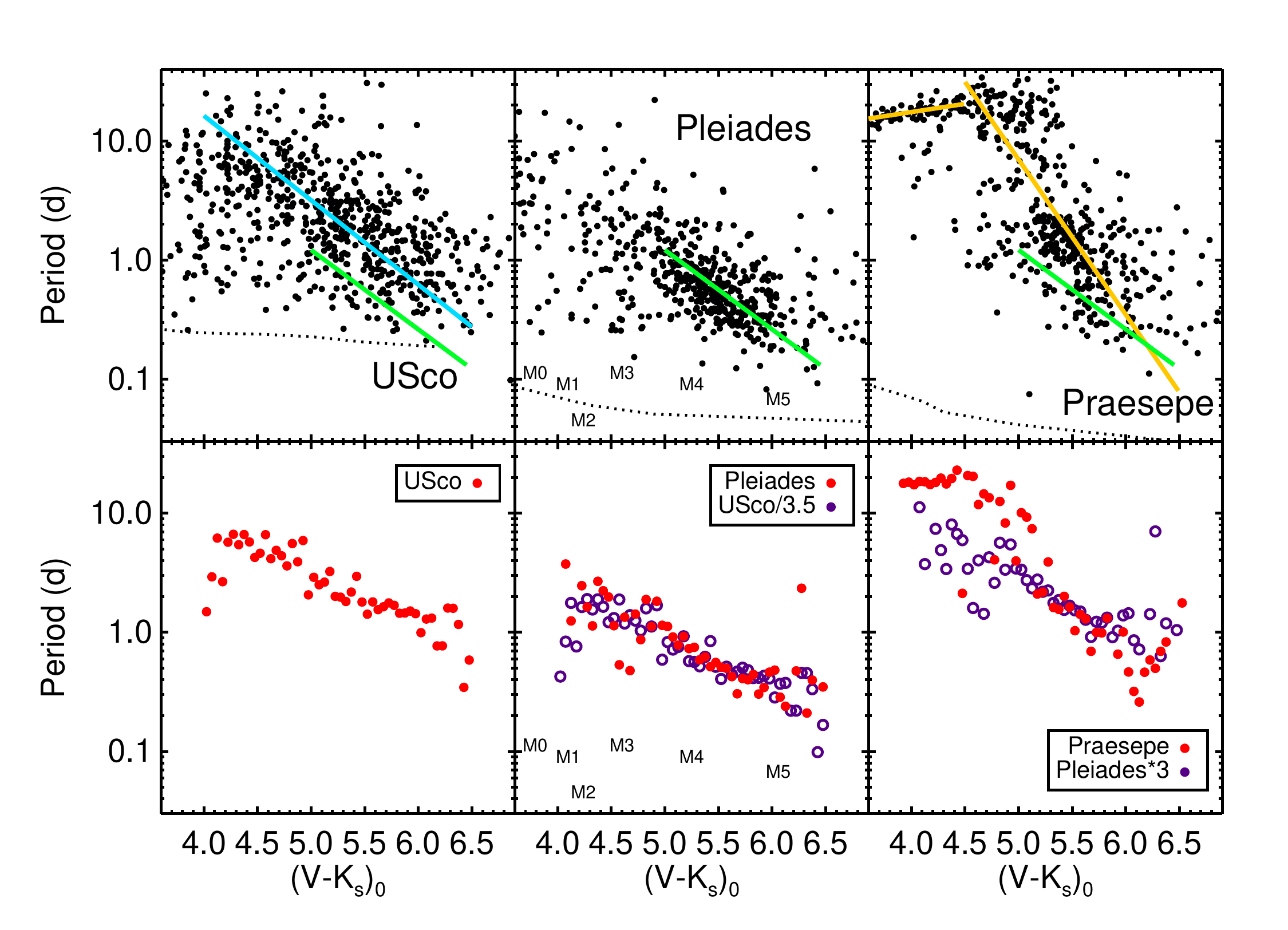}
\caption{$P$ vs.\ \vmkz\ as in Figure~\ref{fig:pvmk1}, but just for M
stars, for USco ($\sim$8 Myr), Pleiades (125 Myr), Praesepe (790 Myr);
unambiguous disk candidates have been removed. Top row: scatter plot
plus linear fits to these data. Light blue line on the first panel:
fit to USco, $\log P = -0.7080(V-K_{\rm s})_0 + 4.0408$; green lines
on all three panels: fit to Pleiades, $\log P = -0.6661(V-K_{\rm s})_0
+ 3.4161$ (since density of Pleiades points falls off dramatically for
\vmkz$<$5, the fit is only shown for 5$<$\vmkz$<$6.5); yellow line on
last panel: fit to Praesepe (from paper IV), where the mid-M stars are
fit by $\log P = -1.303(V-K_{\rm s})_0 + 7.360$. Bottom row: the same
data, plotted as the median period for bins of 0.05 mag in \vmkz. The
Pleiades relationship with that from USco/3.5 is overplotted in purple
open symbols on the center plot; similarly, the Praesepe relationship
is overplotted with Pleiades$\times$3 in purple open symbols. The USco
and Pleiades M stars have an essentially identical slope with the
Pleiades sequence simply spun-up relative to the USco sequence, as the
stars approach the main sequence; Praesepe is much different,
suggesting mass-dependent spin-down on the main sequence. See the text
for more discussion. }
\label{fig:pvmkposter}
\end{figure}

The rotational velocity evolution of FGK stars has been reasonably
well constrained previously using ground-based data. In contrast, the
early evolution of the rotational velocities for the full range of M
stars has not been well-studied using previous data because only K2
has been able to provide rotation periods throughout the 0.5 to 0.1
\msun\ range for such large samples of stars, at a range of ages, and
with only a small fraction of stars without period determinations. We
highlight just the M dwarfs in our three clusters in
Figure~\ref{fig:pvmkposter}, which omits disk candidates. As can be
seen in the top row, in all three clusters, the mean rotation period
decreases with increasing color (decreasing mass). To illustrate in a
qualitative way the evolutionary trends with mass and age, we provide
several lines to guide the eye in the top row of
Figure~\ref{fig:pvmkposter} over the color range where there is a
well-defined locus of points (4$<$\vmkz$<$6.5).  The USco M stars are
fit by a line given by $\log P = -0.7080(V-K_{\rm s})_0 + 4.0408$. The
Pleiades M stars are fit by $\log P = -0.6661(V-K_{\rm s})_0 +
3.4161$. This is a slope that is functionally indistinguishable from
that in USco; the USco relation is about a factor of 3.5 times faster
than that from the Pleiades. That is, $P_{\rm USco} \sim 3.5 \times
P_{\rm Pleiades}$.\footnote{Somers \etal\ 2017 finds a factor of
$\sim$3 rather than $\sim$3.5.}  This implies that M dwarfs spin up
between 8 and 125 Myr by a factor of $\sim$3.5, more or less
independently of mass, if the difference in mapping from $V-K$ to mass
at those two ages is ignored.

In Praesepe, the early M stars compose a large fraction of the slowest
rotating stars in the cluster; by M3, the relationship is falling fast
as color increases.  No simple scaling of the Pleiades line will match
the Praesepe distribution as well as the scaling matched the USco
line.  The early Ms in Praesepe have clearly spun down by a larger
factor and the latest Ms in Praesepe rotate only slightly more slowly
than their Pleiades counterparts.

As a somewhat more quantitative way to compare the M dwarf rotation
period distributions in the three clusters and to justify our claims
for the amount of period evolution between the clusters, we have
computed median periods in bins of 0.05 mag in \vmkz; see the bottom
row of of Figure~\ref{fig:pvmkposter}. The red points in
Fig.~\ref{fig:pvmkposter} are the distributions as observed, but
subtle comparisons between clusters is somewhat difficult. Thus, the
bottom row also includes a scaled version of the prior (younger)
cluster overplotted.  The similarity in the slope of the period-color
trend between USco/3.5 and the Pleiades is evident.  The Pleiades
distribution is multiplied by 3 to match the mid-M Praesepe
distribution; Praesepe has a significantly steeper period-color slope,
again as we had concluded previously.

The K2 rotation data we have presented for these two clusters provide
the basis for a much more quantitative comparison of theoretical
models of angular momentum evolution between 8 and 800 Myr than we
have attempted here.  In Somers \etal\ (2017) we have provided a
preliminary comparison; in a future paper, we plan to
critically assess the ability of the entire K2 clusters period data to
constrain angular momentum losses from the winds of low mass stars.

\subsection{Breakup Velocities}

Figures~\ref{fig:pvmk1} and \ref{fig:pvmkposter} include curves
corresponding to the rotational breakup period using the formula
provided in Maeder (2009, Ch.\ 2), masses, \teff, and radii from
Siess \etal\ (2000) isochrones, and conversions from \teff\ to \vmk\
from Pecaut \& Mamajek (2013).  Several of the most rapidly rotating
high-mass stars in all three clusters can be identified as pulsators.
But there are other short-$P$ high-mass stars that cannot be summarily
categorized as pulsators, at least based on the (sometimes quite
limited) evidence we have besides the very small $P$. However, given
that the period we have would correspond to a rotation rate exceeding
the predicted breakup rate, we assume these periods must be pulsation
periods (or erroneous).

By Praesepe age, even though the most rapidly rotating $\sim$0.1
\msun\ M stars have periods around a quarter day, those periods are
quite far from the predicted breakup speed at 790 Myr. At Pleiades
age, the most rapidly rotating very low mass stars have periods near a
tenth of a day, still about a factor of two slower than the predicted
breakup period for 125 Myr. However, for Upper Sco age, the most
rapidly rotating low mass M stars have rotation periods very near or
coincident with the predicted breakup period at 8 Myr.  For FGK and
early M stars, nearly all stars seem to have solved their ``angular
momentum problem" by USco age (and have periods well removed from the
predicted breakup curve).  However, the USco plot suggests that at
least some of the lowest mass, young M stars may have their rotation
periods set by bumping up against that rotation limit.

If one extrapolates the solid line tracing the locus of $P$ vs.\ color
for USco M stars and the dashed curve for the breakup period to
redder colors, they would meet at \vmkz$\sim$7.0, corresponding
approximately to the main sequence hydrogen burning mass limit. This
could suggest that the rapid and similar rotational velocities of most
brown dwarfs arise because most of them hit their limiting rotational
velocity at an age $\sim$10 Myr.

\clearpage

\begin{figure}[ht]
\epsscale{0.8}
\plotone{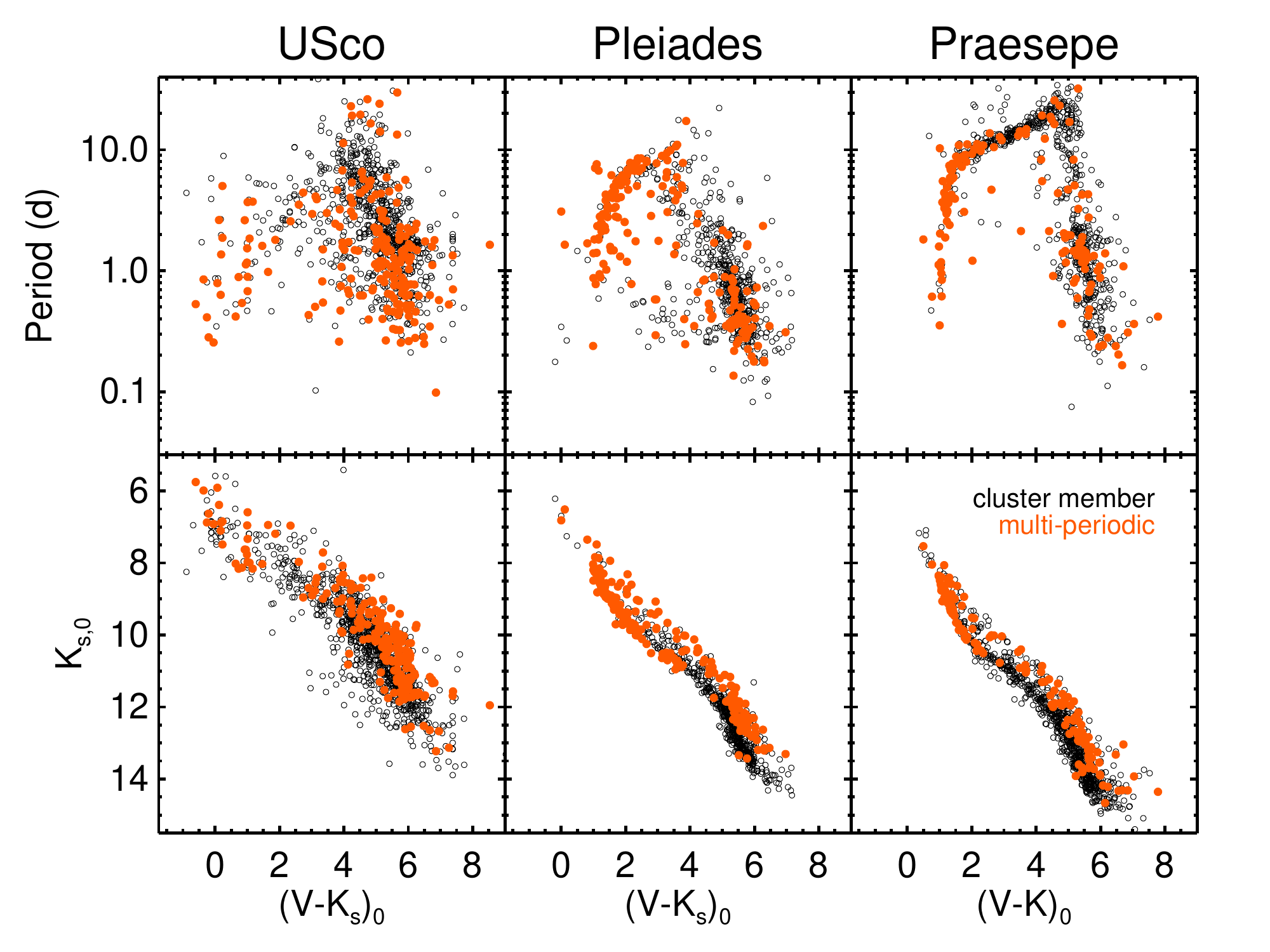}
\caption{Plot of $P$ vs.~\vmkz\ (top) and $K_{s,0}$ vs.~\vmkz\
(bottom) for USco (left), Pleiades (center), and Praesepe (right).  
Multi-periodic sources are highlighted in orange.   For the M stars
(\vmkz$>$3.5),  stars with multiple periods are largely photometric
binaries. 
\label{fig:pvmk3frequsco}  }
\end{figure}

\section{Linkages to Analysis in Papers I-IV}
\label{sec:linkage}

As discussed above, the USco and \rop\ stars and LCs have many
characteristics that make them, importantly, different than the other
clusters we have analyzed with K2 in papers I-IV. However, there are
also many characteristics of the USco and \rop\ LCs that are
comparable to those in Praesepe and the Pleiades, and a comparison of
the clusters using the same approach as in papers I-IV can be
illuminating. We discuss these analyses in this section.

\subsection{Single- and Multi-Periodic Sources: Distribution with
Color}
\label{sec:singleandmulti}
\label{sec:color}

As can be seen in Table~\ref{tab:multipstats}, over all stars,  about
$\sim$20\% of the USco and \rop\ stars with periodic signals also have
at least one plausible additional period. As noted above, this is
quite comparable to the fractions of multiply periodic stars in the
Pleiades and Praesepe.

In the Pleiades and Praesepe, our interpretation of the physical
origin of the multiple periods depended strongly on \vmkz\ color
(hence mass).   For the highest mass stars, in most cases multiple
periods were ascribed to pulsation.  At intermediate mass (spectral
type G or K), where there were two close periods or a complex peak in
the periodogram, we interpreted that as a signature of spot evolution
or latitudinal differential rotation on a single star (this was the
signature seen in $\sim$80\% of the GK stars with multiple periods). A
few GK stars showed two widely separated periods, which we interpreted
as indicative of binary stars.  For the M dwarfs, in most cases, we
interpreted multiple periods as indicative of the rotation periods of
the members of binary or triple systems. We based these
interpretations on the nature of the periodogram morphologies,
placement of the objects in CMDs, and previous studies predicting
significant latitudinal differential rotation for young GK dwarfs but
mostly solid-body rotation for M stars (see paper II). We expect these
same general trends to be present in Upper Sco, but because the CMD
for Upper Sco members has much more scatter than for the older
clusters, it is harder to cleanly separate binary and single stars and
thus harder in particular to determine the primary cause for multiple
periods among the GK dwarfs.   For \vmkz$<$1 (F stars), where the
fraction of stars with multiple periods rises to 63\%, we assume
pulsation dominates.  For \vmkz$>$ 4 (M dwarfs), we expect most of the
multiple period stars to be binaries.

Careful examination of Figure 13 suggests that the fraction of M stars
that are identified as binaries in this way is different for USco than
for the other two clusters.   That is, there appears to be a larger
fraction of red points in the USco plot.  Quantitatively, for the
color range 4.0 $<$ $(V-K$ $<$ 6.5 where our periods are reasonably
complete in all three clusters, the fraction of stars with two
detected K2 periods is 20\% (175 of 709) in USco, 14\% in Pleiades (68
of 501), and 13\% in Praesepe (63 of 496).   This excess of K2
binaries at young ages agrees with other studies that have found a
higher multiplicity fraction at young ages (e.g., Ghez \etal\ 1993;
Prato 2007; Jaehnig \etal\ 2017). There are astrophysical biases,
however, that favor finding more sources with two K2 periods among the
USco M dwarfs compared to the Pleiades and Praesepe samples.  First,
because all three clusters are at similar distance but USco is much
younger, at a given mass its M dwarfs are much brighter; the larger
number of photons should allow us to detect fainter (lower mass)
binary companions and thus make us sensitive to more companions. 
Second, the 8 Myr isochrone is shallower than the ZAMS, and thus at a
given $\Delta$(mass) a companion star will be relatively brighter in
USco than in the other two clusters -- again favoring detection of
lower $q$ systems in USco.  Third, optical light curve amplitudes for
a given mass and period are higher at young, pre-main-sequence ages
than on the ZAMS, which also favors being able to find fainter
companions.   Therefore, our finding a higher fraction of binaries in
USco compared to the other two clusrers may reflect these biases, in
addition to a possibly higher absolute binary fraction.

Assuming that the M stars with multiple periods are in fact multiples,
there are potentially interesting  correlations in our USco data
between multiplicity and disks.  In the
bottom two panels of Figure~\ref{fig:irx2} (which is just the M
stars), we have identified stars with multiple periods (likely
binaries).  Two correlations are evident between rotation and binarity
among the M stars in USco.  First, the stars with disks are less
frequently identified as binaries in this way compared to the
disk-free stars.  Second, at least among the M dwarfs, the
binary fraction is higher among the more rapidly rotating stars. 
Using binomial statistics, for the latest M stars we have here
(mid-M), the probability that the disked stars and non-disked stars
were drawn from a parent population with a common binary fraction is
less than 1 part in a thousand.   The probability that the slowly
rotating half and the rapidly rotating half of the non-disked, mid-M
(latest we have here) stars share the same binary fraction is also
less than 1 part in a thousand. The observed distributions of those
quantities for the early M stars show similar trends, but not at a
statistically significant level.  The correlation between rotation and
binarity for USco and our other K2 clusters will be discussed at
greater length in a future paper (Stauffer \etal\ 2018 in prep).

\subsection{LC and Periodogram Categories}
\label{sec:LCPcats}

In paper II and IV, we classified the LC and periodogram morphologies.
For convenience, the classes we presented in those earlier papers 
are summarized in Appendix~\ref{app:LCPcats}. 

We expect the observed morphologies to reflect trends predicted by
stellar evolution theory. As stars contract towards the main sequence, they will
spin up from momentum conservation and spin down from winds. Disk-related
phenomena should be seen only in the pre-main-sequence. As discussed
earlier, the net evolutionary effects are a spin up from USco to the
Pleiades, followed by a spin down from the Pleiades to Praesepe. 
There is also a convergence in rotation rates for hotter stars as 
they get older. 

We observe phenomena that we interpret as interactions with gas/dust disks
only in the young systems USco and \rop. As described above, there are 
dippers and bursters, and disk-affected light curves, in these two 
young clusters.  There are many of the scallop shell
(and related categories) LCs in these youngest clusters; these are
discussed in Appendix~\ref{app:batwings} and Stauffer \etal\ (2017,
2018).  There are a few of these scallops in the Pleiades, and none in Praesepe.

The older clusters, the Pleiades and Praesepe, have  
LC and periodogram morphologies that also can be found in USco and
\rop.  Table~\ref{tab:countclasses} summarizes the numbers and sample
fractions of the various LC types as defined in the Pleiades and Praesepe. 
Except for \rop, at least 85\% of
the members are periodic. The presence of disks and accretion in the
USco sample apparently does not preclude period derivation at the
same rates as in the older clusters. Of the periodic sample, as noted
above, similar fractions ($\sim$20\%) are multi-periodic. There is
more scatter across the clusters in nearly all of the remaining 
categories, but the distributions are roughly comparable.  
Moving double-dip, shape changers, and complex peaks happen
overall most frequently in Praesepe. Double dips, moving double dips,
and beaters happen least frequently overall in USco and \rop; perhaps
enhanced disk or activity influence in these clusters make it more
likely to find a single sinusoid-family LC, as opposed to a clean
enough ``2 similar dips per cycle'' that defines the double-dip
category.  The fact that the fractions are overall roughly comparable
tells us that there is no large change in the surface phenomena
exhibited by young stars over two orders of magnitude in age. There
are likely to be, however, color dependencies; see next section.

\floattable
\begin{deluxetable}{lccccccccccccc}
\tabletypesize{\scriptsize}
\floattable
\rotate
\tablecaption{Star/Light Curve/Periodogram
Categories\label{tab:countclasses}\label{tab:multipstats}}
\tablewidth{0pt}
\tablehead{\colhead{Category} 
& \multicolumn{3}{c}{{\bf Praesepe}} 
& \multicolumn{3}{c}{{\bf Pleiades}} 
& \multicolumn{3}{c}{{\bf USco}} 
& \multicolumn{3}{c}{{\bf \rop\ }} \\
& \colhead{Number} & \colhead{Frac.~of } & \colhead{Frac.~of } 
& \colhead{Number} & \colhead{Frac.~of } & \colhead{Frac.~of } 
& \colhead{Number} & \colhead{Frac.~of } & \colhead{Frac.~of } 
& \colhead{Number} & \colhead{Frac.~of } & \colhead{Frac.~of }\\
&  & \colhead{sample} & \colhead{periodic sample}& 
& \colhead{sample} & \colhead{periodic sample}&
& \colhead{sample} & \colhead{periodic sample}&
& \colhead{sample} & \colhead{periodic sample}}
\startdata
Periodic & 809 & 0.86 & 1.00  & 759 & 0.92 & 1.00 & 969 & 0.86 & 1.00 &  108 & 0.60 & 1.00   \\
Single period & 645 & 0.69 & 0.80 & 559 & 0.68 & 0.74  &  751 & 0.66 & 0.78 &   86 & 0.48 & 0.80  \\
Multi-period & 164 & 0.17 & 0.20 & 200 & 0.24 & 0.26 &  217 & 0.19 & 0.22 &   22 & 0.12 & 0.20  \\
Double-dip & 163 & 0.17 & 0.20 & 107 & 0.13 & 0.14 &  132 & 0.12 & 0.14 &    6 & 0.03 & 0.06  \\
Moving double-dip  & 121 & 0.13 & 0.15 & 31 & 0.04 & 0.04 &   32 & 0.03 & 0.03 &    0 & 0.00 & 0.00  \\
Shape changer & 297 & 0.32 & 0.37 & 114 & 0.14 & 0.15 &  277 & 0.24 & 0.29 &   48 & 0.27 & 0.44  \\
Orbiting clouds? & 0 & 0 & 0 &  5 & $<$0.01 & $<$0.01 &   28 & 0.02 & 0.03 &    6 & 0.03 & 0.06  \\
Beater &  77 & 0.08 & 0.10 & 135 & 0.16 & 0.18 &  107 & 0.09 & 0.11 &   10 & 0.06 & 0.09  \\
Complex peak &  68 & 0.07 & 0.08 & 89 & 0.11 & 0.12&    8 & 0.01 & 0.01 &    0 & 0.00 & 0.00  \\
Resolved, close peaks &  68 & 0.07 & 0.08 & 126 & 0.15 & 0.17&  150 & 0.13 & 0.15 &   12 & 0.07 & 0.11  \\
Resolved, distant peaks  &  71 & 0.08 & 0.09 & 39 & 0.05 & 0.05 &   84 & 0.07 & 0.09 &   10 & 0.06 & 0.09  \\
Pulsator  &  17 & 0.02 & 0.02 & 8 & 0.01 & 0.01 &   12 & 0.01 & 0.01 &    0 & 0.00 & 0.00  \\
\enddata
%\tablenotetext{a}{}
\end{deluxetable}

%\clearpage

\begin{figure}[ht]
\epsscale{0.8}
\plotone{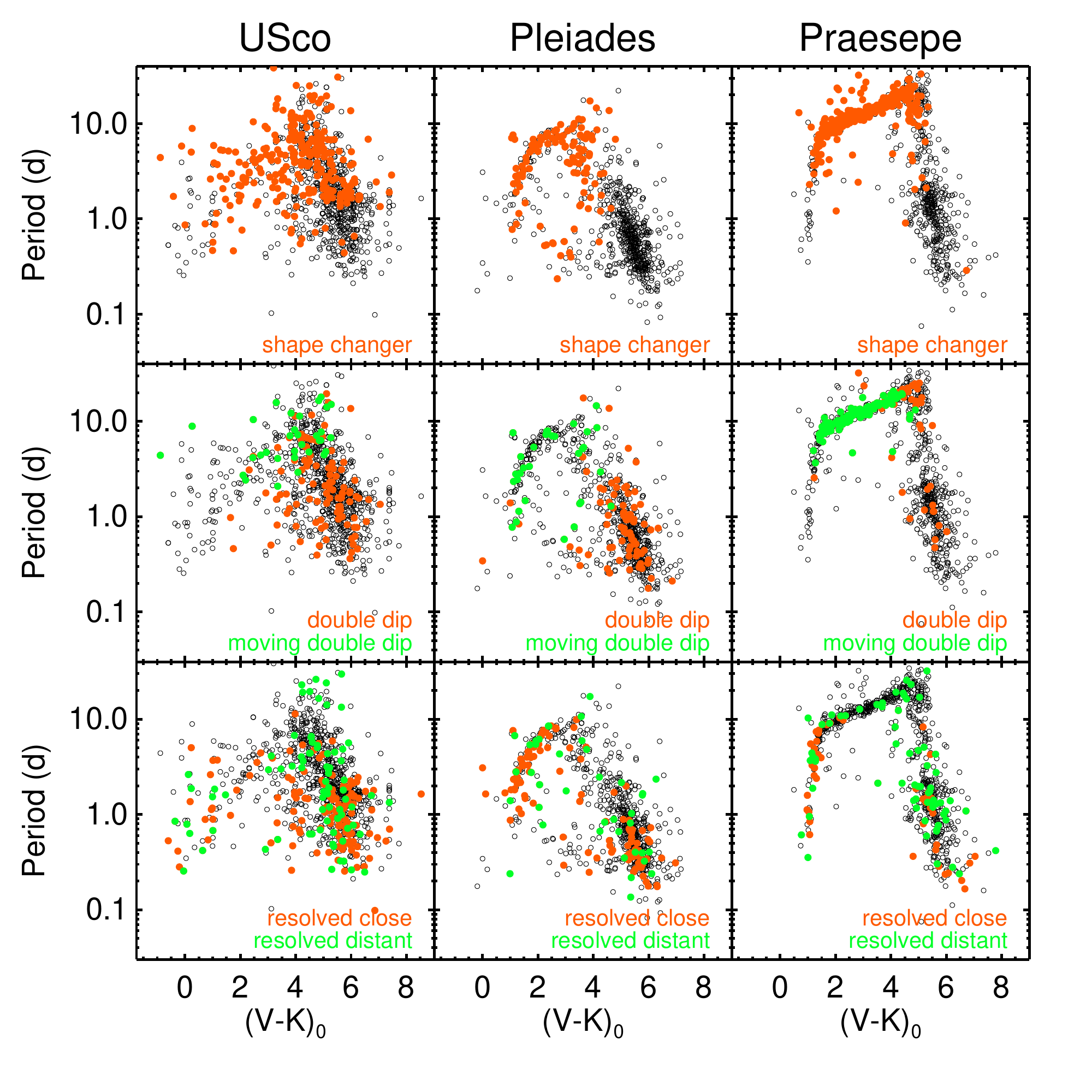}
\caption{Plot of $P$ vs.~\vmkz\  for USco (left), Pleiades (center),
and Praesepe (right), highlighting shape changers (top row),
double-dip stars (middle row), and resolved close (orange) and
resolved distant (green) peaks (bottom row).  The USco shape changers
and to some extent the moving double-dip stars seem to extend to
redder colors in than the other clusters. There are more double-dip M
stars in USco and the Pleiades than Praesepe.  Resolved close peak
sources tend to be quite localized to the earlier types in Praesepe
and the Pleiades; they are more equally distributed in USco. This
sheds light on the rates of spot evolution and/or surface differential
rotation and magnetic field structure stability.
\label{fig:pvmk4classesusco}  }
\end{figure}

\subsection{$P$ vs.\ \vmk: LC Categories}
\label{sec:pvmklcclass}

Figure~\ref{fig:pvmk4classesusco} shows where stars fall in the  $P$
vs.~\vmkz\ diagram for 3 of the LC types where there are differences
across the three most well-populated clusters (USco, Pleiades, Praesepe).

The USco shape changers and to some extent the moving double-dip
stars seem to extend to redder colors than in the other clusters.
Assuming that we have correctly interpreted the shape changers and the
moving double-dip LCs as spot evolution and/or surface differential
rotation, then these phenomena are found on lower mass stars at USco
age compared to the other clusters. 

There are more double-dip M stars in USco and the Pleiades than
Praesepe.  Assuming that double-dip LCs originate from two, stable
spots/spot groups on opposite hemispheres, this implies that the M
star magnetic field structures are more stable over $\sim$70 d to
lower masses in USco than the older clusters.

Resolved close peak sources tend to be quite localized to the earlier
types in Praesepe and the Pleiades; they are more equally distributed
in USco. If these LCs can be interpreted as multiple stable spots/spot
groups and surface differential rotation, then again, these phenomena
may be found on lower mass stars at USco age compared to the other
clusters. On the other hand, these could also be M star binaries; see
Stauffer \etal\ (2018).

\begin{figure}[ht]
\epsscale{1.0}
\plotone{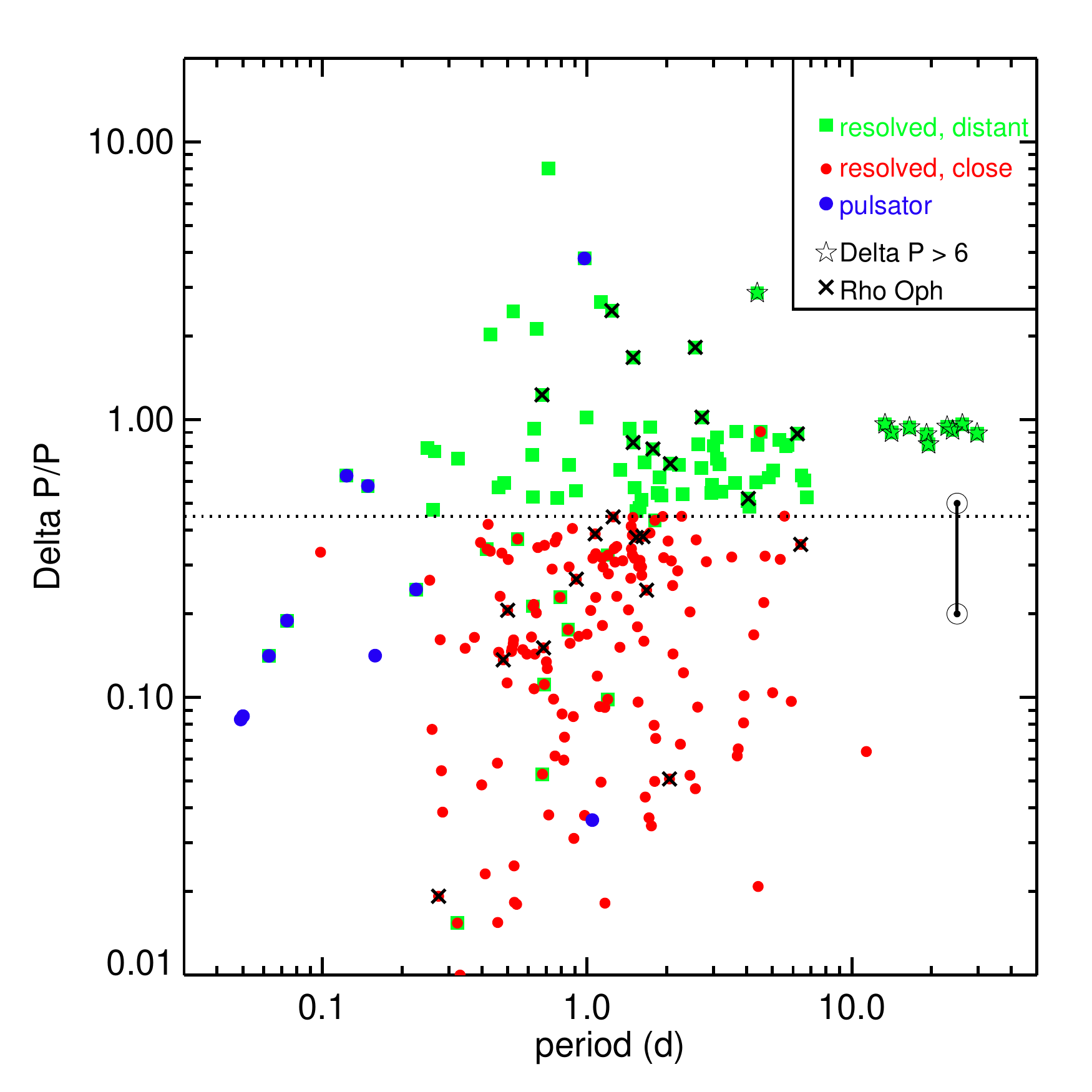}
\caption{Plot of $\Delta P/P_1$ vs.~$P$ for pulsators (blue dots),
resolved distant peaks (green squares), and resolved close peaks (red
dots). An additional black star indicates that $|\Delta P|>$6 d.  An
additional black '$\times$' indicates a \rop\ member.  The range of
possible values for the  Sun is included for reference ($\odot$);  if
one takes as $\Delta P$ the range of periods measured where sunspots
occur, $\Delta P/P_1 \sim 0.1-0.2$, but if one takes the full range of
$\Delta P$, equator to pole, $\Delta P/P_1 \sim 0.5$. The dotted line
is at $\Delta P/P_1$=0.45 and denotes the boundary between close and
distant resolved peaks.  There are far fewer structures in this
diagram as compared to that for the Pleiades; it's possible that there
are not enough complete cycles in the $\sim$70d campaign for
extraction of clear multiple periods, or the disk influence on the LCs
complicates derivation of the periods.}
\label{fig:deltap}
\end{figure}

\subsection{$\Delta P$ distributions}
\label{sec:deltaP}

As for the other clusters, we calculated the $\Delta P/P_{\rm rot}$
metric for stars with resolved multi-period peaks; see
Figure~\ref{fig:deltap}. We take the closest peak to the $P_{\rm
rot}$, subtract the smaller from the larger, and divide by the $P_{\rm
rot}$. The distribution here, like that for Praesepe (paper IV), has
relatively little structure as compared to the Pleiades (paper II). It
might be the case, as we postulated for Praesepe, that the $\sim$70 d
K2 campaigns are not quite long enough to capture enough complete
cycles of the longer periods in USco (and Praesepe) to resolve peaks
in the periodogram. The Pleiades stars are, on average, rotating the
fastest that these stars will ever rotate, so there are enough cycles
(on average) to resolve the peaks and place them on this diagram. 

As for Praesepe (paper IV), there are several M stars with multiple
periods where the difference between the two periods exceeds 6d; these
must be unresolved binaries.   These systems include some of the most
slowly rotating stars in the USco ensemble. They also appear high in
the CMD on average, supporting the idea that they are binaries.

\section{Summary and Conclusions}
\label{sec:concl}

In this paper, we have presented our analysis of rotation rates for
USco ($\sim$8 Myr) and \rop\ ($\sim$1 Myr). While the data reduction
and collection of ancillary data from the literature and archives was
relatively straightforward (and similar to methods we used in papers
I-IV), determining cluster members and dereddening was harder in these
clusters, and as a result, there may be more non-members, and more
scatter in our proxy for mass \vmkz\ in USco and \rop, than there were
in the Pleiades ($\sim$125 Myr; papers I-III) and Praesepe ($\sim$790
Myr; paper IV). Disks are much more important in USco and \rop, both
in terms of their influence on the LC shapes and the rotation rates
themselves; we identified disks via an IR excess at any available
band. Bursters (sudden brightenings) and dippers (sudden fadings) can
both be found in our LCs.

Our analysis places USco (and \rop, though it
has many fewer stars) in context with the Pleiades and Praesepe
(Fig.~\ref{fig:pvmk1}). We now have rotation periods for  $\sim$1000
stars at each of three important ages: $\sim$8 Myr (USco) is about the
time when most primordial circumstellar disks go away; $\sim$125 Myr
(Pleiades) is roughly the age when low mass stars arrive on the MS;
$\sim$790 Myr (Praesepe) is after MS angular momentum loss has
significantly altered the ZAMS rotational velocities.   Given our
results, the basic dependence of period on color (and hence, loosely
speaking, on mass) is already set in place by 8 Myr, but with the
scatter in period at a given mass decreasing as age increases; we note
that some of the scatter in USco may be due to the larger
uncertainties in reddening corrections and membership. For the FGK
stars, the Pleiades $P$ distribution is bimodal, with a dominant
`slow sequence.' The USco distribution is more unimodal and requires
more spin-up for the rapid rotators and more spin-down for the slow
rotators in order to match the Pleiades distribution.  Because of
strong structural evolution, for these stars we caution that a full
interpretation requires theoretical modeling outside the scope of this
paper.

The rotation rates for the M stars have been probed for the first time
using K2. The relationship between color (mass) and log $P$ has the
same slope between USco and the Pleiades; $P_{\rm USco} \sim 3.5 \times
P_{\rm Pleiades}$. M dwarfs spin up between 8 and 125 Myr by a factor
of $\sim$3.5, more or less independently of mass. In contrast, no simple
scaling of the Pleiades or USco relationship matches the Praesepe
distribution as well.  The early Ms in Praesepe have clearly spun down
by a larger factor and the latest Ms in Praesepe rotate only slightly
more slowly than their Pleiades counterparts.

Very few stars in the Pleiades or Praesepe rotate near breakup; in
contrast, in USco, the most rapidly rotating low mass M stars have
rotation periods very near or coincident with the predicted breakup
period at 8 Myr.  Some stars in USco may have their rotation rates set
by this limit.

Disks have a significant influence on the rotation rates in USco.
There are several instances in our set of LCs where dipping behavior
is synchronized with the spot-modulated variations, indicating that
dipping (from the inner disk) is locked to the stellar rotation rate
(from the spot-modulation).  There is a clear pile-up of disked M stars
with rotation rates near 2 d. Stars with disks rotate on average more
slowly than stars without, consistent with results from other young clusters.

The LCs and periodograms from USco and \rop\ can be analyzed using a
similar approach as we used for the Pleiades (papers I-IV) and
Praesepe (paper IV). We can place the objects into the same 
LC/periodogram categories as previously defined. There are very
similar fractions of multi-periodic stars; there are significantly
more of the scallop shell (and related categories) LCs in the youngest
clusters. Other categories appear at roughly the same rates as for the
older clusters. Multi-periodic M stars are likely to be binaries in
all of these clusters.

%\clearpage

\acknowledgments

Thanks to Adric Riedel and Kevin Covey for various versions of
membership lists.

Some of the data presented in this paper were obtained from the
Mikulski Archive for Space Telescopes (MAST). Support for MAST for
non-HST data is provided by the NASA Office of Space Science via grant
NNX09AF08G and by other grants and contracts. This paper includes data
collected by the Kepler mission. Funding for the Kepler mission is
provided by the NASA Science Mission directorate. 
This research has made use of the NASA/IPAC Infrared Science Archive
(IRSA), which is operated by the Jet Propulsion Laboratory, California
Institute of Technology, under contract with the National Aeronautics
and Space Administration.    This research has made use of NASA's
Astrophysics Data System (ADS) Abstract Service, and of the SIMBAD
database, operated at CDS, Strasbourg, France.  This research has made
use of data products from the Two Micron All-Sky Survey (2MASS), which
is a joint project of the University of Massachusetts and the Infrared
Processing and Analysis Center, funded by the National Aeronautics and
Space Administration and the National Science Foundation. The 2MASS
data are served by the NASA/IPAC Infrared Science Archive, which is
operated by the Jet Propulsion Laboratory, California Institute of
Technology, under contract with the National Aeronautics and Space
Administration. This publication makes use of data products from the
Wide-field Infrared Survey Explorer, which is a joint project of the
University of California, Los Angeles, and the Jet Propulsion
Laboratory/California Institute of Technology, funded by the National
Aeronautics and Space Administration. 

\facility{K2, 2MASS, WISE, IRSA, Exoplanet Archive}

\clearpage

\appendix

\section{New Upper Sco Membership Catalog for K2 Campaign 2 Targets}
\label{app:member}

More than a thousand stars were included in the observation list for
K2 campaign 2 (K2C2) because they were candidate or known members of Upper
Sco or $\rho$ Oph.  However, thousands of other K2C2 targets
were simply selected as low mass stars in the direction of Upper Sco
(often with the hope of finding transiting exoplanets); some of these
stars could potentially also be Upper Sco members.   We therefore used
a two step process in order to identify likely USco and $\rho$
Oph members to include in our study of those clusters.   

In the first step, we conducted a literature search to identify all
stars cited as probable members of the two clusters and for which
there was either spectroscopic confirmation of that membership or IR
data indicating the presence of a circumstellar disk.  The papers we
used for this purpose included Preibisch \etal\ (1998, 2001, 2002), 
Wilking \etal\ (2005), Erickson \etal\ (2011), Slesnick \etal\ (2006),
Kraus \& Hillenbrand (2007), Lodieu \etal\ (2011), Luhman \& Mamajek
(2012), and Rizzuto \etal\ (2011, 2012, 2015).   We also accepted
literature members whose Gaia TGAS DR1 proper motions and parallaxes
supported membership.   We then cross-correlated this list with the
list of stars which had been targeted as Upper Sco members during
Campaign 2 of the K2 mission, resulting in a list of $\sim$700
``confirmed" literature members of Upper Sco and $\sim$50 ``confirmed"
members of $\rho$ Oph.

Next, for the entire set of $\sim$13,400 stars observed during K2C2,
we compiled accurate photometry and proper motions from a variety of
all-sky surveys. Specifically, for all these stars, we compiled Gaia
$G$ magnitudes and 2MASS $JHK_s$ magnitudes.  Where Gaia did not
provide a $G$ magnitude, we compiled $I$ magnitudes from the DENIS
survey (Foqu\'e \& Bertin 1995, Epchtein \etal\ 1999)  and/or $r$
magnitudes from the Carlsberg meridian catalog (Mui\~nos \& Evans
2014).  We further compiled proper  motions for all the K2C2 stars from
the GPS1 (Tian \etal\ 2017),  UCAC5 (Zacharias \etal\ 2017) and  PPMXL
(Roeser \etal\ 2010) catalogs.

From the catalog photometry, we derived a measured $G-K_s$ color for
each star,  for those stars which were included in the Gaia DR1 data
release.  From this set of stars, we derived polynomial relations
between $G-K_s$ and $I_{\rm DENIS}-K_s$ and $r_{\rm Carlsberg}-K_s$. 
For those stars not in the Gaia DR1 catalog, we used those polynomial
relations to provide estimates of $G-K_s$.  In this way, we were able
to provide either measured $G-K_s$ colors or estimates of $G-K_s$ for
more than 98\% of the stars in the K2C2 target list.

We also wanted to derive a single proper motion value to attach to
each of the K2C2 stars.  Each of the proper motion surveys had
different magnitude ranges over which they provided data.  For each
catalog, we plotted vector point diagrams for the entire K2C2 set, and
noted the centroid position of the cloud of points at the expected
position for Upper Sco.  Those centroids differ slightly from one
catalog to the next, which we assume is due to small zero-point
offsets between the catalogs. We applied small offsets to the catalog
motions to remove those zero point offsets.  Then, where we had data
from all three catalogs, we checked to see if one of them was
significantly discrepant from the others -- in which case, we did not
consider that proper motion measurement and we averaged the values
from the other two catalogs.  Where we had proper motions from two
catalogs, we just took the average.  This process resulted in our
having relatively homogenous proper motion estimates for $>$98.5\% of
the K2C2 stars.

\begin{figure}[ht]
\epsscale{0.95}
\plotone{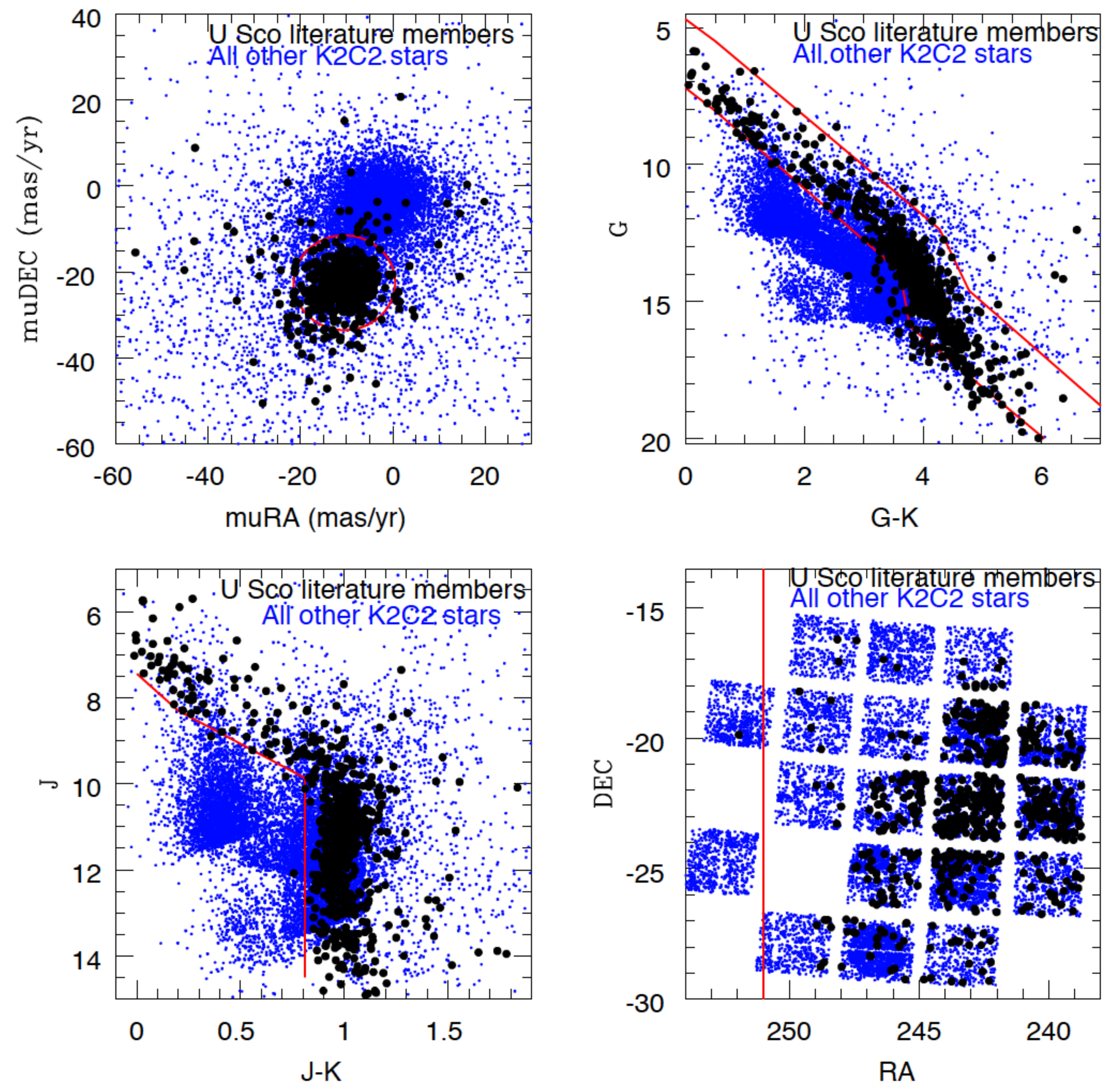}
\caption{Plots used to define regions of parameter space within which ``bronze"
candidate members of Upper Sco must reside.  Black dots are USco literature
members; blue dots are all other stars in K2C2.  Red curves and lines denote
boundaries adopted to separate USco members from field stars.
\label{fig:membercat_select}}
\end{figure}

For USco, we next plotted the ``confirmed" member stars in $G$ vs.\
$G-K_s$ and $J$ vs.\ $J-K_s$\ color-magnitude diagrams, a spatial
plot, and a vector-point diagram, and demarcated regions in these
parameter spaces which enclosed the majority of the literature members
(see  Figure \ref{fig:membercat_select}). We designated the 551
literature members that satisfied all of these criteria as our ``gold"
members of Upper Sco.   Because there is high probability that all of
the literature members are indeed Upper Sco or $\rho$ Oph members, we
designated as ``silver" members of Upper Sco those 71 stars from the
literature sample that had proper motions in our compilation  within
15 mas yr$^{-1}$ from the expected Upper Sco motion.   Finally, we
identified as ``bronze" Upper Sco members the 511 additional stars
from the K2C2 target list that fell within the expected regions in all
four diagrams of Figure \ref{fig:membercat_select}.   The final Upper
Sco catalog includes 1133 stars.  

For \rop, a similar process resulted in 88 stars for our catalog of
candidate members of $\rho$ Oph; 19 gold, 31 silver, and 46 bronze. 
However, the process to this point has excluded many very embedded
\rop\ members. We realized that a pure proper motion and CMD-based
selection will omit embedded stars because the data used for the
proper motions do not exist for the very embedded sources. There are
many such sources  near the cluster ($\alpha$ between 246.14 and
247.43; $\delta$ between $-$25.19 and $-$24, e.g., the box in
Fig.~\ref{fig:where}) that were not already identified as USco members
above, but whose $JHK_s$ suggests high extinction. We additionally
selected sources (assigned to the bronze confidence level) that were in
that spatial box in Fig.~\ref{fig:where}, with inferred \av$>$2.0 or
$J-H>$1.0, and which had $J\leq$14. By imposing the \av\ requirement,
we eliminate all stars foreground to the Oph clouds.We assume that
within that spatial range, the molecular cloud has enough dust column
density to extinct all background stars by enough so that they would
be optically invisible (or at least too faint for a useful K2 light
curve).  Because these stars are very embedded, stars with $J>$14
often have effectively useless optical K2 LCs; those that we omitted
that still have detectable $P$ are included in the non-member catalog
below.   Any field star that happens to have wandered inside the
molecular cloud would get accepted as a member using this approach,
but we assume that is at most a handful of stars -- and they mostly
will not be periodic.

\begin{figure}[ht]
\epsscale{0.95}
\plotone{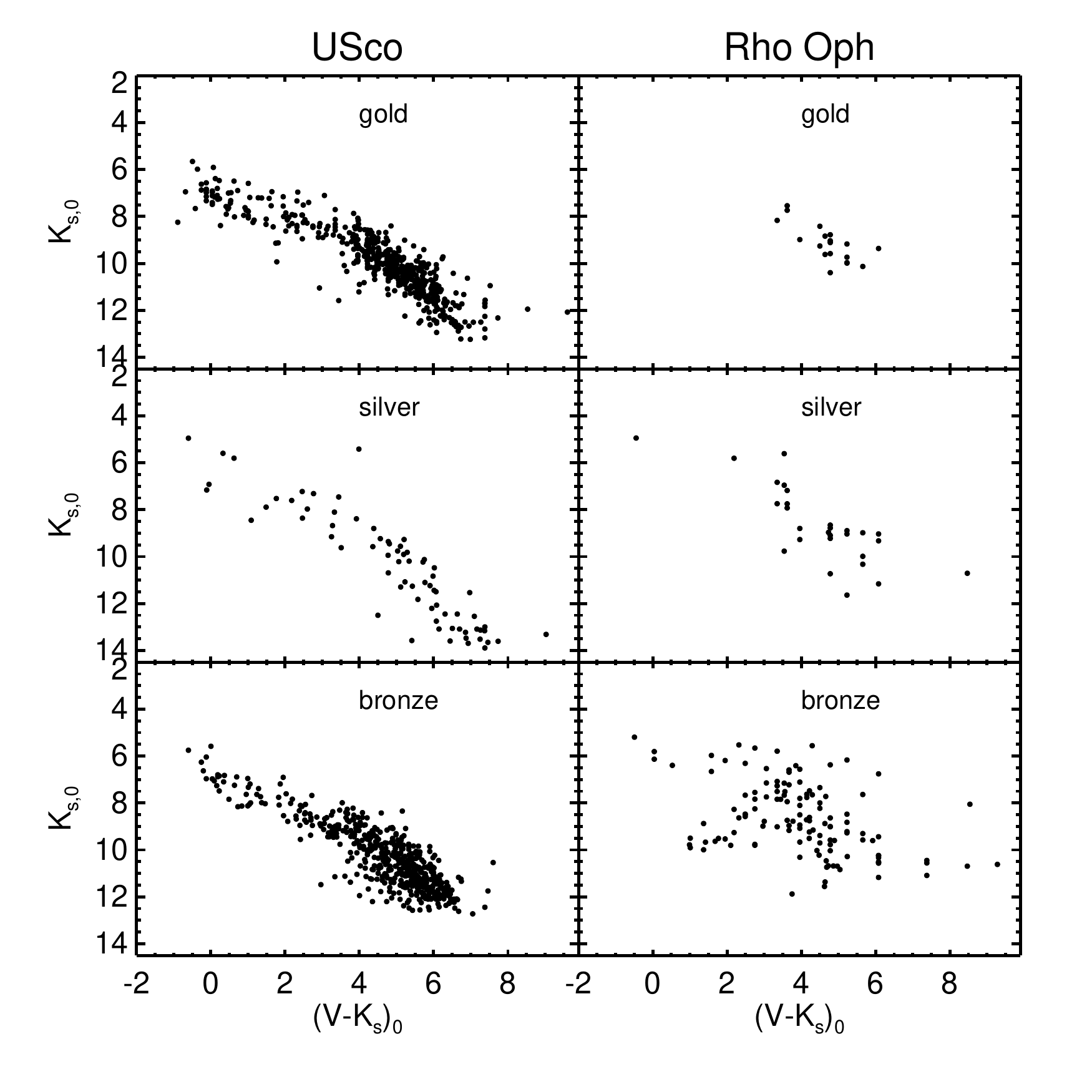}
\caption{CMD with gold/silver/bronze members highlighted for USco and
\rop.
\label{fig:optcmdgsb}}
\end{figure}

\begin{figure}[ht]
\epsscale{0.95}
\plotone{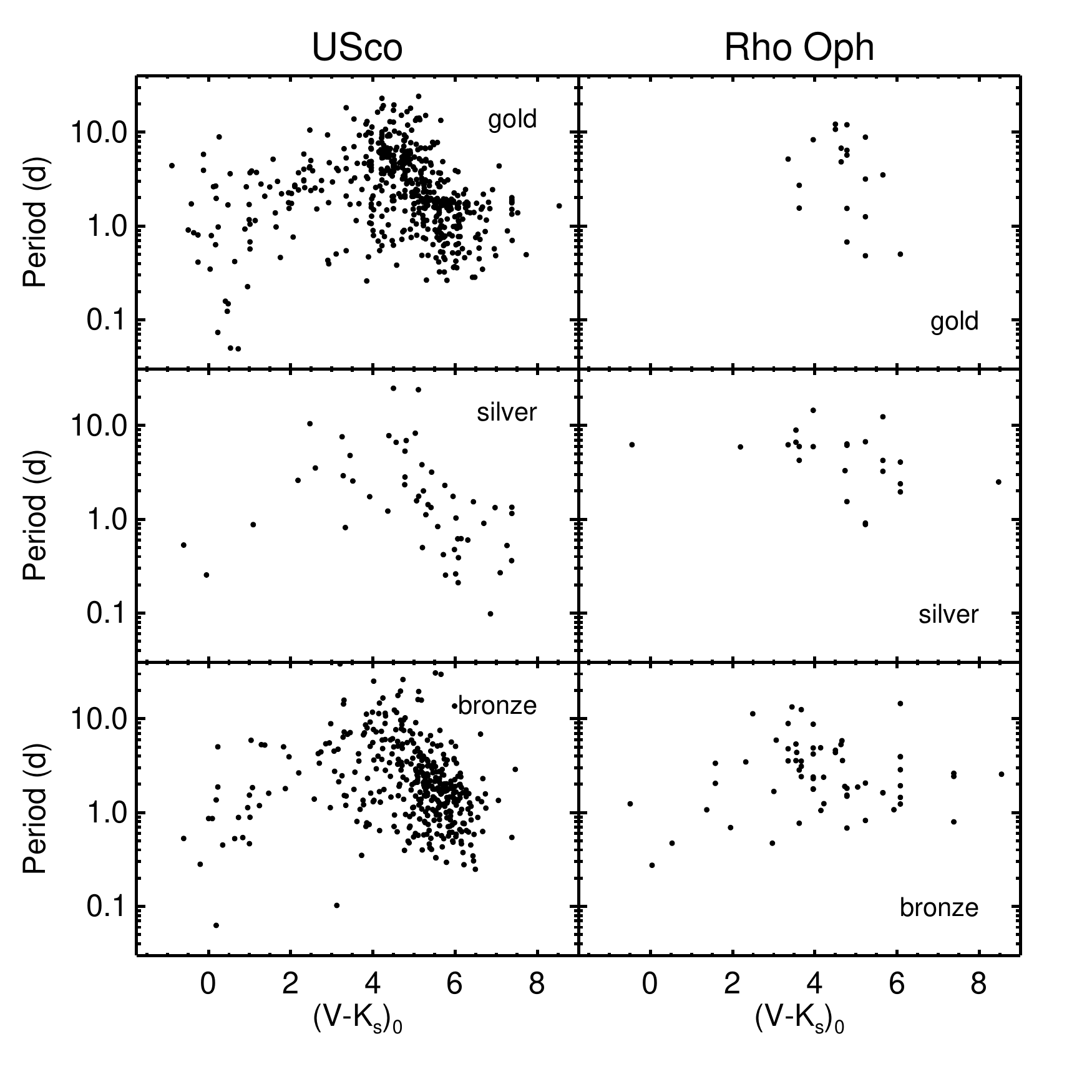}
\caption{$P$ vs.\ \vmkz\ with gold/silver/bronze members highlighted
for USco and \rop.
\label{fig:pvmkgsb}}
\end{figure}

Figures~\ref{fig:optcmdgsb} and \ref{fig:pvmkgsb} show the
distributions of the gold, silver, and bronze samples in the CMD and
the $P$ vs.\ \vmkz\ plots. The gold/silver/bronze membership status is
included in Table~\ref{tab:bigperiods}. For USco, there are 1133 total
members, 551 gold, 71 silver, and 511 bronze, with 94\%, 75\%, and
78\% periodic of the gold, silver, and bronzer samples, rsepectively. 
For \rop, there are 180 total members, 19 gold, 29 silver, and 132
bronze; the periodic fractions are 100\%, 83\%, and 49\%,
respectively. We expect to find a higher fraction of periodic LCs
among the member stars, which is reflected in the higher periodic
fraction of the gold samples, lower periodic fraction in the silver
samples, and still lower periodic fraction in the bronze samples.

\section{Non-Members}
\label{app:nm}

The process of finding probable cluster members leaves 1313 LCs (50\%
of the entire sample) that are tied to objects that we believe are
likely to be NM. However, since we analyzed all of these
light curves in the same way, we provide information on the remaining
1313 LCs here. As membership improves (e.g., Gaia DR2), membership may
be re-evaluated. In this section, we provide in Table~\ref{tab:bignm}
the same contents as in Table~\ref{tab:bigperiods} above (save for the
column on cluster membership), but for NM.  The values
included here should be good on the whole, but more outliers may be
present  because after a certain point, we focused on deep analysis of
the members.

In terms of the various metrics provided in Sec.~\ref{sec:obs},
spectral types are available for $\sim$14\% of the NM. WISE detections
are found for $\sim$30\% of the NM. Nine percent of the NM have clear
disks, and 4\% may have disks. For the entire sample, there are 413
(16\%) disk candidates, leaving 87\% without disks. This lower disk
fraction is consistent with a higher fraction of NMs in this sample.
About 34\% of the NM are periodic; the much lower fraction of periodic
sources is consistent with this sample largely being NM.  About a
third of the periodic LCs from the NMs are sinusoidal and therefore
likely to be starspots;  84\% of the periods are $<$10 d.

\floattable
\begin{deluxetable}{cp{13cm}}
\tabletypesize{\scriptsize}
%\rotate
\tablecaption{Contents of Table: Periods and Supporting Data for
USco and \rop\ Non-Members with K2 Light Curves\label{tab:bignm}}
\tablewidth{0pt}
\tablehead{\colhead{Label} & \colhead{Contents}}
\startdata
EPIC & Number in the Ecliptic Plane Input Catalog (EPIC) for K2\\
coord & Right ascension and declination (J2000) for target \\
othername & Alternate name for target \\
Vmag & V magnitude (in Vega mags), if observed\\
Kmag & \ks\ magnitude (in Vega mags), if observed\\
vmk-obs & \vmk, as directly observed (if $V$ and \ks\ exist), or as inferred (see text)\\
vmk-used & \vmk\ used (observed or inferred; see text)\\
ev-k & $E(V-K_s)$ adopted for this star (see \S~\ref{sec:dereddening}) \\
Kmag0 & dereddened $K_{s,0}$ magnitude (in Vega mags), as inferred (see \S\ref{sec:dereddening})\\
vmk-dered & $(V-K_s)_0$ (dereddened $V-K_s$), as inferred (see \S~\ref{sec:dereddening}; rounded to nearest 0.1 to emphasize the relatively low accuracy)\\
uncertaintycode & two digit code denoting origin of \vmk\ and \vmkz\
(see \S\ref{sec:litphotom} and \ref{sec:dereddening}):
First digit (origin of \vmk): 
1=$V$ measured directly from the literature (including SIMBAD) and $K_s$ from 2MASS; 
2=$V$ from APASS and $K_s$ from 2MASS;
3=\vmk\ inferred from Gaia $g$ and $K_s$ from 2MASS (see \S\ref{sec:litphotom});
4=\vmk\ inferred from Pan-STARRS1 $g$ and $K_s$ from 2MASS (see \S\ref{sec:litphotom});
5=\vmk\ inferred from membership work (see \S\ref{sec:membership}; rare); 
6=$V$ inferred from well-populated optical SED and $K_s$ from 2MASS (see \S\ref{sec:litphotom});
-9= no measure of \vmk.
Second digit (origin of $E(V-K_s)$ leading to \vmkz): 
1=dereddening from $JHK_s$ diagram (see \S\ref{sec:dereddening});
2=dereddening back to \vmkz\ expected for spectral type;
3=used median $E(V-K_s)$=0.7 (see \S\ref{sec:dereddening});
-9= no measure of  $E(V-K_s)$ \\
P1 & Primary period, in days (taken to be rotation period)\\
P2 & Secondary period, in days\\
P3 & Tertiary period, in days\\
P4 & Quaternary period, in days\\
Disk & Whether an IR excess (a disk) is present or not \\
DiskStart & Where the IR excess starts or the limit of our knowledge
of where there is no excess \\
single/multi-P & indicator of whether single or multi-period star\\
dd &  indicator of whether or not it is a double-dip LC (see
\S\ref{sec:LCPcats} and \ref{app:LCPcats})\\
ddmoving & indicator of whether or not it is a moving double-dip LC (see
\S\ref{sec:LCPcats} and \ref{app:LCPcats})\\
shch & indicator of whether or not it is a shape changer (see
\S\ref{sec:LCPcats} and \ref{app:LCPcats})\\
beat & indicator of whether or not the full LC has beating visible (see
\S\ref{sec:LCPcats} and \ref{app:LCPcats})\\
cpeak & indicator of whether or not the power spectrum has a complex,
structured peak and/or has a wide peak (see
\S\ref{sec:LCPcats} and \ref{app:LCPcats})\\
resclose & indicator of whether or not there are resolved close
periods in the power spectrum (see
\S\ref{sec:LCPcats} and \ref{app:LCPcats})\\
resdist & indicator of whether or not there are resolved distant
periods in the power spectrum (see
\S\ref{sec:LCPcats} and \ref{app:LCPcats})\\
pulsator & indicator of whether or not the power spectrum and period
suggest that this is a  pulsator (see
\S\ref{sec:LCPcats} and \ref{app:LCPcats})\\
\enddata
\end{deluxetable}

\clearpage

\section{Timescales}
\label{app:timescales}

As in papers I-IV, some LCs have some repeated patterns that we cannot
identify with certainty as a rotation period. These `timescales' tend
to be longer than most of the rotation periods. Sometimes, there is
not enough data to go $>$1 complete cycle. Table~\ref{tab:timescales}
summarizes the timescales for the stars out of the entire ensemble.
Note that some also appear in the list of periodic stars, but with a
shorter period that we believe to be the rotation period; the
longer-term variability is unlikely to be rotation.

\floattable
\begin{deluxetable}{ccccl}
\tabletypesize{\scriptsize}
\rotate
\tablecaption{Lists of Objects with Timescales\label{tab:timescales}}
\tablewidth{0pt}
\tablehead{\colhead{EPIC} & \colhead{RA, Dec (J2000)}
&\colhead{Timescale (d)}&\colhead{Cluster Membership}&\colhead{Notes} }
\startdata
          205195088&    155533.06-185526.8      &$\sim$35       &\ldots          & \\
          204230552&    155537.64-230910.0      &$\sim$10       &\ldots          & \\
          203284437&    155617.21-263817.1      &$\sim$13       &USco, silver    & \\
          203866225&    155710.94-243753.2      &$\sim$10       &\ldots          &listed as periodic in the master table; could legitimately be timescale instead        \\
          204422391&    155744.90-222351.2      &$\sim$35       &\ldots          & \\
          203799428&    155811.60-245313.2      &$\sim$20       &\ldots          & \\
          204533829&    155813.62-215652.3      &$\sim$9        &\ldots          & \\
          204475702&    155829.62-221111.9      &$\sim$42       &USco, bronze    & \\
          203549979&    155852.11-254538.7      &$\sim$21       &\ldots          & \\
          204081030&    155856.96-234436.0      &$\sim$24       &\ldots          &listed as periodic in the master table; could legitimately be timescale instead        \\
          203382710&    155912.91-261936.7      &$\sim$12       &\ldots          &    \\
          204083945&    160021.12-234354.0      &$\sim$19       &\ldots          &    \\
          204398735&    160142.53-222924.0      &$\sim$32       &\ldots          &    \\
          204539201&    160202.59-215531.5      &$\sim$35       &\ldots          &    \\
          204819741&    160220.29-204306.1      &$\sim$35       &\ldots          &    \\
          204136937&    160257.18-233110.5      &$\sim$11       &\ldots          &    \\
          204083104&    160312.80-234406.0      &$\sim$8        &\ldots          &    \\
          205217672&    160314.92-184823.3      &$\sim$5        &\ldots          &    \\
          205590575&    160642.32-163245.9      &$\sim$35       &\ldots          &    \\
          204824869&    160733.16-204141.5      &$\sim$25       &\ldots          &    \\
          205358744&    160836.58-180249.9      &$\sim$20       &USco gold      &also periodic with another real p                                                        \\
          205661119&    160844.17-160006.5      &$\sim$20       &\ldots          &  \\
          203800848&    160904.61-245254.2      &$\sim$20       &\ldots          &  \\
          205558283&    160920.41-164640.3      &$\sim$5        &USco, bronze    &    \\
          203799913&    160940.18-245307.0      &$\sim$26       &\ldots          &listed as periodic in the master table; could legitimately be timescale instead        \\
          205527458&    160954.85-165936.0      &$\sim$20       &\ldots          & \\
          203401494&    161014.08-261546.9      &$\sim$20       &\ldots          & \\
          202544694&    161023.37-291700.1      &$\sim$24       &\ldots          & \\
          203933869&    161044.64-242113.5      &$\sim$30       &\ldots          & \\
          204902065&    161052.59-202042.4      &$\sim$35       &USco, bronze    & \\
          204908189&    161113.30-201903.2      &$\sim$7        &USco, gold      & \\
          203866727&    161135.20-243745.5      &$\sim$35       &\ldots          & \\
          205660462&    161211.41-160025.5      &$\sim$25       &\ldots          & \\
          205599786&    161240.04-162841.0      &$\sim$20       &\ldots          & \\
          203678376&    161255.87-251858.3      &$\sim$44       &USco, bronze    & \\
          203649180&    161318.96-252503.2      &$\sim$20       &\ldots          & \\
          203329658&    161337.92-262949.9      &$\sim$35       &USco, bronze    & \\
          202742196&    161427.75-283157.4      &$\sim$20       &\ldots          & \\
          205562347&    161510.84-164457.9      &$\sim$20       &\ldots          & \\
          203379868&    161815.72-262012.2      &$\sim$15       &\ldots          & \\
          202639739&    161820.40-285500.1      &$\sim$20       &\ldots          &listed as periodic in the master table; could legitimately be timescale instead        \\
          205471638&    162136.58-172204.6      &$\sim$10       &USco, bronze    &  \\
          205659671&    162136.94-160048.2      &$\sim$30       &\ldots          &  \\
          205438739&    162250.74-173447.4      &$\sim$16       &\ldots          &  \\
          203870911&    162251.13-243645.4      &$\sim$18       &\ldots          &PERIODIC?                                                                              \\
          205498192&    162306.82-171131.5      &$\sim$35       &\ldots          &  \\
          203871153&    162312.58-243641.6      &$\sim$35       &\ldots          &  \\
          202891973&    162352.28-275913.5      &$\sim$7        &USco, bronze    &  \\
          203555168&    162415.50-254434.7      &$\sim$20       &USco, gold      &also periodic with another real p                                                        \\
          203929472&    162446.80-242221.0      &$\sim$15       &ROph, silver    &   \\
          205528856&    162505.96-165903.2      &$\sim$25       &USco, bronze    &   \\
          203927310&    162523.46-242254.1      &$\sim$10       &ROph, bronze    &   \\
          205556278&    162534.87-164732.5      &$\sim$35       &\ldots          &   \\
          203824768&    162550.54-244736.0      &$\sim$7        &\ldots          &   \\
          203899767&    162601.60-242944.9      &$\sim$35       &ROph, bronze    &   \\
          203803804&    162610.91-245215.5      &$\sim$10       &ROph, bronze    &   \\
          204885348&    162626.47-202520.0      &$\sim$10       &\ldots          &   \\
          203954364&    162637.13-241559.9      &$\sim$20       &\ldots          &   \\
          205191051&    162654.80-185643.4      &$\sim$15       &\ldots          &listed as periodic in the master table; could legitimately be timescale instead        \\
          203061783&    162702.52-272326.3      &$\sim$25       &\ldots          &  \\
          203445235&    162725.34-260653.5      &$\sim$35       &\ldots          &  \\
          203870022&    162738.32-243658.5      &$\sim$12       &ROph, silver    &  \\
          204640326&    162751.03-213001.0      &$\sim$23       &\ldots          &  \\
          205157696&    162810.26-190657.9      &$\sim$35       &USco, bronze    &  \\
          203979600&    162812.15-240927.4      &$\sim$25       &\ldots          &  \\
          203925443&    162814.75-242322.5      &$\sim$35       &\ldots          &  \\
          203944388&    162823.30-241833.8      &$\sim$10       &\ldots          &  \\
          203065453&    162823.57-272241.2      &$\sim$26       &\ldots          &  \\
          203905577&    162825.09-242819.6      &$\sim$20       &\ldots          &  \\
          202800225&    162850.18-281906.7      &$\sim$20       &USco, bronze    &  \\
          203844132&    162921.38-244310.5      &$\sim$24       &\ldots          &  \\
          203873636&    162932.33-243606.4      &$\sim$18       &\ldots          &  \\
          203782748&    162936.23-245653.0      &$\sim$20       &USco, silver    &  \\
          202816688&    163001.49-281529.5      &$\sim$10       &USco, bronze    &  \\
          205125047&    163056.32-191648.5      &$\sim$20       &USco, bronze    &  \\
          205073944&    163228.42-193200.0      &$\sim$13       &USco, bronze    &  \\
          205634171&    163235.87-161257.8      &$\sim$35       &\ldots          &  \\
          205074126&    163258.72-193156.3      &$\sim$35       &\ldots          &  \\
          204416606&    163428.77-222512.6      &$\sim$20       &\ldots          &  \\
          204276894&    163439.22-225814.8      &$\sim$35       &\ldots          &  \\
          205140183&    163519.69-191216.3      &$\sim$25       &\ldots          &  \\
          204248248&    163650.90-230456.6      &$\sim$10       &\ldots          &  \\
          203262247&    164019.37-264239.7      &$\sim$35       &\ldots          &  \\
          204478680&    164054.04-221030.2      &$\sim$35       &\ldots          &  \\
          202767998&    164113.58-282611.5      &$\sim$10       &\ldots          &  \\
          203762955&    164634.27-250111.2      &$\sim$20       &\ldots          &  \\
          205035963&    164917.71-194256.7      &$\sim$15       &\ldots          &listed as periodic in the master table; could legitimately be timescale instead        \\
          205011893&    165004.50-194949.8      &$\sim$10       &\ldots          &   \\
          205370897&    165104.97-175846.1      &$\sim$16       &\ldots          &   \\
          205389454&    165116.71-175243.4      &$\sim$35       &\ldots          &   \\
\enddata
\end{deluxetable}

\section{Scallop Shells and Flux Dips}
\label{app:batwings}

Stauffer \etal\ (2017) identified 23 M dwarfs in USco and \rop\ that
have unusually shaped phased K2 light curves.  Eleven of these stars
were described as ``scallop-shells", because their phased light curves
showed multiple scallops or undulations.  The remaining stars show one
or several flux dips in their phased light curves.   With few or no
exceptions, these stars are weak-lined T~Tauris, with no evidence of
on-going accretion or IR excesses from circumstellar dust.  All are
rapid rotators, with most having P $<$ 1 d, and many having P $<$ 0.5
d.  For about half of the group, their light curve morphology is
stable over the duration of the K2 campaign; for many of the others in
the group, small portions of the phased light curve shape change
abruptly during the campaign.   The phased light curve shapes have too
much small-scale structure to be explained by rotational modulation of
photospheric starspots. Instead, Stauffer \etal\ (2017) proposed that
these stars have clumpy tori of gas and dust located at the Keplerian
co-rotation radius, and that variable extinction for the fraction of
such stars where the torus is aligned with our line of sight gives
rise to the observed light curve morphologies.   David \etal\ (2017)
discuss one of these stars in detail. Stauffer \etal\ (2018) identify
an additional 8 stars in USco and \rop\ with similar light curve
morphologies, as well as three additional similar stars in Taurus
using K2 Campaign 13 data.

\begin{figure}[ht]
\epsscale{1.0}
\plotone{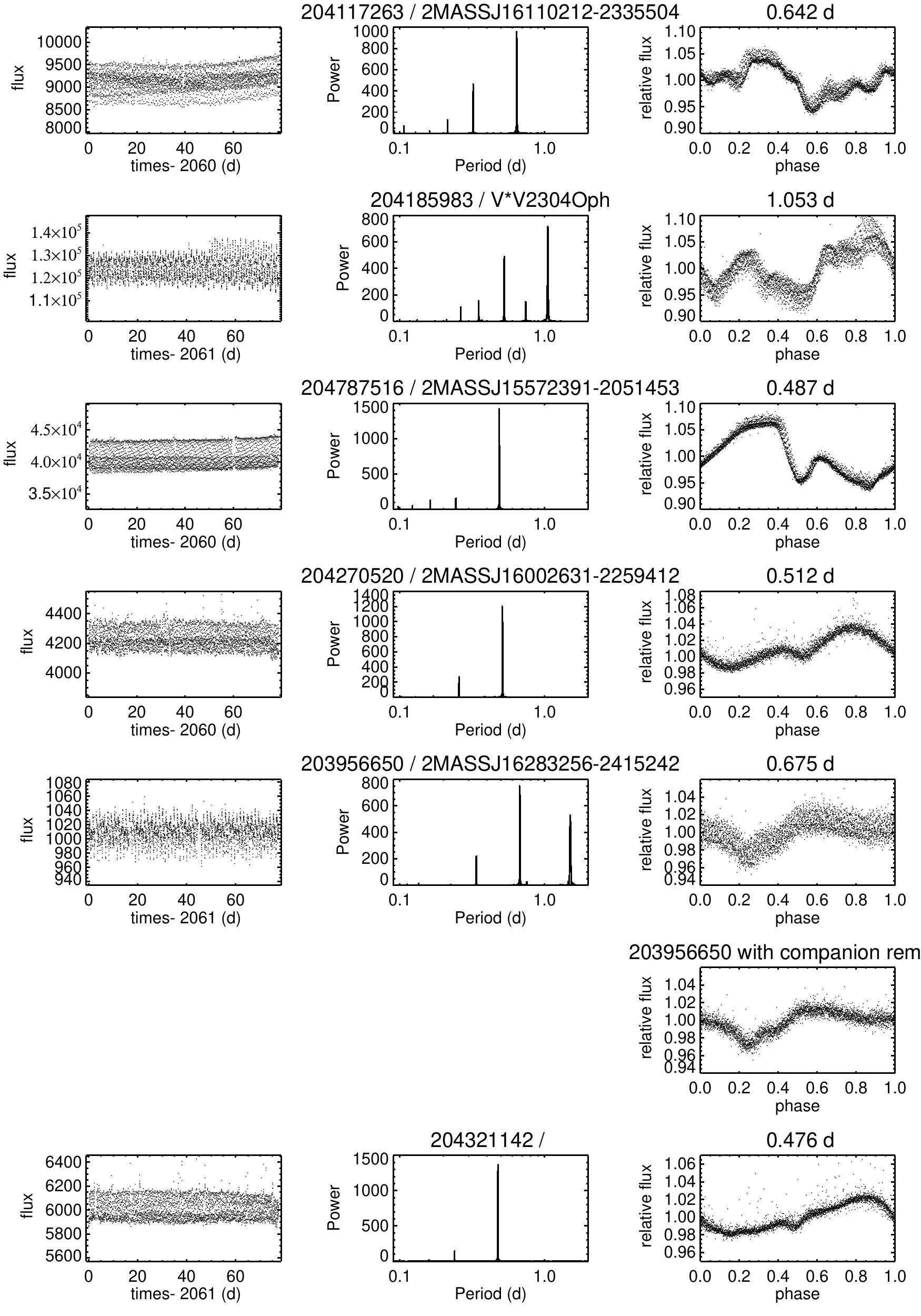}
\caption{Six examples of the scallop shell and flux dip stars. 
204117263/2MASSJ16110212-2335504 (USco `scallop', Stauffer \etal\ 2017),
204185983/V*V2304Oph (\rop\ `scallop', Stauffer \etal\ 2018),
204787516/2MASSJ15572391-2051453 (USco `flux dip' or possible EB, new here),
204270520/2MASSJ16002631-2259412 (USco `flux dip', new here),
203956650/2MASSJ16283256-2415242 (\rop, `flux dip', new here),
followed by a row in which only the scallop star's LC is left here,
having removed the companion's period, then 
204321142 (USco, `flux dip', new here).
\label{fig:batwinglcs}}
\end{figure}

\begin{figure}[ht]
\epsscale{0.7}
\plotone{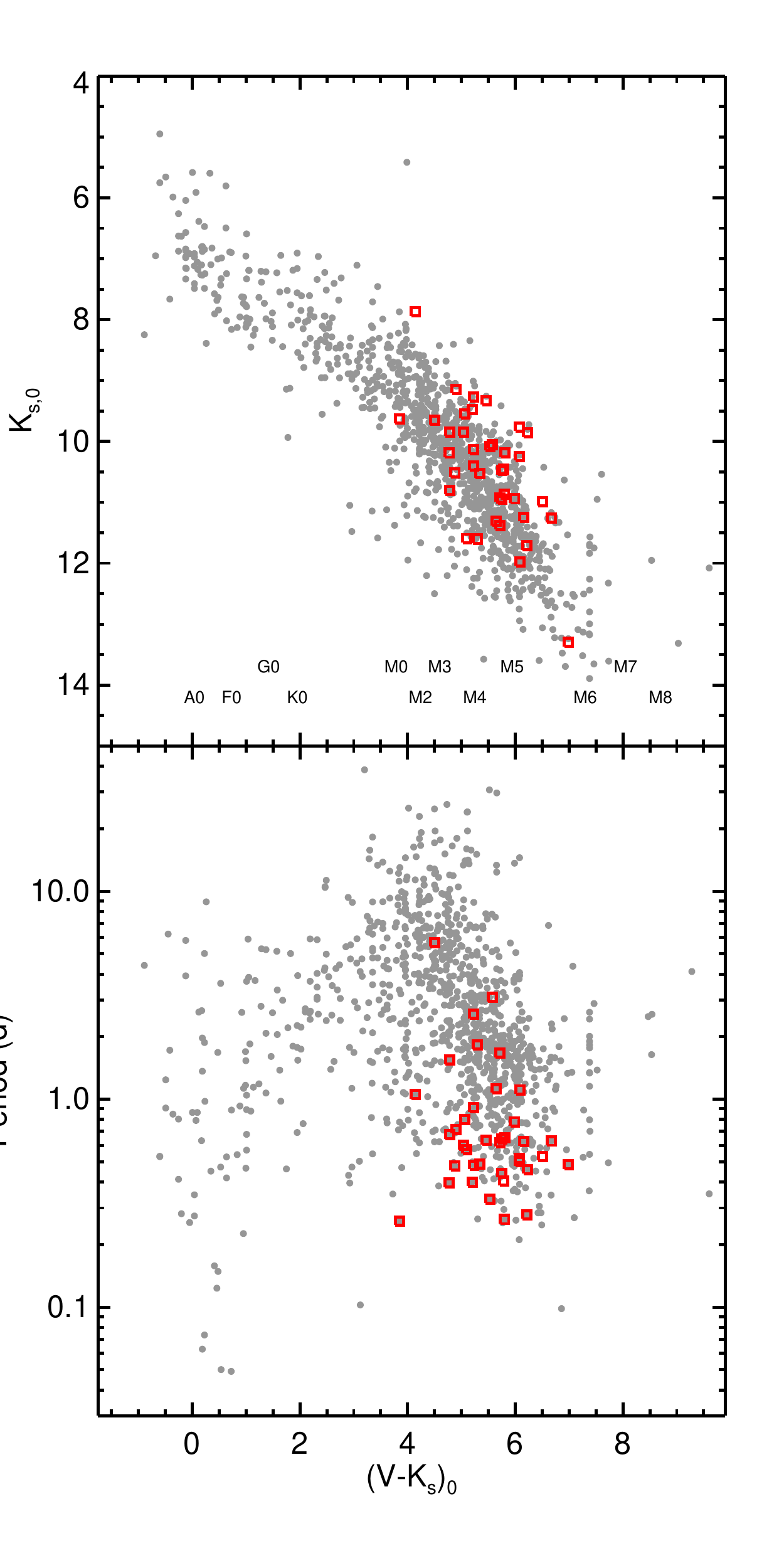}
\caption{Locations of the scallop shell and flux dip stars in the
$K_s,0$/\vmkz\ CMD and in the $P$ vs.~ \vmkz\ space. Grey dots are
members of USco; red boxes are the scallop shell and flux dip stars.
\label{fig:batwingcmd}}
\end{figure}

In Table~\ref{tab:batwingy}, we list all of the USco and \rop\ stars
identified as belonging to this category in Stauffer \etal\ (2017,
2018).  We also list 10 additional stars whose phased light curves
have features that may fall in this category, but where the features
are less obvious or the signal-to-noise ratio of the light curve is
relatively poor or where other K2 artifacts make interpretation of the
light curve difficult.    Figure~\ref{fig:batwinglcs} shows two
examples of the scallop-shell light curve class, plus light curves for
four of the stars we newly identify as possible members of the class
in Table~\ref{tab:batwingy}.  Figure~\ref{fig:batwingcmd} shows a
color-magnitude diagram and a period-color diagram highlighting all of
the stars in Table~\ref{tab:batwingy}.  These diagrams emphasize both
that all of the members of the class are low mass stars and that the
majority of them are very rapid rotators.

%\clearpage

\floattable
\begin{deluxetable}{cclccccccl}
\tabletypesize{\scriptsize}
\rotate
\tablecaption{Scallop Shell and Flux Dip Candidates in USco and \rop\
\label{tab:batwingy}}
\tablewidth{0pt}
\tablehead{\colhead{EPIC} & \colhead{Coordinates (J2000)}
&\colhead{Other name} & \colhead{\ks\ (mag)} & 
\colhead{\vmk\ (mag)} & \colhead{\ks$_0$\ (mag)} &\colhead{\vmkz\ (mag)} 
& \colhead{$P_1$ (d)\tablenotemark{a}} & \colhead{$P_2$ (d)\tablenotemark{a}} & \colhead{Where introduced}
}
\startdata
204882444 & 155505.13-202607.8 & 2MASSJ15550513-2026077 &    9.74 &    5.32 &    9.65 &    4.58 &    0.3829 & \ldots & new here\\
204918279 & 155625.09-201616.2 & 2MASSJ15562511-2016159 &    9.86 &    6.67 &    9.80 &    6.23 &    0.4594 &    0.4665* & Stauffer \etal\ (2017)\\
204787516 & 155723.90-205145.5 & 2MASSJ15572391-2051453 &    9.80 &    5.21 &    9.80 &    5.19 &    0.4868 & \ldots & new here \\
204066898 & 155836.21-234802.1 & UCAC3133-177729 &   10.19 &    5.47 &   10.10 &    4.77 &    0.3956* &    0.5386 & Stauffer \etal\ (2017)\\
203462615 & 155938.05-260323.6 & 2MASSJ15593807-2603233 &   10.25 &    6.24 &   10.23 &    6.08 &    0.5201* &    0.4421 & Stauffer \etal\ (2017)\\
204270520 & 160026.30-225941.4 & 2MASSJ16002631-2259412 &   11.34 &    6.49 &   11.28 &    6.07 &    0.5123 & \ldots & new here\\
204364515 & 160121.55-223726.7 & UCAC222721716 &   10.05 &    5.98 &   10.00 &    5.57 &    3.0863 &    1.4560* & Stauffer \etal\ (2017)\\
204897050 & 160140.97-202208.3 & UScoCTIO56 &   10.86 &    6.55 &   10.77 &    5.80 &    0.2639 & \ldots & Stauffer \etal\ (2017)\\
204099739 & 160839.08-234005.6 & 2MASSJ16083908-2340055 &    9.15 &    5.33 &    9.09 &    4.90 &    0.7158 &    0.7428* & Stauffer \etal\ (2018)\\
202724025 & 160856.94-283557.7 & 2MASSJ16085695-2835573 &    9.63 &    5.67 &    9.40 &    3.85 &    0.2595* &    0.2795 & Stauffer \etal\ (2017)\\
204783273 & 160937.06-205253.2 & 2MASSJ16093706-2052529 &   11.98 &    6.28 &   11.95 &    6.09 &    1.1105 & \ldots & new here\\
203849738 & 160952.87-244153.5 & 2MASSJ16095287-2441535 &   10.93 &    6.17 &   10.87 &    5.72 &    0.6190 & \ldots & Stauffer \etal\ (2017)\\
205024957 & 161010.99-194604.2 & [PBB2002]UScoJ161011.0-194603 &   11.38 &    6.19 &   11.32 &    5.71 &    1.6656 & \ldots & Stauffer \etal\ (2017)\\
205046529 & 161026.38-193951.0 & 2MASSJ16102639-1939513 &   10.40 &    7.77 &   10.08 &    5.23 &    2.5619 &    1.8358* & Stauffer \etal\ (2017)\\
203692610 & 161031.61-251602.1 & EPIC203692610 &   11.60 &    5.71 &   11.55 &    5.29 &    1.8210 & \ldots & Stauffer \etal\ (2017)\\
204060981 & 161056.18-234929.2 & \ldots &    9.47 &    5.80 &    9.40 &    5.20 &    0.3996* &    0.3802* & Stauffer \etal\ (2018)\\
204117263 & 161102.10-233550.7 & 2MASSJ16110212-2335504 &   10.95 &    6.11 &   10.91 &    5.75 &    0.6423 & \ldots & Stauffer \etal\ (2017)\\
205374937 & 161118.12-175728.9 & 2MASSJ16111813-1757286 &    9.33 &    6.03 &    9.26 &    5.46 &    0.6345* &    0.5436 & Stauffer \etal\ (2017)\\
203645396 & 161142.60-252551.4 & 2MASSJ16114261-2525511 &   11.26 &    7.37 &   11.17 &    6.67 &    0.6287* &    0.4928 & new here\\
204367193 & 161154.38-223649.3 & 2MASSJ16115439-2236491 &   13.30 &    7.45 &   13.24 &    6.99 &    0.4835 & \ldots & Stauffer \etal\ (2017)\\
203534383 & 161402.97-254853.3 & 2MASSJ16140298-2548531 &   11.71 &    7.05 &   11.60 &    6.22 &    0.2784* &    0.3234 & Stauffer \etal\ (2017)\\
205110559 & 161519.71-192107.0 & EPIC205110559 &   10.46 &    7.06 &   10.30 &    5.79 &    0.4031 & \ldots & Stauffer \etal\ (2017)\\
204143627 & 161559.25-232936.2 & 2MASSJ16155926-2329363 &   11.31 &    6.00 &   11.27 &    5.65 &    1.1250 & \ldots & Stauffer \etal\ (2017)\\
204082531 & 161620.10-234414.4 & 2MASSJ16162012-2344141 &   10.19 &    5.81 &   10.19 &    5.81 &    0.6513* &    0.4256 & new here\\
205267399 & 161841.87-183240.1 & \ldots &   10.08 &    5.70 &   10.06 &    5.53 &    0.3311 &    0.3344* & Stauffer \etal\ (2018)\\
203636498 & 162105.09-252742.6 & \ldots &   10.94 &    6.34 &   10.90 &    5.99 &    0.7794 & \ldots & Stauffer \etal\ (2018)\\
202873945 & 162139.90-280306.9 & EPIC202873945 &   11.24 &    7.38 &   11.09 &    6.16 &    0.6258 & \ldots & Stauffer \etal\ (2017)\\
204321142 & 162235.29-224743.3 & \ldots &   10.52 &    7.09 &   10.23 &    4.88 &    0.4763 & \ldots & new here\\
205483258 & 162324.52-171727.3 & 2MASSJ16232454-1717270 &    9.65 &    4.98 &    9.59 &    4.50 &    5.6670 & \ldots & Stauffer \etal\ (2017)\\
204296148 & 162451.80-225342.8 & EPIC204296148 &   10.99 &    8.02 &   10.79 &    6.50 &    0.5314* &    0.4717 & Stauffer \etal\ (2017)\\
204185983 & 162552.83-231936.3 & V*V2304Oph &    7.87 &    6.03 &    7.63 &    4.15 &    1.0529 & \ldots & Stauffer \etal\ (2018)\\
203897692 & 162556.09-243014.8 & 2MASSJ16255609-2430148 &    9.76 &    9.17 &    9.23 &    5.00 &    0.5011 &    0.6043* & Stauffer \etal\ (2018)\\
203354381 & 162627.86-262515.1 & \ldots &    9.85 &    5.66 &    9.77 &    5.03 &    0.5993 & \ldots & Stauffer \etal\ (2018)\\
203962559 & 162650.48-241352.2 & 2MASSJ16265048-2413522 &   10.80 &    7.93 &   10.44 &    5.08 &    1.5402 & \ldots & Stauffer \etal\ (2017)\\
203821589 & 162759.96-244819.2 & 2MASSJ16275996-2448193 &    9.27 &    8.15 &    8.84 &    4.83 &    0.9105 &    0.6677* & Stauffer \etal\ (2018)\\
203956650 & 162832.56-241524.4 & 2MASSJ16283256-2415242 &    9.84 &   10.46 &    8.97 &    3.70 &    0.6752* &    1.5052 & new here \\
203927435 & 162843.04-242252.3 & 2MASSJ16284304-2422522 &   10.14 &    8.34 &    9.68 &    4.74 &    0.4820* &    0.4162 & Stauffer \etal\ (2017)\\
203050730 & 163105.78-272546.4 & 2MASSJ16310579-2725460 &   10.53 &    5.62 &   10.49 &    5.34 &    0.4865* &    0.7740 & Stauffer \etal\ (2017)\\
203185083 & 163435.13-265803.2 & 2MASSJ16343514-2658030 &   10.48 &    6.26 &   10.41 &    5.75 &    0.4401 & \ldots & Stauffer \etal\ (2017)\\
202687442 & 163458.64-284410.4 & \ldots &   11.59 &    5.80 &   11.50 &    5.10 &    0.5727 & \ldots & new here\\
204884822 & 163506.25-202528.5 & 2MASSJ16350625-2025282 &    9.55 &    5.72 &    9.46 &    5.07 &    0.7983 & \ldots & new here\\
\enddata
\tablenotetext{a}{An asterisk denotes that, in the case of multiple periods, this is the period that has scallop shell or flux dip candidate properties.}
\end{deluxetable}

\section{LC and Periodogram Categories}
\label{app:LCPcats}

In paper II and IV, we classified the LC and periodogram shapes; we
use these same categorizations here.
Briefly, the classes we presented are summarized here (see papers II
or IV for examples):
\begin{itemize}
\item Single period -- only one period we believe to be real in the
LC; interpreted as arising from spots/spot groups rotating into and out of
view.
\item Multi-period -- more than one period we believe to be real in
the LC; interpretation varies (see other classifications).
\item Double dip -- two peaks appear in the periodogram, but only one
period is real, with the phased LC having two dips or humps (with
different shapes) per cycle; interpreted as arising from two
spots/spot groups rotating into and out of view.
\item Moving double dip -- the phased LC has two dips or humps per
cycle, but a minimum or maximum of one dip/hump moves with respect to
the other; interpreted as arising from surface differential rotation, where
one spot/spot group is moving with respect to another spot/spot group,
or spot/spot group evolution. 
\item Shape changer -- the shape of the LC changes over time;
interpreted as arising from spot/spot group evolution and/or surface
differential rotation.
\item Beaters -- the reduced LC (final fluxes, just prior to period
searching) has signatures of two periods beating (e.g., changing
envelope over the campaign); interpreted as arising from spot/spot
group evolution and/or surface differential rotation, but could also
be nearly synchronized binaries, with one spot per star.
\item Complex peak -- the peak in the periodogram is wider than
expected for that period, or it has multiple maxima within the main
peak; interpreted as arising from spot/spot
group evolution and/or surface differential rotation, but could also
be nearly synchronized binaries, with one spot per star.
\item Resolved, close peaks -- two distinct peaks in the periodogram,
close together; interpreted as arising from binarity (one spot per
star) or from surface differential rotation with spots/spot groups at
different latitudes. (Note that resolved close peaks in M stars are
more likely to be binaries, and resolved close peaks in earlier type
stars are more likely to be surface differential rotataion.)
\item Resolved, distant peaks -- two distinct peaks in the periodogram,
far apart; interpreted as arising from binarity (one spot per
star). 
\item Scallops and flux dips -- narrow flux dips, unusual
but repeatable shapes in their phased LCs, some of which change shape
slightly after a flare; interpreted as arising from orbiting clouds of
material and/or dusty debris near the Keplerian co-rotation radius
(see Stauffer \etal\ 2017, 2018). 
\item Pulsator -- multiple peaks in the periodogram at very short
periods; interpreted as due to pulsation.
\end{itemize}

\end{document}